\documentclass[%
reprint,
amsmath,amssymb,
aps,
]{revtex4-1}

\usepackage{graphicx}% Include figure files
\usepackage{dcolumn}% Align table columns on decimal point
\usepackage{bm}% bold math
\usepackage[colorlinks=true,linkcolor=blue,anchorcolor=blue,citecolor=blue,filecolor=blue,menucolor=blue,runcolor=blue,urlcolor=blue]{hyperref}
\usepackage[usenames, dvipsnames]{color}
\usepackage[T1]{fontenc}

\newcommand{\parr}[2]{\frac{\partial{#1}}{\partial{#2}}}

\newcommand{\txt}[1]{\textrm{#1}}

\def\bnabla{\bm{\nabla}}

\newcommand{\non}{\nonumber}

\def\pdens{\frac{d^3p}{(2\pi)^3}~}

\newcommand{\eps}{\varepsilon}

\begin{document}

%\preprint{APS/123-QED}

\title{Phase transitions and critical behavior in hadronic transport \\with a relativistic density functional equation of state}

%\title{Phase transitions and critical behavior in hadronic transport \\with a relativistic density functional equation of state of dense nuclear matter}

%\title{Relativistic density functional approach to the equation of state of nuclear matter with application to hadronic transport}% Force line breaks with \\

\author{Agnieszka Sorensen}
 \email{agnieszka.sorensen@gmail.com}
\affiliation{%
 Department of Physics and Astronomy, University of California, Los Angeles, CA 90095, US
}%
%\collaboration{}%\noaffiliation

\author{Volker Koch}
\affiliation{%
Lawrence Berkeley National Laboratory, 1 Cyclotron Rd, Berkeley, CA 94720, US
}%
%\collaboration{}%\noaffiliation


\begin{abstract}
We develop a flexible, relativistically covariant parametrization of the dense nuclear matter equation of state suited for inclusion in computationally demanding hadronic transport simulations. Within an implementation in the hadronic transport code \texttt{SMASH}, we show that effects due to bulk thermodynamic behavior are reproduced in dynamic hadronic systems, demonstrating that hadronic transport can be used to study critical behavior in dense nuclear matter, both at and away from equilibrium. We also show that two-particle correlations calculated from hadronic transport simulation data follow theoretical expectations based on the second-order cumulant ratio, and constitute a clear signature of the crossover region above the critical point. 
%
%\begin{description}
%\end{description}
\end{abstract}

\pacs{Valid PACS appear here}% PACS, the Physics and Astronomy
\maketitle

%\tableofcontents

\section{Introduction}
\label{introduction}

Uncovering the phase diagram of QCD matter is one of the major goals of heavy-ion collision research, and the founding reason behind the ongoing Beam Energy Scan (BES) program at the BNL Relativistic Heavy Ion Collider (RHIC). Current understanding of the evolution that QCD matter undergoes at extreme conditions is facilitated by numerous experimental and theoretical advancements to date. The importance of quark and gluon degrees of freedom for the dynamics of very high-energy collisions is strongly supported by comparisons of experiment to theoretical models \cite{Adams:2005dq, Adcox:2004mh}, and suggests that the quark-gluon plasma (QGP) is produced in these events. Collective behavior of matter created in such collisions has been measured \cite{Ackermann:2000tr} %first v2 paper at RHIC
and reproduced in hydrodynamics simulations \cite{Teaney:2000cw, Teaney:2001av}, indicating that for a considerable fraction of a heavy-ion collision's evolution, it can be thought of as a thermal system described by an equation of state (EOS). The exact nature of the transition between the QGP and a hadron gas is studied within a number of approaches. At finite temperature and negligible baryon chemical potential, first-principle calculations in lattice QCD (LQCD) predict a transition of the crossover type \cite{Aoki:2006we}. This result has been further supported with a Bayesian inference approach \cite{Pratt:2015zsa}, where the range of equations of state most consistent with experimental data at high energies has been identified and shown to include the LQCD EOS. On the other hand, numerous chiral effective field theory models predict that at finite baryon number density the transition between hadronic and quark-gluon matter is of the first order \cite{Stephanov:2004wx}. If this is the case, the phase diagram of QCD matter contains a QGP-hadron coexistence line, ending in a critical point. 

The search for signatures of the QCD critical point is premised on the ability to experimentally uncover a number of effects born out in systems of immense complexity. Some of these predicted signatures involve light nuclei production \cite{Sun:2017xrx,Sun:2018jhg}, enhanced multiplicity fluctuations of produced hadrons \cite{Stephanov:1998dy, Stephanov:1999zu, Koch:2008ia}, the slope of the directed flow \cite{Rischke:1995pe, Stoecker:2004qu}, or Hanbury-Brown-Twiss (HBT) interferometry measurements \cite{Hung:1994eq}, and their dependence on the beam energy. Often, the magnitudes of these effects and their interaction with various other experimental signals, as well as the influence of the finite time of the collision or baryon number conservation remain elusive to purely theoretical predictions. In consequence, a clear interpretation of the experimental data will have to be supported by comparisons with results of dynamical simulations of heavy-ion collisions, developed to correctly account for the complex evolution of relevant observables.

Modern heavy-ion collision simulations consist of multiple stages, starting with an initial state model, through relativistic viscous hydrodynamics utilizing a chosen EOS to describe the bulk behavior of QGP from thermalization until particlization, and ending with a hadronic transport code \cite{Petersen:2008dd, Schenke:2020mbo}. Notably, with a few exceptions (see
e.g.\ \cite{Nara:2016hbg}), hadronic afterburners typically neglect hadronic potentials, which means that the role of many-body interactions in the hadronic stage is largely unexplored. This raises the possibility that transport simulations may be missing effects likely to become increasingly important at higher baryon densities, where both the mean-field effects and the time that the system spends in a hadronic state are substantial. In particular, mean-field hadronic interactions may significantly influence the system's evolution, including the diffusion dynamics which is a relevant factor in the propagation of signals for the existence of the critical point \cite{Asakawa:2019kek}. 

Furthermore, since the correct QCD EOS at finite chemical potential is not known from first principles, it needs to be inferred from systematic model comparisons with experimental data. A consistent treatment of the entire span of a hybrid heavy-ion collision simulation requires employing hadronic interactions that reproduce properties of a particular EOS used in the hydrodynamic stage, such as the position of the QCD critical point. While there is a strong theoretical effort to model different variants of the QCD EOS with criticality \cite{Parotto:2020fwu,Karthein:2021nxe}, intended for use in hydrodynamic simulations, often the hadronic part of a heavy-ion collision simulation, if it at all takes hadronic potentials into account, includes only mean-field interactions corresponding to the behavior of ordinary nuclear matter without the possible QGP phase transition \cite{Nara:2016hbg}. As a result, there is a need for a flexible hadronic EOS that on one hand can be easily parameterized to reflect a desired set of properties of the modeled QCD phase transition, and on the other identifies corresponding relativistic single-particle dynamics that can be feasibly implemented in an afterburner.

Here we propose an approach to this problem in which the EOS of nuclear matter and the corresponding single-particle equations of motion are both obtained from a relativistic density functional with fully parameterizable vector-current interactions. Besides the obvious requirements of Lorentz covariance and thermodynamic consistency, the constructed model is constrained to agree with the known behavior of ordinary nuclear matter. Therefore each of the obtained EOSs includes the nuclear liquid-gas phase transition with its experimentally observed properties, in addition to a possible phase transition at high baryon density. The flexibility of the constructed family of EOSs enables systematic studies (e.g.\ using Bayesian analysis) of effects of different dense nuclear matter EOS on final state observables, facilitating meaningful comparisons of simulation results with experimental data.

Furthermore, we implement our mean-field model in the hadronic transport code \texttt{SMASH} \cite{Weil:2016zrk}, and verify that the obtained single-particle equations of motion reproduce bulk behavior expected from the underlying EOS. In particular, we study the evolution of systems undergoing spontaneous separation inside the spinodal region of the phase transition and in the vicinity of the critical point, and we investigate observables carrying signals of collective behavior as well as the effect of finite number statistics on particle number distributions.

This paper is organized as follows: Sections \ref{formalism} and \ref{theoretical_results} give a pedagogical presentation of the model and the corresponding theoretical results. Section \ref{implementation_in_SMASH} briefly reviews the implementation of the model in the hadronic transport code \texttt{SMASH}, while Sec.\ \ref{analysis} discusses the analysis methods used. Section \ref{simulation_results} presents and discusses results of simulations under various conditions. Finally, Sec.\ \ref{summary_and_outlook} provides a summary and an outlook to future developments.

\section{Formalism}
\label{formalism}

\subsection{Background}
\label{background}

Studying nuclear matter requires knowledge of nucleon-nucleon and, more generally, hadronic interactions, which currently cannot be obtained from first principle calculations. In view of this, phenomenological approaches are employed, in which the behavior of nuclear matter is described in terms of effective degrees of freedom. A large class of these approaches uses self-consistent models based on density functional theory (DFT). Such models are a starting point for numerous Skyrme-like potentials of varying degree of complexity which are successfully applied in low-energy nuclear physics \cite{Bender:2003jk}. 

Alternatively, one can employ Landau Fermi-liquid theory \cite{Landau:1957tp}, which can be shown to lead to the same results as various phenomenological models at the mean-field level (see e.g.\ \cite{Matsui:1981ag, Brown:1971zza}), and which combines certain desirable features of other approaches. On one hand, similarly as in DFTs, in Landau Fermi-liquid theory the relevant physics is entirely encoded in the postulated energy density of the system. The theory then allows one to describe the system's deviations from equilibrium (such as energy of an excitation or particle-particle interactions) as well as corresponding bulk properties, encoded in phenomenological parameters. On the other hand, as in many Lagrangian-based, self-consistent approaches at the mean-field level, the main degrees of freedom of the theory are quasiparticles. This means that the role of interactions is embedded in the properties of quasiparticles (which can be thought of as dressed nucleons) and in the quasiparticle distribution function (for a definition of the quasiparticle distribution function as well as its limitations, see Appendix \ref{quasiparticle_distribution_function}).

The Landau Fermi-liquid theory is a very convenient starting point for a phenomenological approach to the nuclear matter EOS, and in particular for applications to hadronic transport simulations, where we want to develop a model that is at the same time flexible and numerically efficient. In constructing our framework, we are additionally guided by the following requirements: First, we need a formalism in which the baryon number density, a natural variable for hadronic transport simulations, is a dynamical variable of the theory (as opposed to theories in which the baryon chemical potential is evolved in time). Moreover, we are guided by the fact that vector-type interactions are more convenient for numerical evaluation of mean-field potentials than, for example, scalar-type interactions, which require solving a self-consistent equation at each point where mean-fields are calculated. Finally, we want to obtain a family of EOSs that on the one hand reproduces the known properties of ordinary nuclear matter, and on the other allows one to postulate and explore critical behavior in dense nuclear matter over vast regions of the phase diagram. The former will ensure that the model takes into the account the known experimental behavior of nuclear matter, while the latter will allow us to meaningfully compare the influence of different EOSs on observables. Such comparisons can be made, among others, through Bayesian analysis \cite{Novak:2013bqa,Bernhard:2016tnd}.

\subsection{Relativistic vector density functional (VDF) model}

With the aforementioned goals in mind, we adopt the relativistic Landau Fermi-liquid theory \cite{Baym:1975va} with vector-density--dependent interactions as the basis for constructing a vector density functional (VDF) model of the dense nuclear matter EOS. Starting from a postulated energy density of the system, we will derive the single-particle equations of motion, the energy-stress tensor, and the corresponding thermodynamic relations. To simplify the notation, we will introduce a VDF model with a single number-current--dependent interaction term; however, it is straightforward to generalize to a model with multiple interaction terms of the same kind, which we do at the end of this subsection. Some of the details of the derivation can be found in Appendix \ref{model_derivations}.

We introduce the energy density $\mathcal{E}_{(1)}(x)$ of a system composed of one species of fermions, interacting through a single mean-field vector interaction term, 
\begin{eqnarray}
&&\mathcal{E}_{(1)}(x) = g \int \pdens \epsilon_{\textrm{kin}} ~ f_{\bm{p}}  + C_1 \big(j_{\mu} j^{\mu}\big)^{\frac{b_1}{2} - 1} \big(j^0\big)^2  \non\\
&& \hspace{20mm}  - g^{00}~C_1 \left( \frac{b_1 - 1}{b_1} \right) \big(j_{\mu}j^{\mu}\big)^{\frac{b_1}{2}} ~,
\label{energy_density_postulated}
\end{eqnarray}
where $g$ is the degeneracy, $\epsilon_{\textrm{kin}}$ is the kinetic energy of a single particle,
\begin{eqnarray}
\epsilon_{\txt{kin}} = \sqrt{\left(\bm{p} - C_1 \big( j_{\mu}j^{\mu}\big)^{\frac{b_1}{2} -1} \bm{j} \right)^2 + m^2}~,
\label{kinetic_energy}
\end{eqnarray}
$\bm{j}$ and $j^0$ are the spatial and temporal component of the number current $j^{\mu}$, given by
\begin{eqnarray}
\bm{j}(x) = g\int \frac{d^3p}{(2\pi)^3} ~ \frac{\bm{p} - C_1 \big(j_{\mu}j^{\mu}\big)^{\frac{b_1}{2} - 1} \bm{j}}{\epsilon_{\txt{kin}}}~ f_{\bm{p}}
\label{current_spatial_postulated}
\end{eqnarray}
and
\begin{eqnarray}
j^{0}(x) = g\int \frac{d^3p}{(2\pi)^3} ~ f_{\bm{p}}~,
\label{current_temporal_postulated}
\end{eqnarray}
respectively, $m$ is the particle mass, $f_{\bm{p}}$ is the quasiparticle distribution function, and finally $C_1$ and $b_1$ are constants specifying the interaction, as of yet undetermined. The energy density, Eq.\ (\ref{energy_density_postulated}), is constructed as the $00$ component of the energy-momentum tensor and transforms accordingly. The interaction terms depend both on the local frame number density $j_0$ and the relativistic invariant $j_{\mu}j^{\mu} = n^2$, where $n$ denotes the rest frame number density. The quasiparticle energy, defined in the Landau Fermi-liquid theory as the functional derivative of the energy density, is given by (see Appendix \ref{quasiparticle_energy_derivation})
\begin{eqnarray}
\eps_{\bm{p}} \equiv \frac{\delta \mathcal{E}_{(1)}}{\delta f_{\bm{p}}} = \epsilon_{\txt{kin}} + C_1 \big(j_{\mu} j^{\mu} \big)^{\frac{b_1}{2} - 1} j_0 ~.
\label{quasiparticle_energy}
\end{eqnarray} 
Note that the quasiparticle energy is equivalent to the single-particle Hamiltonian, $\eps_{\bm{p}} = H_{(1)}$.

To simplify the notation, we introduce a vector field,
\begin{eqnarray}
A^{\lambda} (x; C_1, b_1) \equiv C_1 \big( j_{\mu} j^{\mu}\big)^{\frac{b_1}{2} - 1} j^{\lambda}~.
\end{eqnarray}
In the following derivation we will suppress the dependence on $C_1$ and $b_1$ and refer to this variable simply as $A^{\lambda} (x)$, which allows us to concisely write
\begin{eqnarray}
\eps_{\bm{p}} = \sqrt{\big(\bm{p} - \bm{A}\big)^2 + m^2} + A_0
\label{quasiparticle_energy_short}
\end{eqnarray}
and
\begin{eqnarray}
\mathcal{E}_{(1)}(x) = g \int \pdens \eps_{\bm{p}} ~  f_{\bm{p}} - g^{00}~ \left( \frac{b_1 - 1}{b_1} \right) A_{\lambda} j^{\lambda} ~.
\label{energy_density_short}
\end{eqnarray}

Given Eq.\ (\ref{quasiparticle_energy_short}), the equations of motion follow immediately from Hamilton's equations,
\begin{eqnarray}
&& \frac{dx^i}{dt} \equiv - \parr{H_{(1)}}{p_i} = - \parr{\eps_{\bm{p}}}{p_i} = \frac{p^i - A^i}{\epsilon_{\txt{kin}}}~,
\label{equation_of_motion_x} \\
&& \frac{dp^i}{dt} \equiv \parr{ H_{(1)} }{x_i} = \parr{ \eps_{\bm{p}} }{x_i} = \frac{(p^k - A^k)}{ \epsilon_{\txt{kin}}} \parr{A_k}{x_i} + \parr{A_0}{x_i} 
\label{equation_of_motion_p}~. 
\end{eqnarray}
Inserting Eqs.\ (\ref{equation_of_motion_x}) and (\ref{equation_of_motion_p}) into the Boltzmann equation gives
\begin{eqnarray}
&& \parr{f_{\bm{p}}}{t} - \parr{\eps_{\bm{p}}}{p_i}  \parr{f_{\bm{p}}}{x^i} +  \parr{ \eps_{\bm{p}} }{x_i} \parr{f_{\bm{p}}}{p^i} =  \mathcal{I}_{\txt{coll}}~,
\label{Boltzmann_equation}
\end{eqnarray} 
where $ \mathcal{I}_{\txt{coll}}$ is the collision term. Multiplying both sides of Eq.\ (\ref{Boltzmann_equation}) by $X = \{1, \eps_{\bm{p}}, p^{j}  \}$ and integrating over $g \int \frac{d^3p}{(2\pi)^3}$ yields the conservation laws for particle number ($X=1$), energy ($X = \eps_{\bm{p}}$), and momentum ($X = p^j$). In particular, one notices that the particle number conservation,
\begin{eqnarray}
\hspace{-5mm}\parr{}{t} ~g\int \frac{d^3p}{(2\pi)^3} ~ f_{\bm{p}} + \partial_i  ~g\int \frac{d^3p}{(2\pi)^3} ~ \frac{p^i - A^i}{\epsilon_{\textrm{kin}}} ~ f_{\bm{p}}= 0 ~,
\end{eqnarray}
confirms that the baryon number current and density, Eqs.\ (\ref{current_spatial_postulated}) and (\ref{current_temporal_postulated}), are correctly defined. The obtained conservation laws for energy and momentum allow us to identify the energy-momentum tensor, whose components are density and flux of energy and momentum in spacetime,
\begin{eqnarray}
&& T^{00} = \mathcal{E}_{(1)} ~, \\
&& T^{0i} = g\int \pdens \eps_{\bm{p}} \frac{p^i - A^i}{\epsilon_{\txt{kin}}}~  f_{\bm{p}} ~, \\
&& T^{i0} = g\int \pdens p^i ~  f_{\bm{p}}~, \\
&& T^{ij} = g\int \pdens p^i \frac{p^j - A^j}{\epsilon_{\txt{kin}}}~ f_{\bm{p}}  \non \\
&& \hspace{15mm} +~  g^{ij} \left( \mathcal{E}_{(1)} - g \int \pdens \eps_{\bm{p}} ~  f_{\bm{p}}  \right)~.
\end{eqnarray}
One can show that $T^{\mu\nu}$ has the correct transformation properties under a Lorentz boost (details of this calculation, for a general case of the relativistic Landau Fermi-liquid theory without a specified form of the interactions, can be found in \cite{Baym:1975va}). Additionally, energy and momentum conservation, $\partial_{\nu} T^{\mu\nu} = 0$, is ensured by construction. Using Eq.\ (\ref{current_spatial_postulated}), it can be readily verified that $T^{0i} = T^{i0}$.

Having derived the properties of the VDF model with one interaction term, we can easily extend the formalism to an arbitrary number of interaction terms. Here, we are dealing with multiple vector fields labeled by the index $n$,
\begin{eqnarray}
A_n^{\lambda} (x; C_n, b_n) \equiv C_n \big( j_{\mu} j^{\mu}\big)^{ \frac{b_n}{2} - 1} j^{\lambda} ~,
\end{eqnarray}
in terms of which the energy density is given by
\begin{eqnarray}
&&\mathcal{E}_{(N)}(x) = g \int \pdens \eps_{\bm{p}}^{(N)} ~  f_{\bm{p}}  \non\\
&& \hspace{20mm} - g^{00}~ \sum_{n=1}^N \left( \frac{b_n - 1}{b_n} \right) A_n^{\lambda} j_{\lambda} .
\label{energy_density_generalized}
\end{eqnarray}
We note that taking $N=1$, $b_1 = 2$ leads to the form of the vector interaction known well, e.g., from the Walecka model \cite{Walecka:1974qa, Chin:1974sa}, corresponding to the mean-field approximation of a two-particle interaction mediated by a vector meson. (In fact, an alternative description of the mean-field approximation to the Walecka model in terms of the relativistic Landau Fermi-liquid theory is given in \cite{Matsui:1981ag}.) Similarly, evaluating (\ref{energy_density_generalized}) in the rest frame and taking $N=2$, $b_1 = 2$, and $b_2 = 3$ ($b_2 = \frac{13}{6}$) results in the interaction of the same form as a commonly used stiff (soft) parametrization of the Skyrme model (see e.g.\ \cite{Kruse:1985hy}). Indeed, in postulating the form of the energy density, Eq.\ (\ref{energy_density_postulated}) or Eq.\ (\ref{energy_density_generalized}), we took inspiration from the form of the energy density in models mentioned above, and we made sure that our expression reproduces the terms appearing in these models when particular coefficients and powers of the interaction terms are used. In contrast to these approaches, however, our model allows for arbitrary interaction parameters, including the number of interaction terms as well as powers of number density characterizing the interactions, that remain unspecified until a later time when we fit them to match chosen properties of nuclear matter. 

The generalization of the remaining parts of the VDF model is straightforward, and in particular we arrive at the quasiparticle energy,
\begin{eqnarray}
\eps_{\bm{p}}^{(N)} = \sqrt{\bigg(\bm{p} - \sum_{n=1}^N\bm{A}_n\bigg)^2 + m^2} + \sum_{n=1}^N A^0_n ~,
\label{quasiparticle_energy_generalized}
\end{eqnarray}
and the equations of motion,
\begin{eqnarray}
&& \frac{dx^i}{dt} = \frac{p^i - \sum_{n=1}^N (A_n)^i}{\epsilon^{(N)}_{\txt{kin}}}~,
\label{equation_of_motion_x_generalized} \\
&& \frac{dp^i}{dt}  = \frac{\Big(p^k - \sum_{n=1}^N (A_n)^k\Big)}{ \epsilon^{(N)}_{\txt{kin}}} \bigg( \sum_{n=1}^N \parr{(A_n)_k}{x_i}\bigg)  \non\\
&& \hspace{15mm} +~  \sum_{n=1}^N \parr{A_n^0}{x_i}  ~. 
\label{equation_of_motion_p_generalized}
\end{eqnarray}
We stress that the generalization to $N$ interaction terms preserves the conservation laws and the relativistic covariance of the $T^{\mu\nu}$ tensor.

Finally, the equations of motion, Eqs.\ (\ref{equation_of_motion_x_generalized}) and (\ref{equation_of_motion_p_generalized}), can be rewritten in a manifestly covariant way.  First, we rewrite Eq.\ (\ref{quasiparticle_energy_generalized}) as
\begin{eqnarray}
\hspace{-5mm}\eps_{\bm{p}}  - \sum_{n=1}^N A_n^0 = p^0 - A^0= \sqrt{ \Big(\bm{p} - \sum_{n=1}^N \bm{A}_n \Big)^2  + m^2 } ~.
\label{quasi_energy_generalized_2}
\end{eqnarray}
It is then natural to define a quantity known as the kinetic momentum $\Pi^{\mu}$ \cite{Blaettel:1993uz},
\begin{eqnarray}
\Pi^{\mu} \equiv p^{\mu} - \sum_{n=1}^N A_n^{\mu}~,
\label{kinetic_momentum}
\end{eqnarray}
which by construction satisfies
\begin{eqnarray}
\Pi^0 = \sqrt{\bm{\Pi}^2 + m^2} ~.
\end{eqnarray}
Using the kinetic momentum, one can rewrite the equations of motion as (see Appendix \ref{EOMs_covariant} for details)
\begin{eqnarray}
&& 
\frac{dx^{\mu}}{dt}  = \frac{\Pi^{\mu}}{\Pi_0}   ~, 
\label{EOM_covariant_formulation_x}\\
&&
\frac{d \Pi^{\mu}}{dt} = \sum_{\nu} \frac{\Pi_{\nu}}{\Pi_0} \sum_{n=1}^N \Big( \partial^{\mu} (A_n)^{\nu} - \partial^{\nu} (A_n)^{\mu}  \Big)  ~.
\label{EOM_covariant_formulation_p}
\end{eqnarray}
We note that the force term in Eq.\ (\ref{EOM_covariant_formulation_p}) has a form analogous to that known from the covariantly formulated electrodynamics, except that in our case there are multiple vector fields.

\subsection{Thermodynamics and thermodynamic consistency}

Let us consider the thermodynamic properties of the VDF model. Taking the entropy density to have the same functional dependence on the distribution function, $f_{\bm{p}}$, as in the case of the ideal Fermi gas leads to $f_{\bm{p}}$ having the Fermi-Dirac form (for details, see Appendix \ref{form_of_the_quasiparticle_distribution_function}),
\begin{eqnarray}
f_{\bm{p}} = \frac{1}{ e^{\beta ( \eps_{\bm{p}} -   \mu)} + 1} ~,
\end{eqnarray}
where $\beta = 1/T$ and $\mu$ is the chemical potential, with $T$ denoting the temperature.

In the rest frame the energy-momentum tensor has the form $T^{\mu\nu} = \txt{diag} \big(\mathcal{E}, P, P, P \big)$, and the spatial components of the current vanish, $j^i = 0$, while $j_{\mu}j^{\mu} = n^2$. Then the pressure is given by
\begin{eqnarray}
P_{(N)} &=&  \frac{1}{3} \sum_k T^{kk} \bigg|_{\substack{\text{rest} \\ \text{frame}}} \\
&=& g\int \pdens T ~ \ln \Big[ 1 + e^{-\beta (\eps_{\bm{p}} - \mu)}  \Big]   \non\\
&& \hspace{5mm} +~  \sum_{i=1}^N C_i \frac{b_i - 1}{b_i} n^{b_i} ~.
\label{pressure_generalized}
\end{eqnarray}
We note that in an equilibrated system, vector-density--dependent interactions can be described in terms of a shift of the chemical potential $\mu_B$. Using Eq.\ (\ref{quasiparticle_energy_generalized}), we can always write
\begin{eqnarray}
\eps_{\bm{p}} - \mu_B = \sqrt{ \bm{p}^2 + m^2}   - \mu^* = \epsilon_{\txt{kin}} - \mu^*~,
\label{effective_chemical_potential}
\end{eqnarray}
where we have introduced the effective chemical potential, $\mu^* = \mu_B -  \sum_{i=1}^N A_n^0$. Consequently, the dependence of the thermal part of the pressure, Eq.\ (\ref{pressure_generalized}), on temperature $T$ and effective chemical potential $\mu^*$ is just like that of an ideal Fermi gas.

The grand canonical potential is related to the pressure through $\Omega(T,\mu, V) = - PV$, and we can immediately calculate the entropy density,
\begin{eqnarray}
\hspace{-15mm} s &\equiv& - \frac{1}{V}  \left( \frac{d \Omega}{dT} \right)_{V,\mu} =  \\
&=& g \int \pdens \bigg( \ln \Big[ 1 + e^{-\beta (\eps_{\bm{p}} - \mu)}  \Big] +    \frac{\eps_{\bm{p}} - \mu }{T} ~  f_{\bm{p}} \bigg) ~, \hspace{5mm}
\end{eqnarray}
and the number density,
\begin{eqnarray}
n \equiv - \frac{1}{V} \left( \frac{d \Omega}{d\mu} \right)_{V,T} = g\int  \pdens f_{\bm{p}}~,
\end{eqnarray}
where the latter equation proves the correct normalization of our distribution function. Calculating the energy density using $\mathcal{E} \equiv sT - P + \mu n$ yields Eq.\ (\ref{energy_density_generalized}) evaluated in the rest frame, thus confirming that the model is thermodynamically consistent.

\section{Theoretical results}
\label{theoretical_results}

\subsection{Parametrization}
\label{parametrization}

To apply the VDF model to studies of heavy-ion collisions, it needs to describe hadronic matter whose phase diagram contains two first-order phase transitions. The first of these is the experimentally observed low-temperature, low-density phase transition in nuclear matter, sometimes known as the nuclear liquid-gas transition. The second is a postulated high-temperature, high-density phase transition that is intended to correspond to the QCD phase transition. 

We want to stress that while the latter may, in principle, coincide with the location of the phase transition in the real QCD phase diagram, its nature is fundamentally different. This is because within Landau Fermi-liquid theory, unlike in QCD, the degrees of freedom do not change across the phase transition. This is also the case in some other approaches to the QCD EOS, for example in models based on quarkyonic matter \cite{McLerran:2018hbz}, where the active degrees of freedom at the Fermi surface remain hadronic even after quark degrees of freedom appear; however, to which extent such dynamics may be captured in the VDF model remains to be seen. The nature of the phase transition that we can simulate in the VDF model is that of going from a less organized to a more organized state. This is easily visualized in the case of the transition from gas to liquid (nucleon gas to nuclear drop). In the case of the high-temperature, high-density phase transition, we may think of it as a transition from a fluid to an even more dense, and more organized, fluid (nuclear matter to quark matter). This interpretation is supported by the functional dependence of entropy per particle on the order parameter, which decreases across the phase transition from a less dense to a more dense state (for an extended discussion, see \cite{Hempel:2013tfa}).

For brevity, in the following we will refer to the high-temperature, high-density phase transition within the VDF model as ``QGP-like'' or ``quark-hadron'' phase transition, with the expectation that it is understood as a useful moniker rather than a statement on the nature of the described transformation. In addition, we emphasize that the degrees of freedom present in the VDF model agree with those expected after hadronization. Since ultimately we intend to use the VDF model in the hadronic afterburner stage of a heavy-ion collision simulation, the issue of hadronic degrees of freedom present above the QGP-like phase transition will never arise in realistic calculations. At the same time, in parts of the phase diagram close to the critical region, the hadronic systems studied will display behavior typical for systems approaching a phase transition.

In the present, rather simplified version of the VDF model, we chose the degrees of freedom to be those of isospin symmetric nuclear matter, that is nucleons with nucleon mass $m_N = 938$ MeV and degeneracy factor $g_N=4$. In the case where thermally induced $\Delta$ resonances are included as well (which can be easily done through a substitution $g f_{\bm{p}} \to g_N f^{(N)}_{\bm{p}} + g_{\Delta} f^{(\Delta)}_{\bm{p}}$, where $g_N$, $g_{\Delta}$, $f^{(N)}_{\bm{p}}$, and $f^{(\Delta)}_{\bm{p}}$ are the degeneracy factors and distribution functions corresponding to the nucleons and Delta resonances, respectively), their mass is taken to be $m_{\Delta} = 1232$ MeV and the degeneracy factor is $g_{\Delta} = 16$. We note that the model can be easily extended to arbitrarily many baryon resonances, however, we leave the study of the corresponding effects for a future work. In a system that undergoes two first-order phase transitions, the pressure exhibits two mechanically unstable regions (known as spinodal regions), defined by the condition that the first derivative of the pressure with respect to the order parameter is negative \cite{Landau_Stat, Chomaz:2003dz}. In a minimal model realizing such behavior, the pressure needs to be a four-term polynomial in the order parameter, and thus we adopt a version of the VDF model in which we utilize four interaction terms. (We note that to describe only one of the phase transitions mentioned above, it is enough to adopt a model with two interaction terms. In the case of the nuclear liquid-gas phase transition, the resulting model will be not unlike many Skyrme-based parametrizations of the EOS.) 

The energy density, Eq.\ (\ref{energy_density_generalized}), is easily adapted to include $N=4$ interaction terms. In the rest frame, 
\begin{eqnarray}
\mathcal{E}\big|_{\substack{\text{rest} \\ \text{frame}}} = g \int \pdens \epsilon_{\txt{kin}} ~  f_{\bm{p}} + \sum_{i=1}^4 \frac{C_i}{b_i} n_B^{b_i}~,
\end{eqnarray}
where $n_B \equiv \sqrt{j_{\mu} j^{\mu}}$ is the rest frame baryon number density. As mentioned in the introduction to the VDF model (Sec.\ \ref{background}), our goal is to construct an EOS with a general QGP-like phase transition properties while ensuring that the known properties of ordinary nuclear matter are well reproduced. To that end, we choose the following constraints to fix the eight free parameters $\{b_1, b_2, b_3, b_4, C_1, C_2, C_3, C_4\}$ in the VDF model:\\
1) the position of the minimum of the binding energy of nuclear matter at the saturation density $n_B = n_0$,
\begin{eqnarray}
\frac{d \left( \frac{\mathcal{E}_{(4)} }{n_B} - m_N \right)}{dn_B} \bigg|_{\substack{T =0 \\ n_B=n_0 }} = 0~,
\label{minimum}
\end{eqnarray}
2) the value of the binding energy at the minimum,
\begin{eqnarray}
\frac{\mathcal{E}_{(4)} }{n_B}\bigg|_{\substack{T =0 \\n_B=n_0  }} - m_N = E_0~,
\label{binding_energy}
\end{eqnarray}
3, 4) the position of the critical point $\big(T^{(N)}_c, n^{(N)}_c\big)$ for the nuclear liquid-gas phase transition,
\begin{eqnarray}
&& \frac{dP}{dn_B} \Big(T =T^{(N)}_c,n_B= n^{(N)}_c\Big) = 0~, 
\label{nuclear_CP_1} \\
&&\frac{d^2P}{dn_B^2} \Big(T = T^{(N)}_c, n_B = n^{(N)}_c\Big) = 0~,
\label{nuclear_CP_2}
\end{eqnarray}
5, 6) the position of the critical point $ \big(T^{(Q)}_c, n^{(Q)}_c\big)$ for the quark-hadron phase transition,
\begin{eqnarray}
&& \frac{dP}{dn_B} \Big(T=T^{(Q)}_c, n_B=n^{(Q)}_c\Big) = 0~,
\label{QGP_like_CP_1} \\ 
&& \frac{d^2P}{dn_B^2} \Big(T=T^{(Q)}_c, n_B=n^{(Q)}_c\Big) = 0~,
\label{QGP_like_CP_2} 
\end{eqnarray}
7, 8) the position of the lower (left) and upper (right) boundaries of the spinodal region, $\eta_L$ and $\eta_R$, for the quark-hadron phase transition at $T=0$,
\begin{eqnarray}
&& \frac{dP}{dn_B} \Big(T=0, n_B=\eta_L\Big) = 0~,
\label{spinodal_1} \\
&& \frac{dP}{dn_B} \Big(T=0, n_B=\eta_R\Big) = 0~.
\label{spinodal_2}
\end{eqnarray}
The set of quantities $(n_0 , E_0, T_c^{(N)}, n_c^{(N)}, T_c^{(Q)}, n_c^{(Q)} , \eta_L, \eta_R)$ is referred to as the characteristics of an EOS.

We choose the properties of the ordinary nuclear matter, encoded in conditions (\ref{minimum}-\ref{nuclear_CP_2}), based on experimentally determined values \cite{Bethe:1971xm,Elliott:2013pna}:
\begin{eqnarray}
&& n_0 = 0.160~ \txt{fm}^{-3}~, \hspace{5mm} E_0 = - 16.3 ~\txt{MeV}~, \label{enforced_saturation_density_and_binding_energy} \\
&& T^{(N)}_c = 18~ \txt{MeV}~, \hspace{5mm} n^{(N)}_c = 0.06 ~ \txt{fm}^{-3} ~. 
\label{enforced_nuclear_critical_point}
\end{eqnarray}
On the other hand, the properties of dense nuclear matter, $n_B \gg n_0$, are only weakly constrained by experiment at this time. We are then in a position to create a family of possible EOSs based on a number of different postulated characteristics (\ref{QGP_like_CP_1}-\ref{spinodal_2}), while ensuring that nuclear matter properties are preserved. The resulting family of EOSs encompasses QGP-like phase transition characteristics spanning vast regions of the dense nuclear matter phase diagram. This allows for a systematic comparison with experimental data, with the goal of constraining the number of allowed EOSs to a small subfamily with qualitatively similar properties.

In the remainder of this paper, we illustrate properties of the VDF model by discussing key results for a few representative EOSs which reproduce sets of the QGP-like phase transition characteristics $\big( T_c^{(Q)}, n_c^{(Q)}, \eta_L, \eta_R\big)$ listed in Table \ref{example_characteristics}. The corresponding parameter sets can be found in Appendix \ref{parameter_sets}.
	\begin{table}%[htbf]
		\caption{Example characteristics $\big( T_c^{(Q)}, n_c^{(Q)}, \eta_L, \eta_R\big)$ of the QGP-like phase transition: critical temperature $T_c^{(Q)}$, critical baryon number density $n_c^{(Q)}$, and the boundaries of the spinodal region at $T =0$, $\eta_L$ and $\eta_R$. The corresponding parameter sets can be found in Appendix \ref{parameter_sets}. Characteristics in sets I-V are obtained based on systems composed only of nucleons, while in set VI we consider a system composed of nucleons and thermally produced $\Delta$-resonances. We also show the incompressibility at saturation density and zero temperature, $K_0$, calculated for the parameterized EOSs.}
		\label{example_characteristics}
		\begin{center}
			\bgroup
			\def\arraystretch{1.4}
			\begin{tabular}{c c c c c c c}
				\hline
				\hline
				%\cline{2-6}
				set & $T_c^{(Q)}[\txt{MeV}]$ & $n_c^{(Q)}[n_0]$ & 
				$\eta_L[n_0]$ &  $\eta_R[n_0]$ & species & $K_0[\txt{MeV}]$ \\ 
				\hline
				 I &50  & 3.0 & 2.70 & 3.22 & N &260 \\  
				 II & 50  & 3.0 & 2.85 & 3.12 & N & 279\\  
				 III & 50  & 4.0 & 3.90 & 4.08 & N & 280\\  
				 IV & 100& 3.0 & 2.50 & 3.32 & N & 261\\  
				 V & 100& 4.0 & 3.60 & 4.28 & N & 271\\  
				 VI & 125 & 4.0 & 3.60 & 4.28 & N + $\Delta$& 277\\
				\hline
				\hline
			\end{tabular}
			\egroup
		\end{center}
	\end{table}

\subsection{Results: Pressure, the speed of sound, and energy per particle}

\begin{figure*}[t]
	\includegraphics[width=\textwidth]{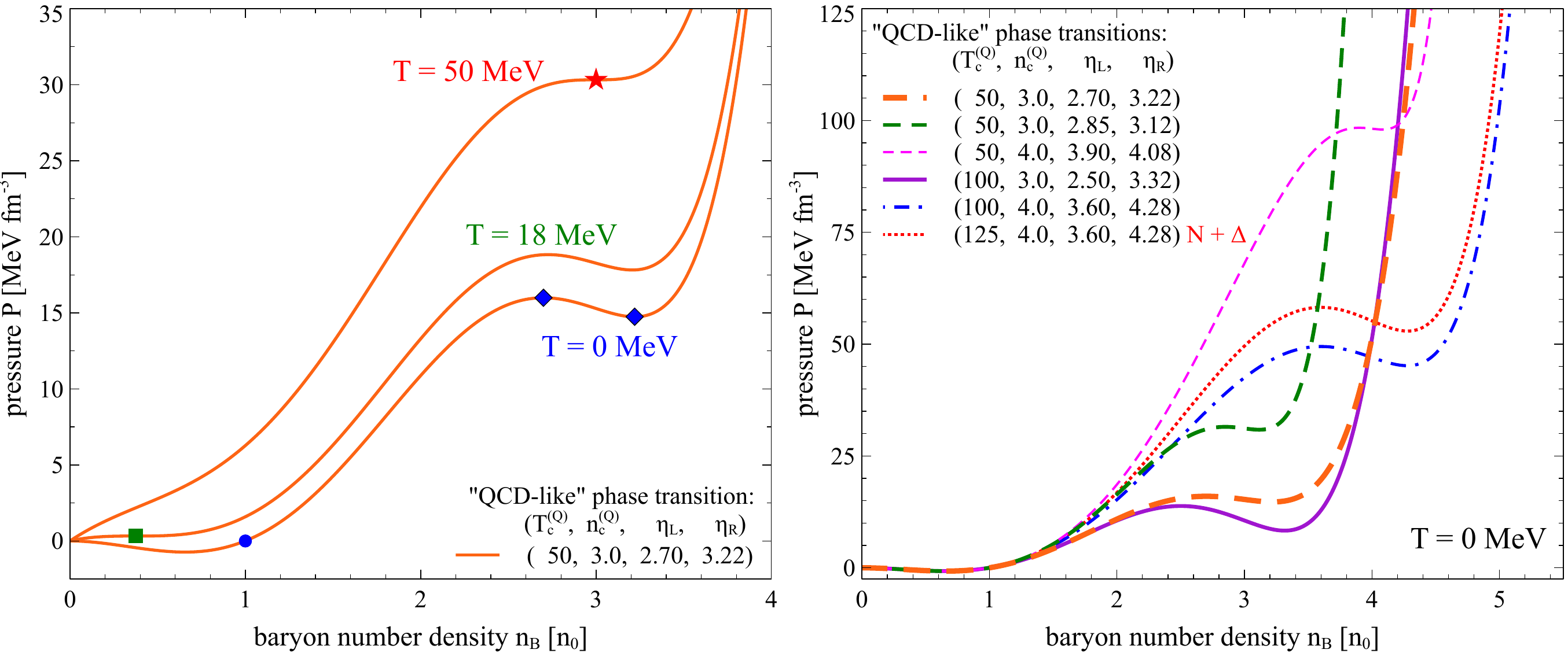}
	\caption{(Color online) Pressure \textit{versus} baryon number density. In the legend, the critical temperature of the QGP-like phase transition $T_c^{(Q)}$ is given in MeV, while the critical density, $n_c^{(Q)}$, and the boundaries of the spinodal region at $T=0$, $\eta_{L}$ and $\eta_R$, are given in units of saturation density, $n_0 = 0.160 \ \txt{fm}^{-3}$. Left panel: Illustration of the fitting procedure. Pressure is plotted at three significant temperatures ($T=0$, nuclear critical temperature $T_c^{(N)}$, and quark-hadron critical temperature $T_c^{(Q)}$) for an EOS with characteristics from set I, see Table \ref{example_characteristics}. Specific points at which the parameters of the EOS are fixed are indicated on the plot as follows: a blue dot represents the equilibrium point of ordinary nuclear matter; blue diamonds denote the left and right boundary of the QGP-like spinodal region; a green square denotes the critical point of the nuclear phase transition; a red star denotes the critical point of the QGP-like phase transition. Right panel: Pressure is plotted at temperature $T=0$ for all sets of characteristics listed in Table \ref{example_characteristics}. All obtained EOSs describe the same physics in the region $n_B \lesssim 1.5 n_0$, where the behavior of nuclear matter is relatively well known. The hardness of the EOSs is noticeable for densities above the quark-hadron transition regions, and is a consequence of employing interaction terms with high powers ($b_i > 2$) of baryon number density $n_B$ (see text for details). }
	\label{pressures}
\end{figure*}

The left panel in Fig.\ \ref{pressures} shows pressure \textit{versus} baryon number density at three significant temperatures ($T = 0$, nuclear critical temperature $T_c^{(N)}$, and quark-hadron critical temperature $T_c^{(Q)}$) for an EOS with characteristics from set I (see Table \ref{example_characteristics}). On the same plot, we also indicate the location of key points that determine the fit parameters. At temperature $T = 0$, conditions (\ref{minimum}) and (\ref{binding_energy}) are applied at the saturation density of nuclear matter, denoted with a blue circle. Also at $T =0$, conditions (\ref{spinodal_1}) and (\ref{spinodal_2}) fix the positions of the lower (left) and upper (right) boundary of the high density spinodal region, $\eta_L$ and $\eta_R$; these are denoted with blue diamonds. At the critical point of nuclear matter, $T = T_c^{(N)}$ and $n_B = n_c^{(N)}$, denoted with a green square, conditions (\ref{nuclear_CP_1}) and (\ref{nuclear_CP_2}) are enforced. Finally, conditions (\ref{QGP_like_CP_1}) and (\ref{QGP_like_CP_2}) are applied to set the position of the QGP-like critical point $\big(T_{c}^{(Q)}, n_c^{(Q)} \big)$, denoted with a red star.

The right panel in Fig.\ \ref{pressures} shows pressure \textit{versus} baryon number density at zero temperature, where the curves correspond to all sets of characteristics listed in Table \ref{example_characteristics}. While most of the results are calculated in the presence of nucleons only, the thin dotted red line shows pressure for a system with both nucleons (protons and neutrons) and thermally excited $\Delta$ resonances. As already emphasized, all of the EOSs display the same behavior for baryon number densities corresponding to ordinary nuclear matter, and only start differing from each other in regions currently not constrained by experimental data, $n_B \gtrsim 1.5 n_{0}$.

\begin{figure}[t]
	\includegraphics[width=\columnwidth]{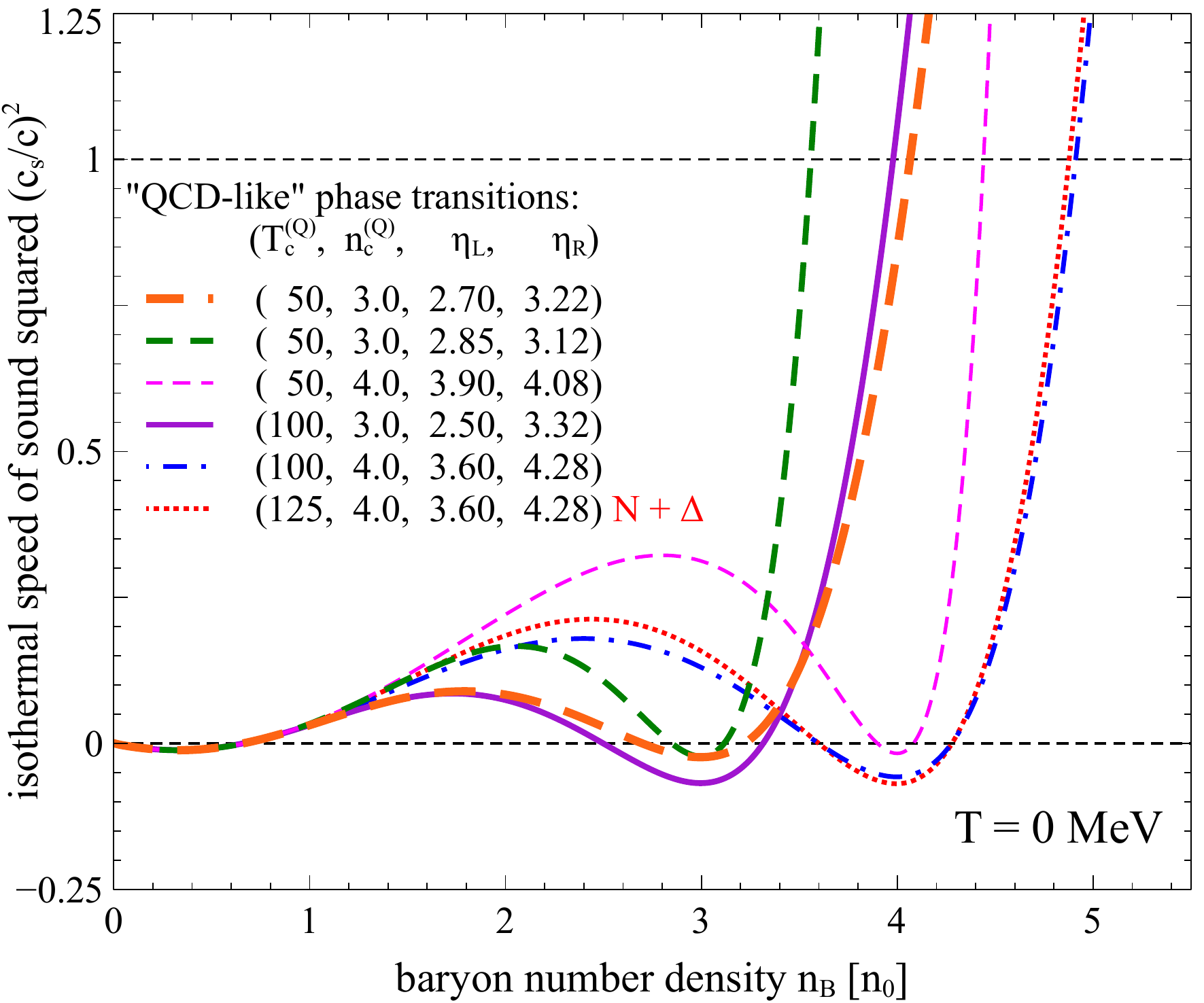}
	\caption{(Color online) The isothermal speed of sound squared at $T=0$ \textit{versus} baryon number density, plotted for all sets of characteristics listed in Table \ref{example_characteristics}. In the legend, the critical temperature of the QGP-like phase transition $T_c^{(Q)}$ is given in MeV, while the critical density, $n_c^{(Q)}$, and the boundaries of the spinodal region at $T=0$, $\eta_{L}$ and $\eta_R$, are given in units of saturation density, $n_0 = 0.160 \ \txt{fm}^{-3}$. It is apparent that the speed of sound becomes acausal for relatively large baryon number densities above the quark-hadron transition region, which is a consequence of the hardness of the equation of state in the same region (see the right panel on Fig.\ \ref{pressures}). This pathological behavior of the EOS is expected outside of the region in which its parameters are fitted, and it does not pose an issue for uses in afterburner simulations: by construction, these deal with systems below the quark-hadron phase transition, where the behavior of the speed of sound is typical (for more details, see text).}
	\label{speed_of_sound} 
\end{figure}
\begin{figure}[t]
	\includegraphics[width=\columnwidth]{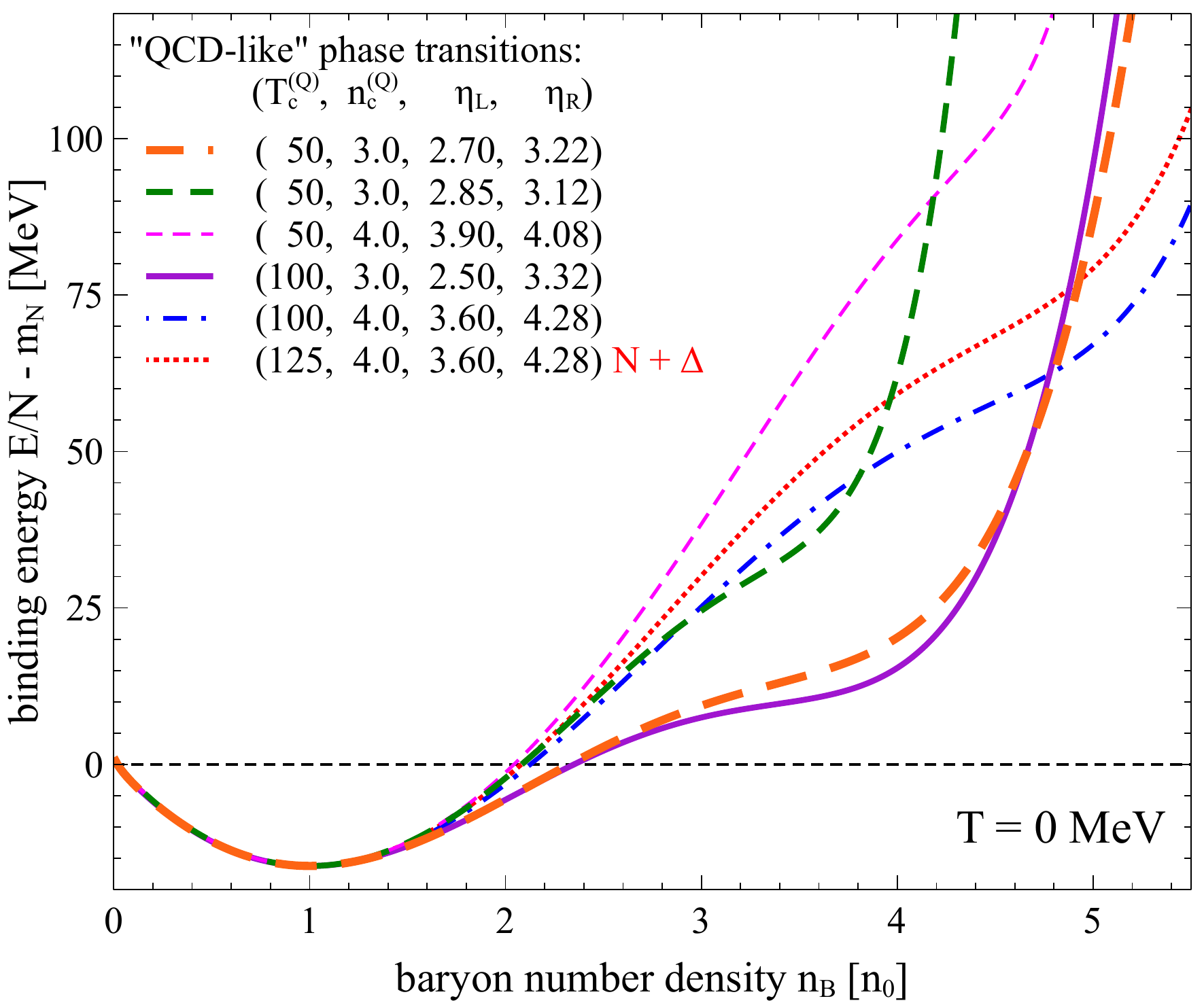}
	\caption{(Color online) The binding energy at $T=0$ \textit{versus} baryon number density, plotted for all sets of characteristics listed in Table \ref{example_characteristics}. In the legend, the critical temperature of the QGP-like phase transition $T_c^{(Q)}$ is given in MeV, while the critical density, $n_c^{(Q)}$, and the boundaries of the spinodal region at $T=0$, $\eta_{L}$ and $\eta_R$, are given in units of saturation density, $n_0 = 0.160 \ \txt{fm}^{-3}$. As shown also in previous figures, all obtained EOSs describe the same physics in the region $n_B \lesssim 1.5 n_0$, where the behavior of nuclear matter is relatively well known; in particular, all curves reproduce the value of the chosen binding energy at nuclear matter saturation as well as the location of the saturation density. The degree of the softening in energy per particle at high baryon number density is directly related to the width of the spinodal region of a given EOS (see text for more details).}
	\label{binding_energy_plots} 
\end{figure}

A few regularities are apparent in the behavior of the pressure curves at zero temperature in regions corresponding to the QGP-like phase transition. Let us focus on the value of the pressure at the lower boundary of the spinonal region $P(\eta_L)$ (which is directly related to the average value of the pressure across the transition region), and compare its values for sets of characteristics between which only one property of the QGP-like phase transition changes substantially. First, $P(\eta_L)$ increases with critical baryon number density $n_c^{(Q)}$, which can be seen by comparing the pressure curves for the second and third sets of characteristics (delineated with medium dashed green and thin dashed magenta lines, respectively). Second, $P(\eta_L)$ decreases with critical temperature $T_c^{(Q)}$, as evidenced by pressure curves for the first and fourth sets of characteristics (delineated with thick dashed orange and solid purple lines, respectively). Third, $P(\eta_L)$ decreases with the width of the spinodal region, $\Delta \eta = \eta_R - \eta_L$, which can be seen by comparing pressure curves for the first and second sets of characteristics (thick dashed orange and medium dashed green lines, respectively). Furthermore, the magnitude of the drop in the pressure across the spinodal region, $\Delta P = P(\eta_R) - P(\eta_L)$, increases with the critical temperature, as seen by comparing curves for the first and fourth sets of characteristics (thick dashed orange and solid purple lines, respectively). These features, in fact, create a physical bound on which QGP-like transitions are allowed in the VDF model. A transition with a wide spinodal region, with a critical point at a relatively low baryon number density but a relatively high critical temperature can often be excluded, as it leads to such a significant drop in the pressure across the spinodal region that the pressure becomes negative in some parts of the quark-hadron coexistence region, which would correspond to an unphysical ``QGP bound state''. This is because at $T=0$ the pressure is given by
\begin{eqnarray}
P \equiv n_B^2 \frac{d}{dn_B} \left(\frac{\mathcal{E}}{n_B} \right) ~,
\label{pressure_Tzero}
\end{eqnarray}  
and locally negative pressure implies that there exists a baryon density for which $\frac{d}{dn_B} \left(\frac{\mathcal{E}}{n_B} \right) = 0$ and $\frac{d^2}{dn_B^2} \left(\frac{\mathcal{E}}{n_B} \right)> 0$, corresponding to a local minimum in energy per particle, $\frac{\mathcal{E}}{n_B}$. While such a minimum is in fact expected in the region of the phase diagram corresponding to ordinary nuclear matter, where $\frac{d}{dn_B} \left(\frac{\mathcal{E}}{n_B} \right) = 0$ at the nuclear saturation density, it is forbidden for large baryon number densities, where it would correspond to a metastable or even stable state of QGP. For example, most obtained phase transitions with $n_c^{(Q)} = 2.5n_0$ and $T_c^{(Q)} \geq 125 \  \txt{MeV}$ are rejected based on this argument.

Next, it is easy to notice that the pressure rises rapidly after leaving the quark-hadron transition region. This hardness of the EOS is a general feature of models based on high powers of baryon number density (specifically, with exponents higher than 2), and is ubiquitous among various Skyrme-type models (see e.g.\ \cite{Dutra:2012mb}). In fact, it can be shown that any relativistic Lagrangian with vector-type interactions leading, in the mean-field approximation, to terms of the form $n_B^\alpha$, where $\alpha>2$, results in acausal phenomena at high baryon number densities \cite{Zeldovich:1962emp}. Indeed, Fig.\ \ref{speed_of_sound} shows the isothermal speed of sound squared $\left(\frac{c_T}{c}\right)^2$ at $T=0$ for the chosen sets of phase transition characteristics (Table \ref{example_characteristics}). (We note that at $T=0$, the isothermal and isentropic speeds of sound are identical.) The speed of sound squared is negative within the spinodal region, as expected for a first-order phase transition \cite{Chomaz:2003dz}, while for large baryon number densities above the quark-hadron phase transition it eventually becomes acausal. Although this behavior is non-ideal, it is entirely to be expected that a fitted function will behave pathologically outside of the region in which it is constrained. Moreover, because we intend to use the VDF model in a hadronic afterburner, its main application is for matter at densities below the quark-hadron coexistence region, where this problem does not arise (though in some of the studied phase transitions the conformal bound of $\left(\frac{c_T}{c}\right)^2 \leq \frac{1}{3}$ can still be violated; it is presently unclear if this bound is satisfied in dense nuclear matter; see for example \cite{Bedaque:2014sqa, McLerran:2018hbz, Annala:2019puf, Fujimoto:2019hxv}). With this issue in mind, in creating parameter sets we make sure that the speed of sound preserves causality for all baryon number densities below the upper boundary of the quark-hadron coexistence region.

Finally, in Fig.\ \ref{binding_energy_plots} we show the binding energy at $T=0$, which is the energy per particle minus the rest mass $\mathcal{E}_{(4)}/n_B - m_N$,  \textit{versus} baryon number density, obtained for EOSs corresponding to all sets of characteristics listed in Table \ref{example_characteristics}. As expected, all curves reproduce the value of the chosen binding energy at nuclear matter saturation as well as the location of the saturation density, Eq.\ \eqref{enforced_saturation_density_and_binding_energy}. On the other hand, at high densities the binding energy displays a softening related to the postulated QGP-like phase transition, which is different for each considered EOS. We note that the extent of this softening is directly related to the width of the spinodal region of a given EOS. This can again be seen from the fact that at zero temperature the pressure is given by Eq.\ \eqref{pressure_Tzero}, from which it immediately follows that the curvature of the energy density, $\left(\frac{d^2 \mathcal{E}}{dn_B^2} \right)$, must be negative in the spinodal region; consequently, the region over which $(d^2 \mathcal{E}/dn_B^2) <0$ holds is related to $(\eta_L, \eta_R)$.

Although we have only shown results corresponding to a few possible QGP-like phase transitions, arbitrarily many versions of the dense nuclear matter EOS can be obtained in the VDF model. While they vary widely in the high baryon density region, by construction they all reproduce the same physics in the range of baryon number densities corresponding to ordinary nuclear matter. In fact, fitting the VDF model to reproduce the experimental values of the saturation density, the binding energy, and the nuclear critical point gives a remarkably good prediction for the value of pressure at the nuclear critical point, $P_c$, and the value of incompressibility, $K_0$, as compared with experiment and against other models (summarized in Table \ref{table_comparison_vdf_vs_other_models}). This is partially expected, as the value of the incompressibility $K_0$ depends strongly on critical temperature \cite{Kapusta:1984ij}. Nevertheless, it is noteworthy that the minimal VDF model, based on a few characteristics taken at their experimentally established values (here $n_0$, $E_0$, $T_c^{(N)}$, $n_c^{(N)}$), leads to predictions for other properties of nuclear matter agreeing remarkably with experimental data. Apparently, constraining four properties of the EOS is enough to reproduce the thermodynamic behavior of nuclear matter in the fitted region. The same could be true in the case of nuclear matter at high baryon number density. We may be hopeful that postulating QGP-like phase transition characteristics that happen to lay close to their true QCD values will lead to a VDF model parametrization correctly describing other properties of dense nuclear matter in the transition region. We expect that this correct description would manifest itself through agreement of simulation results with experimental data.

%\begin{widetext}
\begin{table}[b]
	\caption{Comparison of values of the nuclear phase transition critical temperature $T_c^{(N)} \ [\txt{MeV}]$, the critical baryon number density $n_c^{(N)} \ [\txt{fm}^{-3}]$, pressure at the critical point $P_c \ [\txt{MeV fm}^{-3}]$, and incompressibility $K_0 \ [\txt{MeV}]$ as obtained in experiment \cite{Elliott:2013pna} and in various models, where ``W'' denotes the Walecka model \cite{Walecka:1974qa}, ``QVdW'' denotes the quantum Van der Waals model \cite{Poberezhnyuk:2017yhx}, ``VDF N'' denotes the VDF model with nuclear phase transition only (two interaction terms), and ``VDF N+Q'' denotes the VDF model with both nuclear and quark-hadron phase transitions (four interaction terms). For the last case, the values of $P_c$ and $K_0$ are given as averages calculated across all obtained EOSs for quark-hadron critical temperatures $T_c^{(Q)} \in \{ 50, 100, 125 \} \  [\txt{MeV}]$ and critical baryon number densities $n_c^{(Q)} \in \{ 3.0, 4.0, 5.0\}\  [n_0]$. Values marked with an asterisk are input parameters of the models.}  
	\label{table_comparison_vdf_vs_other_models}
	\begin{center}
		\bgroup
		\def\arraystretch{1.4}
		\begin{tabular}{ l c c c c c}
			\hline
			\hline
			& Experiment & W & 
			QVdW & 
			VDF N& VDF N+Q \\ 
			\hline
			$T_c^{(N)}$ & $17.9 \pm 0.4$& 18.9 & 19.7 & 18* & 18*\\  
			$n_c^{(N)}$ & $0.06 \pm 0.01$ & 0.070 & 0.072 & 0.06* & 0.06* \\  
			$P_c $ & $0.31 \pm 0.07$  & 0.48 & 0.52 & 0.311 & $0.3066 \pm 0.0014$ \\  
			$K_0 $ & 230-315 & 553 & 763 & 282 & $273.5 \pm 5.1$\\
			\hline
			\hline
		\end{tabular}
		\egroup
	\end{center}
\end{table}
%\end{widetext}

\begin{figure*}[t]
	\includegraphics[width=\textwidth]{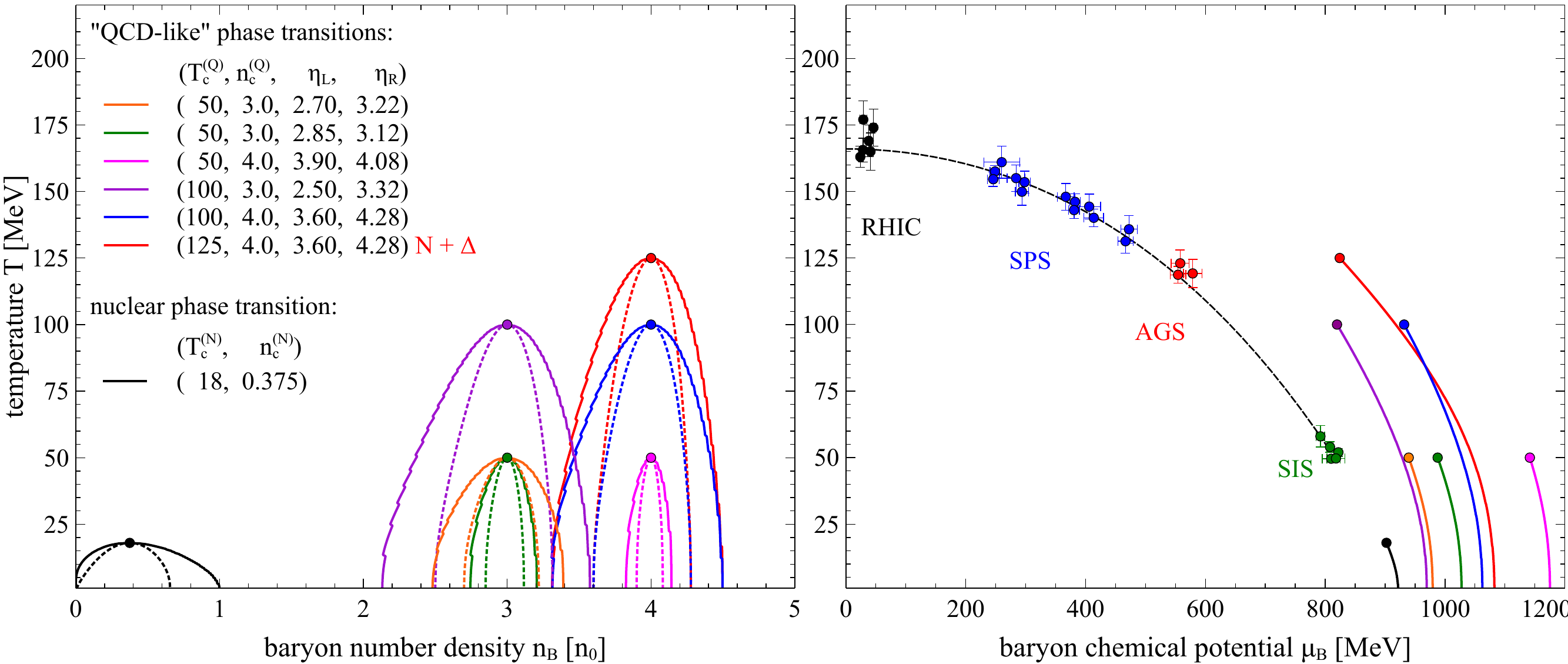}
	\caption{(Color online) Phase diagram in the $T$-$n_B$ (left panel) and $T$-$\mu_B$ (right panel) planes for sets of characteristics listed in Table \ref{example_characteristics}. Solid and dashed lines represent the boundaries of the coexistence and spinodal regions, respectively. In the legend, the critical temperature of the QGP-like phase transition $T_c^{(Q)}$ is given in MeV, while the critical baryon number density $n_c^{(Q)}$ and the boundaries of the spinodal region, $\eta_{L}$ and $\eta_R$, are given in units of saturation density, $n_0 = 0.160 \ \txt{fm}^{-3}$. The coexistence and spinodal regions of the nuclear phase transition, depicted with solid black and dashed black lines, respectively, are common to all sets of characteristics. Also shown are chemical freeze-out points obtained in experiment and a parametrization of the freeze-out line from \cite{Cleymans:2005xv}.}
	\label{phase_diagram} 
\end{figure*}

\subsection{Results: Phase diagrams}
\label{phase_diagrams}

The phase diagrams for the EOSs corresponding to the characteristics listed in Table \ref{example_characteristics} are shown in Fig.\ \ref{phase_diagram}. Solid and dashed lines represent the boundaries of the coexistence and spinodal regions, respectively. The coexistence and spinodal regions of the nuclear phase transition, depicted with black lines, are common for all used EOSs by construction.

It is immediately apparent that the QGP-like coexistence curves on the phase diagrams all look alike. This is a consequence of our choice to employ only interactions depending on vector baryon number density, as in this case the dependence of the thermal part of the pressure on temperature $T$ and effective chemical potential $\mu^*$ is just like that of an ideal Fermi gas, as can be seen from Eq.\ (\ref{effective_chemical_potential}). Consequently, all VDF EOSs display similar behavior with increasing temperature $T$. This can be especially easily seen on the $T$-$\mu_B$ phase diagram (right panel of Fig.\ \ref{phase_diagram}), where the coexistence lines exhibit the exact same curvature. An exception from this behavior shown on the plot is the curve calculated for a system with both nucleons and thermally produced $\Delta$ resonances (denoted with a red line), which bends more forcefully towards the $\mu_B = 0$ axis as the temperature increases. This is to be expected as including an additional baryon species lowers the value of the baryon chemical potential for a given baryon number density. Including more baryon species would strengthen this effect.

Another feature, easily discerned on the $T$-$n_B$ phase diagram (left panel of Fig.\ \ref{phase_diagram}), is that the spinodal regions $[\eta_L, \eta_R]$ (and likewise the coexistence regions $[n_L, n_R]$) are always approximately centered around the critical baryon number density, $n_c^{(Q)}$. This is again an effect related to having only the ideal-gas--like contribution to the thermal pressure in case of vector-like interactions (for details see Appendix \ref{symmetric_spinodal_regions}). As a result, the critical baryon number density, $n_c^{(Q)}$, and the boundaries of the spinodal region, $\eta_L$ and $\eta_R$, are not independent. In consequence, we have effectively one less free parameter. For example, once we set the ordinary nuclear matter properties, the critical point of the quark-hadron phase transition, and the lower spinodal boundary at $T=0$, $\eta_L$, the upper spinodal boundary at $T=0$, $\eta_R$, is practically fixed. 

We expect that all these regularities in the behavior of the spinodal and coexistence lines would not be as prominent if other types of interactions were included, rendering the thermal part of the pressure non-trivial. In particular, we expect that adding scalar-type interactions would allow us to obtain coexistence regions bending towards the $n_B = 0$ axis in the $T$-$n_B$ plane, which would correspond to an even stronger tendency to bend towards the $\mu_B=0$ axis in the $T$-$\mu_B$ plane. This expectation is based on the fact that, typically, scalar interactions result in a small effective mass, which in addition decreases with temperature, and that in turn produces a relatively larger thermal contribution to the pressure for a given $n_B$ and $T$. As a result, such phase transitions would more significantly affect the region of the phase diagram covered by the BES program. Extensions of the VDF model leading to such effects are planned for the near future.

\begin{figure*}
	\includegraphics[width=0.96\textwidth]{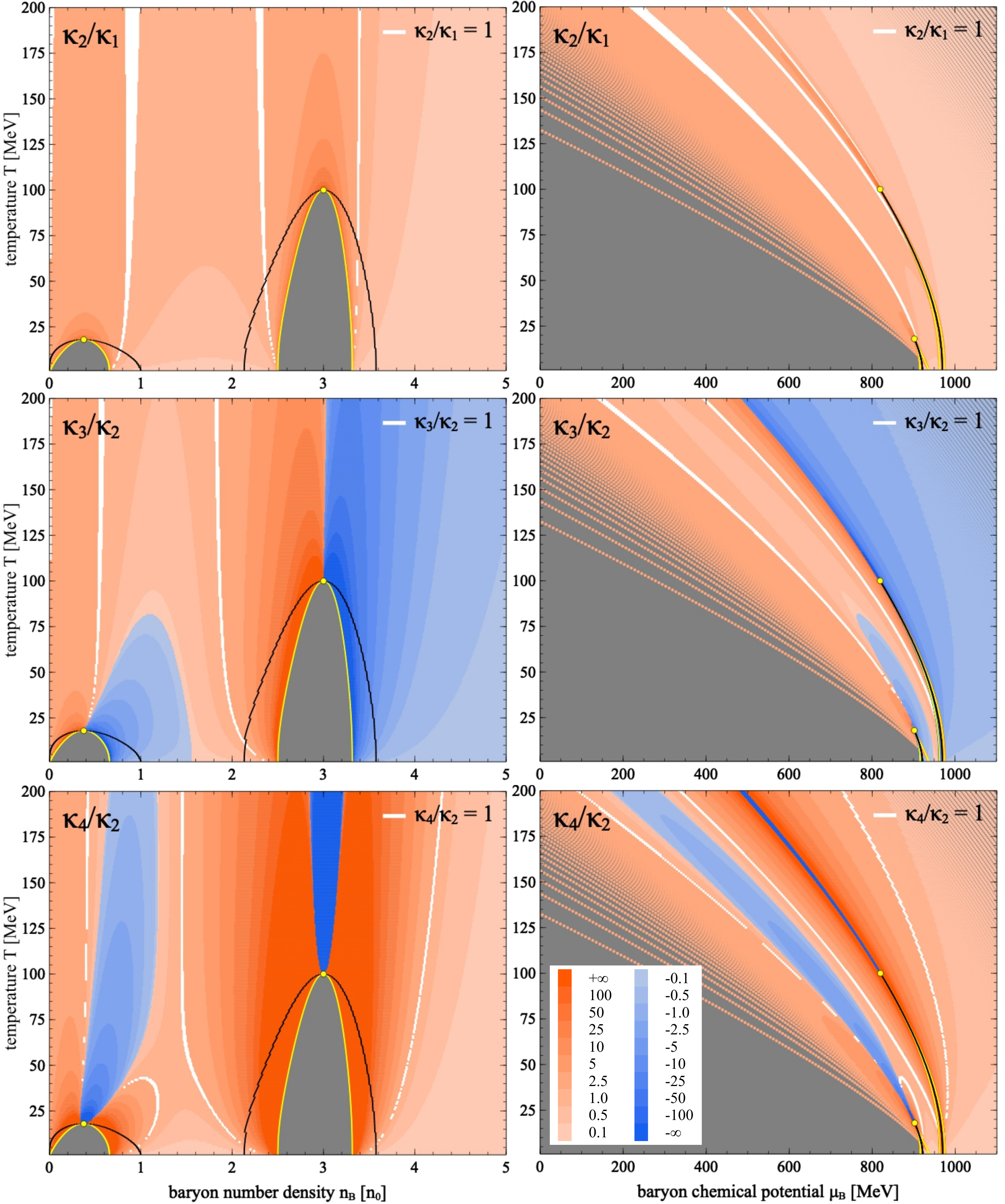}
	\caption{Contour plots of cumulant ratios $\kappa_2/\kappa_1$ (upper row), $\kappa_3/\kappa_2$ (middle row), and $\kappa_4/\kappa_2$ (lower row) in both $T$-$n_B$ and $T$-$\mu_B$ plane (left and right column, respectively), for the EOS identified by the fourth (IV) set of characteristics listed in Table \ref{example_characteristics}. Black lines denote coexistence regions, while yellow lines denote spinodal regions; critical points are indicated with yellow dots. White regions correspond to values of cumulant ratios close to the Poissonian limit, $\kappa_i/\kappa_j= 1 \pm 0.03$. Grey color signifies regions of the phase diagram in which either the cumulant calculation is invalid (left column: inside the spinodal region, which is unstable), or where data has not been produced (right column: regions with extremely small values of the baryon number density $n_B$). The legend entries denote upper (lower) boundaries of ranges of positive (negative) values of cumulant ratios.}
	\label{Cumulants_diagrams}
\end{figure*}

\subsection{Results: Cumulants of baryon number}
\label{Cumulants_of_baryon_number}
 
In analyses of heavy-ion collision experiments, considerable attention has been paid to cumulants of the baryon number distribution. In the grand canonical ensemble, the $j$th cumulant of the baryon number, $\kappa_j$, can be calculated from
\begin{eqnarray}
\kappa_j = T^j \frac{d^j}{d \mu_B^j} \ln \mathcal{Z} (T,V,\mu_B)~,
\label{cumulant_definition}
\end{eqnarray}
where $\mathcal{Z} (T,V,\mu_B)$ is the grand canonical partition function. Because the logarithm of the partition function is related to the pressure $P$ through
\begin{eqnarray}
\ln \mathcal{Z} (T,V,\mu_B) = \frac{PV}{T}~ ,
\end{eqnarray}
we can also write Eq.\ (\ref{cumulant_definition}) as
\begin{eqnarray}
\kappa_j = VT^{j-1} \frac{d^j P}{d \mu_B^j} ~.
\end{eqnarray}
The explicit volume dependence of the cumulants, which is typically divided out in theoretical calculations, is difficult to control in experiment. %In experiment, this is not really volume; the questions is how does the data behave under a superposition of independent sources
Therefore, it is customary to consider ratios of cumulants, most commonly
\begin{eqnarray}
\frac{\sigma^2}{\mu} = \frac{\kappa_2}{\kappa_1}~ , \hspace{5mm}  S \sigma= \frac{\kappa_3}{\kappa_2}~, \hspace{5mm}  \kappa \sigma^2 = \frac{\kappa_4}{\kappa_2}~,
\label{cumulant_ratios}
\end{eqnarray}
where $\mu$ denotes the mean, $\sigma^2$ denotes variance, $S$ denotes skewness, and $\kappa$ denotes excess kurtosis.

The values of cumulants are expected to be influenced by enhanced fluctuations of conserved charges in the vicinity of the critical point, rendering them a signal for the existence of the critical point and a first-order phase transition in QCD \cite{Stephanov:1998dy, Stephanov:1999zu, Koch:2008ia}. In particular it is argued that, for systems crossing the phase diagram close to and above the critical point, the sign of the third-order cumulant, $\kappa_3$, will change \cite{Asakawa:2009aj}, while the fourth-order cumulant, $\kappa_4$, will exhibit a nonmonotonic behavior \cite{Stephanov:2011pb}. Because cumulants of the baryon number distribution can be measured in experiment, they provide one of the strongest links between theoretical predictions and experimental data. Preliminary results from the Beam Energy Scan indeed suggest that the fourth-order cumulant ratio, $\frac{\kappa_4}{\kappa_2}$, exhibits non-monotonic behavior with the collision energy \cite{Adam:2020unf}.

In this as well as in the following sections, we will focus on results for the fourth (IV) set of characteristics listed in Table \ref{example_characteristics}. The choice of this set is arbitrary and does not reflect any preference for the location of the QCD critical point, but simply serves as an illustration of the properties of the VDF model which are qualitatively comparable for all obtained EOSs. In Fig.\ \ref{Cumulants_diagrams}, we plot the cumulant ratios (\ref{cumulant_ratios}) in the $T$-$n_B$ and $T$-$\mu_B$ planes. Dramatic increase in magnitudes of cumulant ratios as well as sudden changes in sign, observed in regions close to and above the critical point, agree with the expectations mentioned above. Interestingly, the effects of the nuclear phase transition are clearly present even at very high temperatures (as has been also observed in \cite{Vovchenko:2016rkn}). This raises the question to what extent the presence of the nuclear phase transition affects the interpretation of experimental data, either by damping the signal originating at the QGP phase transition, or by acting as an imposter. Such questions could be answered by comparing outcomes of simulations utilizing a VDF EOS with either nuclear phase transition only, or both nuclear and quark-hadron phase transitions. Studies of this type are planned for future research.

\section{Implementation in \texttt{SMASH}}
\label{implementation_in_SMASH}

We implemented the VDF equations of motion, Eqs.\ (\ref{EOM_covariant_formulation_x}) and (\ref{EOM_covariant_formulation_p}), in the hadronic transport code \texttt{SMASH} \cite{Weil:2016zrk}, version 1.8 \cite{smash_version_1.8}, where simulating hadronic non-equilibrium dynamics is achieved through numerically solving the Boltzmann equation, in this context often also called the Vlasov equation, the Boltzmann-Uehling-Uhlenbeck (BUU) equation, or the Vlasov-Uehling-Uhlenbeck (VUU) equation. The specification comes from solving the Boltzmann equation for the time evolution of the phase-space density $f(t, \bm{x},\bm{p})$ in the presence of the mean-field $U(\bm{x},\bm{p})$,
\begin{eqnarray}
\bigg[ \parr{}{t} + \parr{H_{(1)}}{\bm{p}} \bnabla_{\bm{p}} - \parr{H_{(1)}}{\bm{x}} \bnabla_{\bm{p}} \bigg]  f(t, \bm{x},\bm{p}) =   I_{\txt{coll}}~,
\label{BUU_equation}
\end{eqnarray} 
where the single-particle Hamiltonian is given by $H_{(1)} = \sqrt{ \bm{p}^2 + m^2} + U(\bm{x},\bm{p})$, and $I_{\txt{coll}}$ denotes the collision integral. Usually, the term Vlasov equation is reserved for the case with no collisions, $I_{\txt{coll}} = 0$. 

The time evolution in hadronic transport is realized within a numerical approach known as the method of test particles \cite{Wong:1982zzb}, where the continuous phase-space distribution of a system of $A$ particles, $f(t,\bm{x},\bm{p})$, is approximated by the distribution of a large number $N$ of discrete test particles with phase space coordinates $\big(\bm{x}_i (t), \bm{p}_i(t)\big)$,
\begin{eqnarray}
f(t, \bm{x}, \bm{p}) \approx \frac{1}{N_T} \sum_{i =1}^{N} \delta \Big( \bm{x} - \bm{x}_i (t)  \Big) \delta \Big( \bm{p} - \bm{p}_i (t)  \Big)~.
\label{test_particle_approximation}
\end{eqnarray}
Here, $N_T$ is the number of test particles per nucleon and $N = N_T A$. Each test particle carries a charge of the corresponding real particle divided by $N_T$ (for example, the baryon number of a ``nucleon test particle'' is $\frac{1}{N_T}$), so that the total charge in the simulation equals that of a system of $A$ particles. Propagating the test particles according to equations of motion governing the system, together with performing decays and particle-particle collisions, effectively solves Eq.\ (\ref{BUU_equation}). In \texttt{SMASH}, the equations of motion propagate the kinetic momentum of particles; see Eq.\ (\ref{kinetic_momentum}). An alternative approach, in which the canonical momenta are propagated, is possible \cite{Ko:1988zz}. For more technical details on the method of test particles, see Appendix \ref{the_method_of_test_particles}.
 
In practice, there exist two ways of realizing the method of test particles in hadronic transport. Within the first approach, one initializes a system with $N_TA$ test particles, which are then propagated according to the equations of motion. Scatterings are performed according to cross sections that are scaled as $\sigma/N_T$, where $\sigma$ is the physical cross section, which ensures that an average number of scatterings is the same as in a system of $A$ particles. Because each test particle carries a fraction $1/N_T$ of the charge of a corresponding real particle, the resulting mean field will be a smoothed out version of the mean field corresponding to $A$ particles. This approach is sometimes referred to as the ``full ensemble''.
 
An alternative approach is known as ``parallel ensembles'' \cite{Bertsch:1988ik}. In this paradigm, $N_T$ instances of a system of $A$ particles are created. Particles in each instance are propagated according to the equations of motion, and scatterings are performed using the physical cross section $\sigma$. Each test particle carries a fraction $1/N_T$ of the charge of a corresponding real particle, and the test particle densities (and consequently the mean fields) are calculated by summing contributions from all $N_T$ instances of the system. Evolving the $N_T$ systems with mean fields calculated in this fashion means that the systems are not in fact independent, and their evolution due to the mean fields is shared. At the same time, this approach is computationally much more efficient, as collision searches are performed only within individual instances of the system, thus reducing the numerical cost by a factor of $N_T^2$. 
 
It can be checked that these two simulation paradigms lead to the same results in typical cases \cite{Xu:2016lue}. In this study we utilized the full ensemble approach to the test particle method.

\section{Analysis}
\label{analysis}

In this paper we investigate simulations of nuclear matter in \texttt{SMASH} \cite{Weil:2016zrk} realized in a box with periodic boundary conditions. Such studies are best suited for testing the thermodynamic behavior following from equations of motion with mean-field interactions, as well as for exploring observables sensitive to critical phenomena in a scenario in which matter is allowed to equilibrate. While admittedly systems considered here cannot be reproduced in the laboratory, insights gained in this study will provide a useful stepping stone to understanding results of simulations of heavy-ion collisions utilizing the VDF EOS, planned for future work. 

In contrast to heavy-ion collision experiments, semiclassical hadronic transport simulations have an access to the positions of individual particles. Consequently, observables that can be used as a measure of the collective behavior of the system include the spatial pair correlation function and the distribution of particles in coordinate space. We describe the details of extracting these observables below.

\subsection{Pair distribution function}
\label{pair_correlation_function}

The radial distribution function $g(r)$ gives the probability of finding a particle at a distance $r$ from a reference particle. While in select simple cases it can be calculated analytically, in practice, given a distribution of particles, $g(r)$ is obtained by determining the distance between the reference particle and all other particles and constructing a corresponding histogram. Thus for finding the radial distribution about the $i$th (reference) particle at a given distance $r$, we count all particles within an interval $\Delta r$ around $r$, which can be written as
\begin{eqnarray}
\hspace{-5mm}g_i (r, \Delta r) = \sum_{\substack{j=1 \\ j\neq i}}^N \theta\Big(  r + \Delta r - \mathcal{R}_{ij}  \Big) \theta\Big( \mathcal{R}_{ij} - ( r - \Delta r )   \Big) ~.
\label{radial_distribution_function}
\end{eqnarray}
Here, the sum is performed over all particles (with the exception of the $i$th particle) which we index by $j$, $N$ is the total number of particles, $\theta$ is the Heaviside theta function, and $\mathcal{R}_{ij} = | \bm{r}_i - \bm{r}_j|$, where $\bm{r}_i$ is the position of the reference particle and $\bm{r}_j$ is the position of the $j$th particle. The role of the Heaviside theta functions is to only allow contributions from particles whose positions are within a distance $\mathcal{R}_{ij} \in (r - \Delta r, r + \Delta r)$ from the reference particle. The obtained histogram is then normalized with respect to an ideal gas, whose radial distribution histogram is that of completely uncorrelated particles, $g_{0}(r) \propto n ~ 4\pi r^2~ dr$, where $n$ denotes density. 

We can also define the radial distribution function of all distinct pairs in the system (which we also call the pair distribution function), 
\begin{eqnarray}
\hspace{-5mm}\widetilde{g} (r, \Delta r) &=& \mathcal{N}~ \sum_{i=1}^N g_i (r, \Delta r)  \non  \\
&&  \hspace{-15mm}=\frac{\mathcal{N}}{2} \sum_{i=1}^N \sum_{ \substack{j=1\\ j \neq  i}}^N \theta\Big(  r + \Delta r - \mathcal{R}_{ij}   \Big) \theta\Big( \mathcal{R}_{ij} - ( r - \Delta r )   \Big) ~,
\label{pair_distribution_function}
\end{eqnarray}
where the factor of $1/2$ appears to avoid counting any of the particle pairs twice, and where $\mathcal{N}$ is a normalization factor, so far unspecified (as already mentioned above, in practice the radial distribution function is compared to that of an ideal gas, in which case the normalization factors cancel out). The pair distribution function in an ideal gas, $\widetilde{g}_0(r)$, is related to $g_0(r)$ through $\widetilde{g}_0(r) \approx (N/2) g_0(r)$, where $N$ is the total number of particles in the system, which stems directly from the fact that the total number of distinct pairs in the system is equal $N(N-1)/2$. For simulations in a box with periodic boundary conditions, however, this relationship becomes more complicated for distances $r > L/2$, where $L$ is the side length of the box, due to geometry effects (see below). For this reason and because in simulations presented in this work we initialize the systems uniformly, in our analysis we use the $t=0$ histogram as the reference pair distribution function, $\widetilde{g}_0 = \widetilde{g}(t=0)$.

We stress that taking the pair distribution function of a uniform system as the reference ensures that the normalized pair distribution function, $\widetilde{g}/\widetilde{g}_0$, is sensitive to density fluctuations in the system. A prominent example here is the spinodal breakup, where a spontaneous separation into two coexistent phases with different densities occurs. If the system is confined to some constant volume $V$, then the average density of the system is the same before and after the spinodal decomposition takes place. %However, the pair distribution function is proportional to an average, calculated over a series of points, of density at these points weighted by the number of particles in the vicinity of these points, and as such will reflect the density fluctuations. 
However, local fluctuations in the number of particles will be visible in the pair distribution function, as more particle pairs reside inside a high density region as compared to a low density region.
	
While the spinodal decomposition is the most obvious example of a situation where $\widetilde{g}/\widetilde{g}_0 \neq 1$, the normalized pair distribution function deviates from unity for any system in which the interactions between the particles affect their collective behavior. In particular, at small $r$, the normalized pair distribution function satisfies $\widetilde{g}/\widetilde{g}_0 > 1$ for correlated particles and $\widetilde{g}/\widetilde{g}_0 < 1$ for anti-correlated particles (see Appendix \ref{pair_distribution_function_and_the_second-order_cumulant} for details), which corresponds to attractive and repulsive interactions between the particles, respectively. Since the number of particles and thus the number of pairs is conserved, one sees an opposite trend at intermediate to large distances. 

We note that in our simulations the range of $r$ over which $\widetilde{g}(r)/\widetilde{g}_0(r)$ deviates from 1 significantly is related to the range of the interaction, which is determined by the smearing range in the density calculation (for more details see Appendix \ref{the_method_of_test_particles}).

Importantly, for a system with periodic boundary conditions the radial distance between two particles $\mathcal{R}$ is not uniquely defined. This is because for any reference particle the distance to any other particle can be calculated using the position of that other particle in the original box or in any of its 26 equivalent images. We adopt a prescription in which the smallest distance between particles is used in calculating the pair distribution function $\widetilde{g}$ (known as the minimum image criterion). This smallest distance can range from $\mathcal{R}_{\txt{min}} = 0$ to $\mathcal{R}_{\txt{max}} = \frac{\sqrt{3}L}{2}$, where $L$ is the side length of the box. That said, even for a uniform and uncorrelated system the geometry of the problem affects the number of particles that can be encountered at the maximal distance $\mathcal{R}_{\txt{max}}$. Specifically, the only points for which it is possible to have $\mathcal{R} = \mathcal{R}_{\txt{max}}$ are points on the diagonal of the box; for any points separated by $\mathcal{R}_{\txt{max}}$ that are not on the diagonal, there exists a smaller $\mathcal{R}$ obtained by using the position of the second particle from one of the equivalent box images. This problem also affects, to a proportionally lesser extent, inter-particle distances $\mathcal{R}$ in the range  $\frac{L}{2} < \mathcal{R} < \mathcal{R}_{\txt{max}}$.  Only in the case of particles which are $\frac{L}{2}$ or less apart the geometry of the box never affects the pair distribution function.

This influence of finite size effects can be clearly seen in the left panel of Fig.\ \ref{spinodal_decomposition_nuclear}, which shows the pair distribution function for a box of side length $L = 10 \  \txt{fm}$ at initialization ($t=0$), when the system is uniform and the particles are uncorrelated. In infinite matter, the pair distribution function of uncorrelated particles grows like $r^2$. However, finite geometry effects described above introduce an effective cut on the distribution starting at $\frac{L}{2} = 5 \ \txt{fm}$, explaining the shape of the presented distribution. Similarly, geometry and periodic boundary conditions play a role in the shape of the normalized pair distribution function for $r > \frac{L}{2}$ at $t > 0$. In our simulations, nuclear spinodal decomposition at $T =1 \ \txt{MeV}$ results in a nuclear drop surrounded by a nearly perfect vacuum. (Here we note that the number of drops that form during spinodal decomposition depends on the size of the box, and the size of a drop depends on the smearing range used in density calculation; for more details on the latter, see Appendix \ref{the_method_of_test_particles}.) The diameter of the nuclear drop turns out to satisfy $D > \frac{L}{2}$, which means that for some of the particles belonging to that drop, the smallest distance to some of the other particles in that same drop will be ``across the vacuum'', to one of the equivalent mirror images of these particles. This explains the rise in the normalized distribution function for $r > \frac{L}{2}$ on the right panel in Fig.\ \ref{spinodal_decomposition_nuclear}. The magnitude of this effect depends on the drop diameter $D$.

The artifacts produced by the geometry of the problem and periodic boundary conditions do not present a significant complication in analyzing critical behavior if we resolve to only probe the system at length scales $\frac{L}{2}$ or smaller.

One may ask whether calculating a pair distribution function in hadronic transport is justified in view of the fact that the BUU equation explicitly evolves a one-body distribution function which does not carry any information about the two-body distribution, usually employed in the description of two-particle correlations. While this may appear to be problematic, a closer look reveals that such analysis is correct. First, one needs to note that hadronic transport simulations only solve the Boltzmann equation exactly in the limit of an infinite number of test particles per particle $N_T$. The finite number of test particles employed in simulations leads to intrinsic numerical fluctuations. These numerical fluctuations are of statistical nature, similarly to variances of microscopic observables, and likewise, through both scattering and mean fields, they can become a seed for collective behavior such as spontaneous spinodal decomposition. Such effects have been described, e.g., in \cite{Bonasera:1994iam} (see also \cite{Bonasera:1990ikj,Bonasera:1992zz}), where fluctuation observables calculated using hadronic transport with the method of test particles agree with both theoretical predictions and experimental results. Additionally, it was established that for large enough $N_T$ (which the authors of that particular study found to be $N_T \gtrsim 40$) the numerical noise intrinsic to the method of test particles is negligible, while the correct statistical fluctuations are preserved. 

It is possible to construct a Boltzmann-Langevin extension of the standard BUU equation, which ensures that the simulated fluctuations are physically correct (see, e.g., \cite{Burgio:1991ej}). However, it has been found that, for example, in the case of the nuclear spinodal fragmentation the source of the noise seeding the spinodal decomposition is not essential, and it is possible to develop good approximations to the Boltzmann-Langevin equation that are numerically favorable, including the method of test particles \cite{Chomaz:2003dz}. 

We note here that a particular problem that arises in the method of test particles is that the fluctuations in the events, simulating the evolution of $N_T N_B$ test particles, are suppressed by a factor of $N_T$. The authors of \cite{Bonasera:1994iam} dealt with this issue by employing the method of parallel ensembles at final simulation times, that is \textit{a posteriori}, which allows one to obtain events with the number of test particles corresponding to the physical baryon number $N_B$ (we briefly describe this method in Sec.\ \ref{implementation_in_SMASH}, while Appendix \ref{parallel_ensembles_in_SMASH} explains the \textit{a posteriori} application of the method).

Based on the above it is apparent that the distribution function obtained through hadronic transport simulations, and in particular through the method of test particles, contains information not only about the mean of the distribution function $\langle f_{\bm{p}} \rangle$, but also about its fluctuations. Consequently, calculating fluctuation observables such as the pair distribution function is well-defined in hadronic transport. Some questions regarding the quantitative behavior of fluctuation observables obtained in simulations using the number of test particles $N_T > 1$ remain, in particular regarding the specific methods used to connect fluctuations in systems evolving $N_T N_B$ particles as compared to systems evolving $N_B$ particles. For that reason we refrain from making quantitative statements at this time, and focus on the qualitative behavior of the pair distribution functions. Future work will be devoted to a quantitative analysis of this problem, and in section \ref{effects_of_finite_number_statistics} we give a short overview of the effects due to this issue.

\subsection{Number distribution functions}
\label{number_distribution_functions}

A complementary method of analyzing the collective behavior in a simulation utilizes coordinate space number distribution functions. To calculate number distribution functions, we divide the simulation box into $C$ cells of side length $\Delta l$ (also referred to as cell width), and construct a histogram of the number of cells in which the number of particles lies in a given interval $N_i \pm \Delta N$, where $N_i$ is the central value of the $i$th bin. We note that we scale the entries by the total number of cells $C$ so that the resulting histogram is a properly normalized representation of the corresponding probability distribution. We also note that in the subsequent parts of the paper we scale the histogram entries by the volume of the cells $(\Delta l)^3$ in order to obtain the histogram as a function of number density.

The test-particle evolution in \texttt{SMASH} is governed by the mean field, which depends on the underlying continuous baryon number density for a given baryon number $N_B$,
\begin{eqnarray}
n_B(\bm{x}; N_B) = g  \int \frac{d^3p}{(2\pi)^3} ~  f(\bm{x}, \bm{p})~.
\label{continuous_baryon_number_density_distribution_function}
\end{eqnarray}
Formally, hadronic transport can give access to $n_B(\bm{x}; N_B)$ through solving the Boltzmann equation, Eq.\ (\ref{BUU_equation}), in the limit of infinitely many test particles per particle, and substituting the obtained quasiparticle distribution function $f (\bm{x}, \bm{p})$ in Eq.\ \eqref{continuous_baryon_number_density_distribution_function}. Below, we present three number distribution functions accessible in practice given the finite number of test particles used.

\subsubsection{Test-particle-number distribution function}

Hadronic transport simulations of nuclear matter are realized through evolving $N = N_B N_T$ test particles in space and time (where $N_B$ is the baryon number in the simulation and $N_T$ is the number of test particles per particle), giving a direct access to a discrete test-particle-number distribution function. This distribution can be written as a probability of obtaining a cell contributing to the $i$th bin of the histogram with a center value $N_i$ (that is, a cell with $N \in (N_i - \Delta N, N_i + \Delta N)$ test particles),
\begin{eqnarray}
P_N (N_i) &=& P \Big(N_i, N (N_B, N_T), \Delta l\Big) = 
\label{test_particle_number_distribution_function_def}
\\
&=& \frac{N^{(i)}_c \Big(N( N_B, N_T), \Delta l\Big)}{C} ~.
\label{test_particle_number_distribution_function}
\end{eqnarray}
Here, $C$ is the total number of cells used and $N^{(i)}_c$ is the number of cells containing a number of test particles $N$ within the range $N_i \pm \Delta N$. We note that the number of test particles in any given cell depends both on the baryon number evolved in the simulation, $N_B$, and the number of test particles per particle, $N_T$. We also stress that the distribution $P_N$ depends on the scale (chosen cell width $\Delta l$) at which the system is analyzed.

\subsubsection{Continuous baryon number distribution function}
\label{Continuous_baryon_number_distribution_function}

The discrete test particle distribution function, Eq.\ \eqref{test_particle_number_distribution_function_def}, can be thought of as having been obtained through sampling from the underlying continuous baryon number distribution function with a finite number $N_T N_B$ of test particles. Given access to the underlying baryon number distribution, one could use it directly to create a corresponding histogram. Indeed, the number of baryons at a cell at position $\bm{x}_k$ is given by the integral of the continuous baryon number density, Eq.\ (\ref{continuous_baryon_number_density_distribution_function}), 
\begin{eqnarray}
B ( \bm{x}_k) = \int_{V_k = (\Delta l)^3} dV~  n_B (\bm{x}, N_B) ~,
\end{eqnarray}
where $k$ indexes the histogram cells. Adding contributions from all cells yields the total baryon number in the system, $B$. We can then construct a probability distribution function for encountering a cell with a given number of baryons $N_i$,
\begin{eqnarray}
P_B (N_i) = \frac{N_{c}^{(i)}  \Big(N_i, B, \Delta l\Big)  }{C}~,
\end{eqnarray}
where $N_{c}^{(i)}$ is the number of cells containing a number of baryons $N$ within the range $N_i \pm \Delta N$. 

For a large number of test particles per particle $N_T$, statistical observables calculated using the test-particle-number distribution, with the number of test particles in a given sample scaled by $\frac{1}{N_T}$, are a very good approximation to the underlying continuous baryon number distribution \cite{Steinheimer:2017dpb}. That is, it can be shown that
\begin{eqnarray}
P_B (N_i) = \lim_{N_T \to \infty} P \bigg(N_i , \frac{N(N_B, N_T)}{N_T}, \Delta l\bigg) ~.
\label{baryon_number_distribution_function}
\end{eqnarray}
Given that in our simulations we use sufficiently large numbers of test particles per particle $N_T$, we will refer to histograms constructed through the prescription on the right-hand side of Eq.\ (\ref{baryon_number_distribution_function}) as the continuous baryon number distribution function (or just baryon number distribution function) $P_B(N_i)$, with the understanding that it is only exact in the limit $N_T \to \infty$.

\subsubsection{Physical baryon number distribution function}
\label{Physical_baryon_number_distribution_function}

Both the test particle and the continuous baryon number distribution functions, Eqs.\ (\ref{test_particle_number_distribution_function_def}) and (\ref{baryon_number_distribution_function}), are markedly different from the physical baryon number distribution function corresponding to a discrete baryon number $N_B$. Here we can intuitively think of the physical baryon number distribution function as obtained through sampling from the underlying continuous baryon number distribution with $N_B$ test particles, 
\begin{eqnarray}
P_{N_B}(N_i) = P \Big(N_i, N ( N_B, N_T = 1), \Delta l\Big) ~.
\label{physical_baryon_number_distribution_function}
\end{eqnarray}

Strictly speaking, the physical baryon number distribution function could be obtained in transport by solving the Boltzmann equation in the limit of infinitely many test particles per particle, thus obtaining the underlying continuous baryon number distribution function, Eq.\ \eqref{continuous_baryon_number_density_distribution_function}, and sampling $n_B(\bm{x}, N_B)$ with $N_B$ particles. Naturally, this is a numerically feasible but tedious approach. Alternatively, one can turn to the concept of parallel ensembles (introduced in Sec.\ \ref{implementation_in_SMASH}). It can be shown that the test particle distribution obtained within a parallel ensembles mode serves as a proxy for the physical baryon number distribution. To reiterate, within the concept of parallel ensembles, a simulation corresponding to $N_B$ baryons with $N_T$ test particles per baryon is divided into $N_T$ events with $N_B$ test particles each. These $N_T$ events are not independent, as they share a common mean field. Nevertheless, at the end of the simulation we have access to $N_T$ events with the test particle number exactly corresponding to the baryon number in the ``real'' system. That is, each of the $N_T$ events is described by the probability distribution function $P_{N_B}(N_i) = P\Big(N_i; N (N_B, N_T = 1); \Delta l\Big)$. Observables calculated using $P_{N_B}(N_i)$ are probably the closest to those one would find in an experiment if one could measure positions of the  particles. We postpone a rigorous derivation of this result and corresponding investigations to a future work.

\section{Infinite matter simulation results}
\label{simulation_results}

\begin{figure}[t]
	\includegraphics[width=\columnwidth]{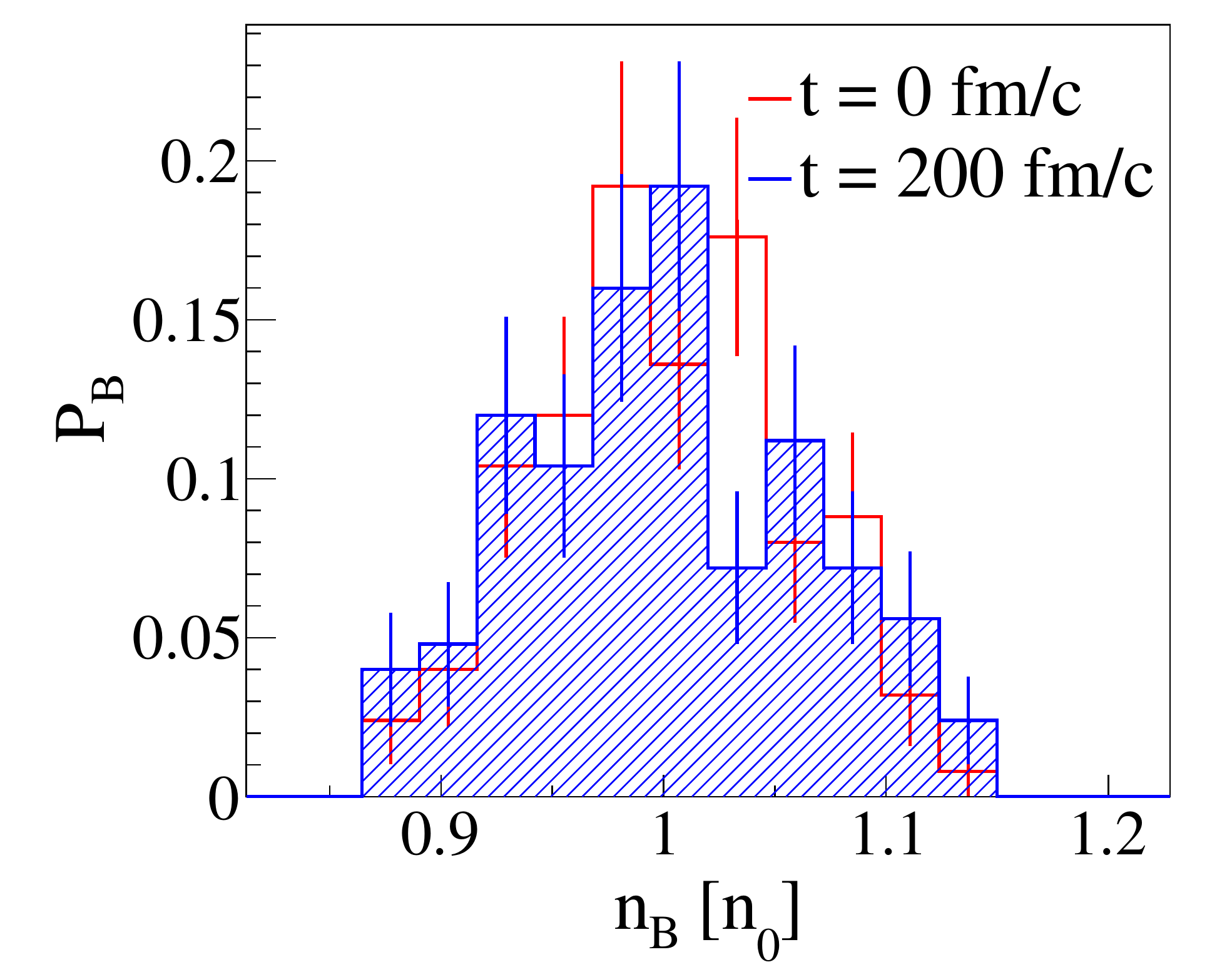}
	\includegraphics[width=\columnwidth]{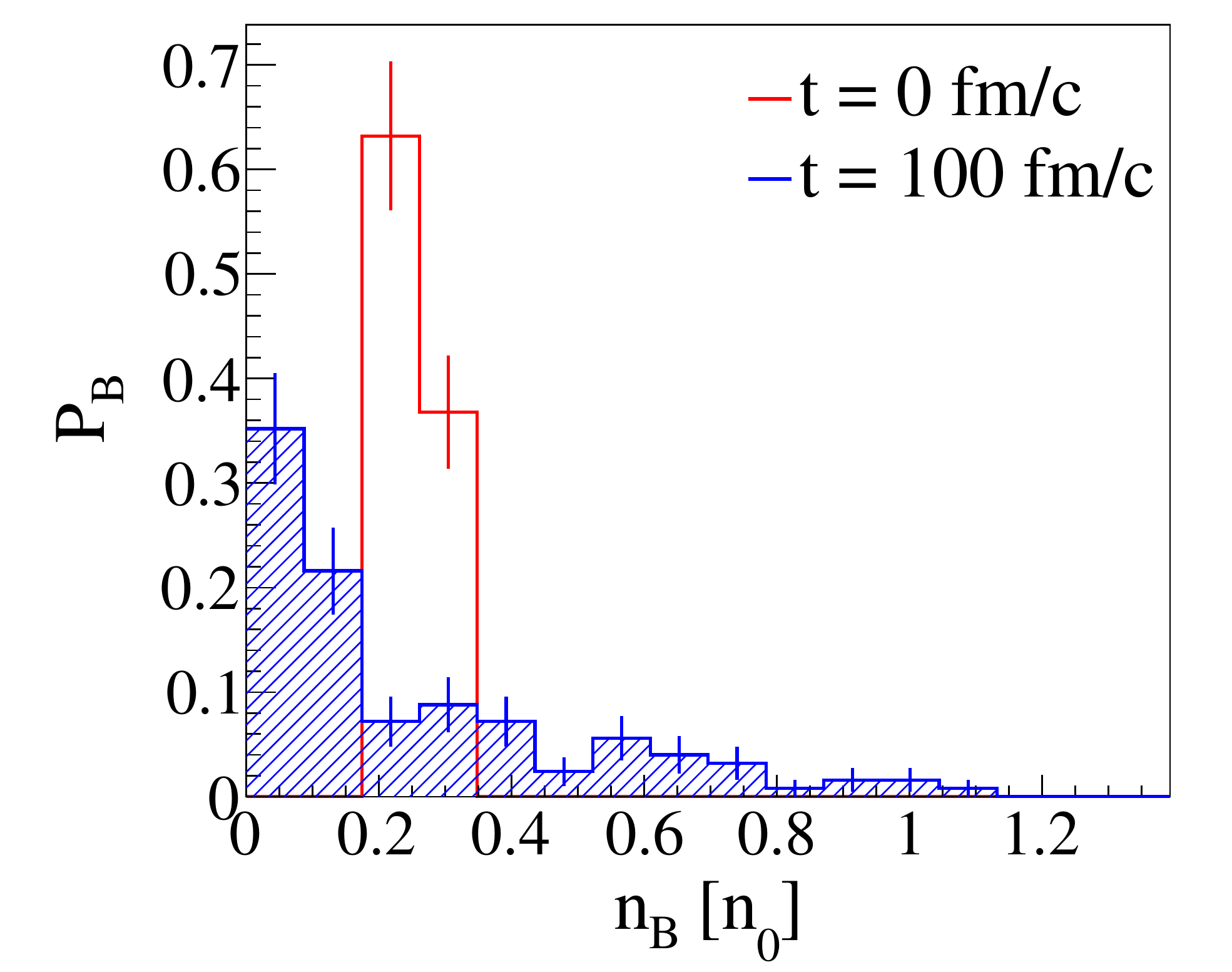}
	\caption{(Color online) Baryon number distribution, scaled by the volume of the cell and shown in units of the saturation density of nuclear matter, $n_0 = 0.160\  \txt{fm}^{-3}$. The cell width is chosen at $\Delta l = 2\ \txt{fm}$. Histograms delineated with red curves correspond to distributions at initialization ($t = 0$), while histograms delineated and shaded with blue curves correspond to distributions at the end of the evolution, $t = t_{\txt{end}}$. Upper panel: Nuclear matter initialized at the saturation density $n_0$ and temperature $T = 1 \ \txt{MeV}$, evolved until $t_{\txt{end}} = 200\ \txt{fm}/c$. The system, initialized at equilibrium, remains in equilibrium at $t_{\txt{end}}$. Lower panel: Nuclear matter initialized inside the spinodal region of the nuclear phase transition, at baryon number density $n_B = 0.25 n_0$ and temperature $T = 1\ \txt{MeV}$, evolved until $t_{\txt{end}} = 100\ \txt{fm}/c$. The system, initialized in a mechanically unstable region of the phase diagram, undergoes a spontaneous separation into a (very dilute) nucleon gas and a nuclear liquid drop with a central density $n_B \approx n_0$.}
	\label{cell_particle_number_distribution}
\end{figure}

\begin{figure*}[t]
	\includegraphics[width=\textwidth]{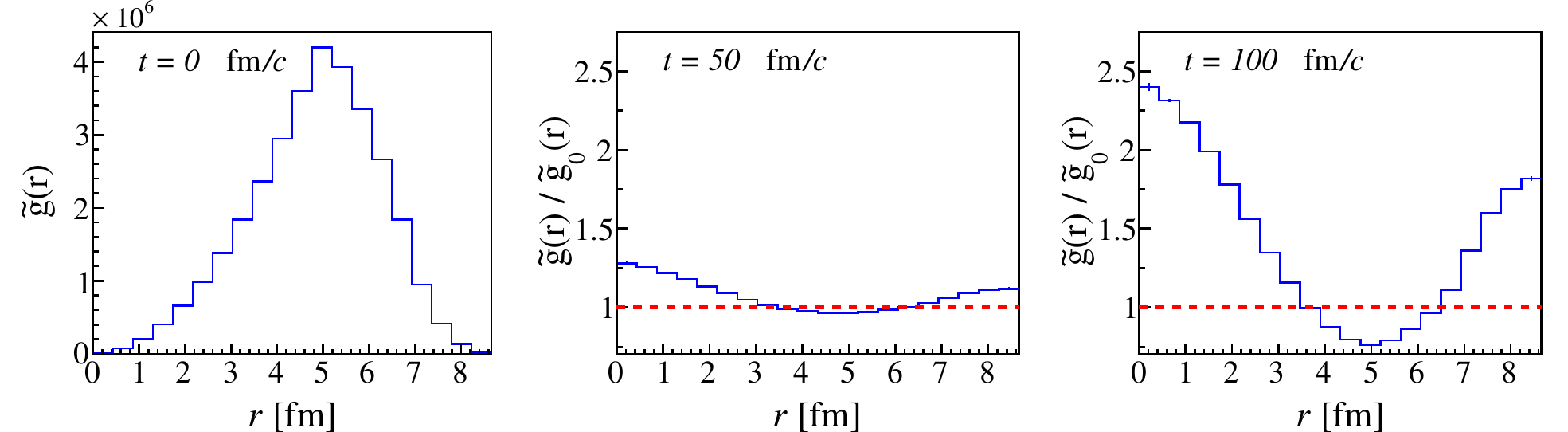}
	\caption{(Color online) Time evolution of the pair distribution function for a system initialized inside the nuclear spinodal region (at baryon number density $n_B = 0.25\ n_0$ and temperature $T = 1 \ \txt{MeV}$). The $t =0$ plot (left) shows the distribution at initialization, while plots at $t = 50$ and $100 \ \txt{fm}/c$ (middle and right, respectively) show normalized distributions. Spontaneous spinodal decomposition occurs at $t > 0$ and leads to a formation of a nuclear drop surrounded by a near-perfect vacuum, resulting in a strong correlation between particles clustered within the drop. See Sec.\ \ref{pair_correlation_function} for a discussion of the influence of finite-size effects on the shape and large-distance behavior of the pair distribution function.}
	\label{spinodal_decomposition_nuclear}
\end{figure*}

To simulate isospin-symmetric infinite nuclear matter, we initialize equal numbers of proton and neutron test particles in a box with periodic boundary conditions. The side length of the box is taken to be $L = 10 \ \txt{fm}$; this is informed by the fact that with periodic boundary conditions, the box can be kept relatively small with no significant finite-size effects. The time step used in the simulation needs to be small enough to resolve all gradients occurring during the evolution (intuitively speaking, a test particle should not "jump over" a potential gradient within a single time step). We found that a time step of $\Delta t = 0.1 \ \txt{fm}/c$ is small enough to satisfy this condition, and it correctly solves the equations of motion, Eqs.\ (\ref{EOM_covariant_formulation_x}) and (\ref{EOM_covariant_formulation_p}), using the leapfrog algorithm. The mean-field is calculated on a lattice with lattice spacing $a = 1 \ \txt{fm}$, which has been tested to be sufficiently fine for accurately resolving mean-field gradients. To ensure smooth density and density gradient calculations, we utilize a large number of test particles per particle, specifically, we use $N_T = 200$ for ordinary nuclear matter (Sec.\ \ref{nuclear_phase_transition}) and $N_T = 50$ for dense nuclear matter (Secs. \ref{quark-hadron_phase_transition} and \ref{effects_of_finite_number_statistics}). Using different numbers of test particles in these two cases is justified by the fact that smooth density and density gradient calculations are ensured when the average number of test particles encountered in a cell of the lattice, $N_{\txt{avg}}$, is large enough. As an example, within the described setup, this number will be equal to $N_{\txt{avg}} = 8$ for ordinary nuclear matter at $n_B = 0.25\ n_0$, and equal to $N_{\txt{avg}} = 24$ for dense nuclear matter at $n_B = 3\ n_0$. We choose $N_{\txt{avg}}$ to be bigger in the case of dense nuclear matter as mean-fields encountered in that region of the phase diagram are significantly larger and require an even more smooth gradient computation.

For studying the thermodynamic behavior of nuclear matter, we are simulating systems in which all collision and decay channels are turned off. We have checked that the thermodynamic effects described here persist when collisions are allowed, and in this work we choose to omit them because our goal is to study mean-field dynamics. As in Sec.\ \ref{Cumulants_of_baryon_number}, we are considering only one of the many EOSs accessible within the VDF model, namely, the one corresponding to the fourth (IV) set of characteristics listed in Table \ref{example_characteristics}. The choice of this set is arbitrary and serves as an illustration of the properties of the VDF model which are qualitatively comparable for all obtained EOSs.

\subsection{Ordinary nuclear matter}
\label{nuclear_phase_transition}

We investigate the behavior of systems initialized at temperatures and baryon number densities specific to ordinary nuclear matter to validate the implementation of the VDF model in \texttt{SMASH} \cite{Weil:2016zrk}. For illustrative purposes, we discuss results for a single simulation run, that is one event. Remarkably, the thermodynamic behavior of the system is apparent already for this minimal statistics. This is a consequence of the large number of test particles per particle used ($N_T = 200$), as well as the fact that the investigated effects are characterized by large fluctuations, which result in clear signals.

To start, we initialize symmetric nuclear matter at saturation density $n_B = n_0$, which for the box setup described above corresponds to the number of protons and neutrons $N_p = N_n = 80$, and at temperature $T = 1 \ \txt{MeV}$. Except for a slight increase in temperature from the degenerate limit, which is not significant enough to introduce any relevant changes, this is the equilibrium point of nuclear matter. We let the simulation evolve until $t_{\txt{end}} = 200 \  \txt{fm}/c$ and investigate whether the equilibrium is preserved by hadronic transport. To address this question, we examine the continuous baryon number distribution function (for details, see Sec.\ \ref{number_distribution_functions}), which we calculate using the cell width $\Delta l = 2 \ \txt{fm}$; we scale the histogram entries by the volume of the cell to obtain the distribution in units of the baryon number density, and further scale the results to express them in units of the saturation density, $n_0 = 0.160 \ \txt{fm}^{-3}$. As expected for matter in equilibrium, the baryon number distribution remains unchanged throughout the evolution, as can be seen in the upper panel of Fig.\ \ref{cell_particle_number_distribution}. We also find that throughout the simulation, the binding energy per particle agrees with the theoretically obtained value within $0.1\%$ (for more details on energy evolution, see Appendix \ref{energy_evolution}). An in-depth discussion of the mean-field response to fluctuations around nuclear saturation density, comparing the results from several transport codes including \texttt{SMASH} utilizing the VDF model, can be found in \cite{Colonna:2021xuh}.

\begin{figure*}[t]
	\includegraphics[width=\textwidth]{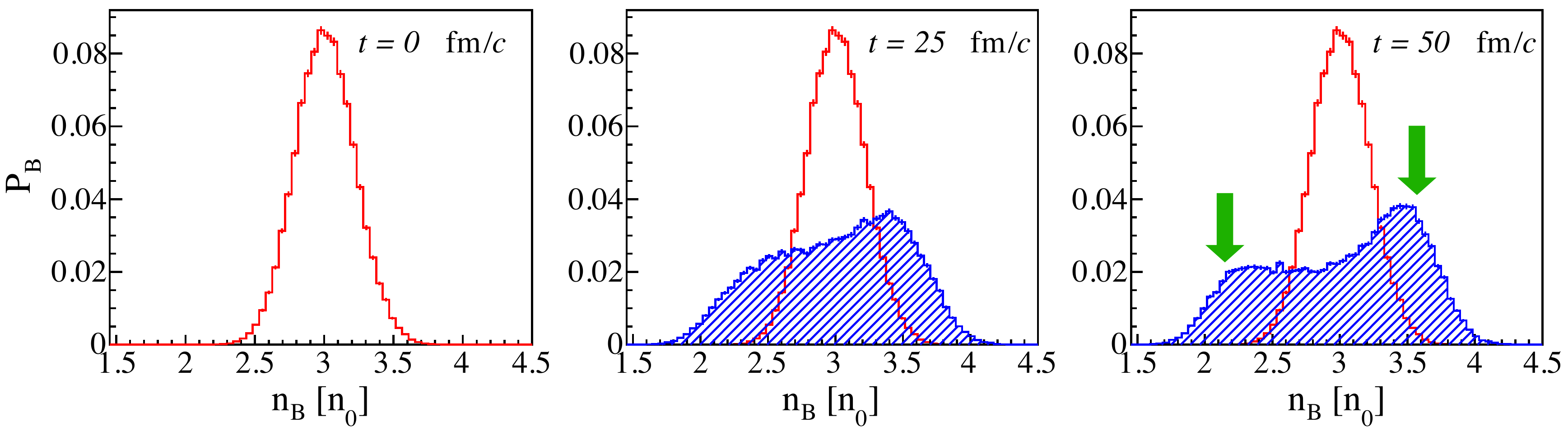}
	%{Spinodal_decomposition_hadron_histograms_full.pdf}
	\caption{(Color online) Time evolution of the baryon number distribution, scaled by the volume of the cell and shown in units of the saturation density of nuclear matter, $n_0 = 0.160\  \txt{fm}^{-3}$, for a system initialized inside the quark-hadron spinodal region (at baryon number density $n_B = 3\ n_0$ and temperature $T = 1 \  \txt{MeV}$), averaged over $N_{\txt{ev}} = 500$ events. The cell width is chosen at $\Delta l = 2\ \txt{fm}$. Histograms delineated with red curves correspond to the baryon distribution at initialization ($t = 0$), while histograms delineated and shaded with blue curves correspond to baryon distributions at a chosen time during the evolution ($t = \{25, 50\}\ \txt{fm}/c$). The system, initialized in a mechanically unstable region of the phase diagram, undergoes a spontaneous separation into a less dense and a more dense nuclear liquid (see Sec.\ \ref{parametrization} for more discussion), resulting in a double-peaked baryon number distribution. The green arrows point to values of baryon number densities corresponding to the boundaries of the coexistence region at $T=1\ \txt{MeV}$, $n_L = 2.13\ n_0$ and $n_R = 3.57\ n_0$.}
	\label{spinodal_decomposition_hadron_histograms_full}
\end{figure*}

Next, we model nuclear matter inside the spinodal region of the nuclear phase transition. Specifically, we initialize the system with the number of protons and neutrons $N_p = N_n = 20$, corresponding to a baryon number density $n_B = 0.25\ n_0$, at temperature $T = 1 \ \txt{MeV}$. We let the system evolve until $t_{\txt{end}} = 100 \  \txt{fm}/c$. The spinodal region is both thermodynamically and mechanically unstable, and so we expect that local density fluctuations will drive the matter to separate into two coexisting phases: a dense phase, also known as a nuclear drop, and a dilute phase which is a nucleon gas. That this indeed happens can be seen on the lower panel in Fig.\ \ref{cell_particle_number_distribution}, which shows the change in the baryon number distribution function due to the system's separation into two coexisting phases. The distribution, initially centered at $n_B = 0.25 \ n_0$, at the end of the evolution has a large contribution at $n_B \approx 0$ and a long tail reaching out to $n_B \approx n_0$, which corresponds to the center of the nuclear drop.

We then proceed to calculate the pair distribution function (for details, see Sec.\ \ref{pair_correlation_function}) for the system initialized in the spinodal region of nuclear matter. The results are shown in Fig.\ \ref{spinodal_decomposition_nuclear}. Here, the three panels correspond to three time slices of the evolution: $t = 0, 50, 100 \  \txt{fm}/c$. The $t =0$ plot (left) shows the pair distribution function, Eq.\ (\ref{pair_distribution_function}), at initialization $\widetilde{g}_0(r, \Delta r)$, while plots at $t = 50$ and $100 \ \txt{fm}/c$ (middle and right, respectively) show normalized pair distribution functions $\widetilde{g}(r, \Delta r)/\widetilde{g}_0(r, \Delta r)$. The time evolution of the pair distribution function shows that during the spinodal decomposition the test particles cluster into the nuclear drop. The half width at half maximum of the pair distribution function is about $2\ \txt{fm}$, which corresponds to the density smearing range used (see Appendix \ref{the_method_of_test_particles} for more details). The influence of the periodic boundary conditions on the shape and behavior of the pair distribution function at large inter-particle distances is discussed in Sec.\ \ref{pair_correlation_function}.

All of the results presented above demonstrate that the VDF equations of motion implemented in \texttt{SMASH} reproduce the expected bulk behavior of ordinary nuclear matter.

\subsection{Dense nuclear matter and the QGP-like phase transition}
\label{quark-hadron_phase_transition}

For simulations of critical behavior in dense symmetric nuclear matter, we run $N_{\txt{ev}} = 500$ events and average the results, calculated event-by-event. We first initialize the system at $n_B = 3\ n_0$, which corresponds to the number of protons and neutrons $N_p = N_n = 240$, and at temperature $T = 1 \ \txt{MeV}$. It can be seen in Figs.\ \ref{phase_diagram} and \ref{Cumulants_diagrams} that this corresponds to initializing dense nuclear matter inside the spinodal region of the QGP-like phase transition described by the EOS employed (the fourth (IV) set of characteristics listed in Table \ref{example_characteristics}). We evolve the system until $t_{\txt{end}} = 50 \ \txt{fm}/c$, which is sufficient for reaching equilibrium after a spinodal decomposition at high baryon number densities, since due to considerably larger values of the mean-field forces on test particles the density instabilities develop more rapidly.

\begin{figure*}[t]
	\includegraphics[width=\textwidth]{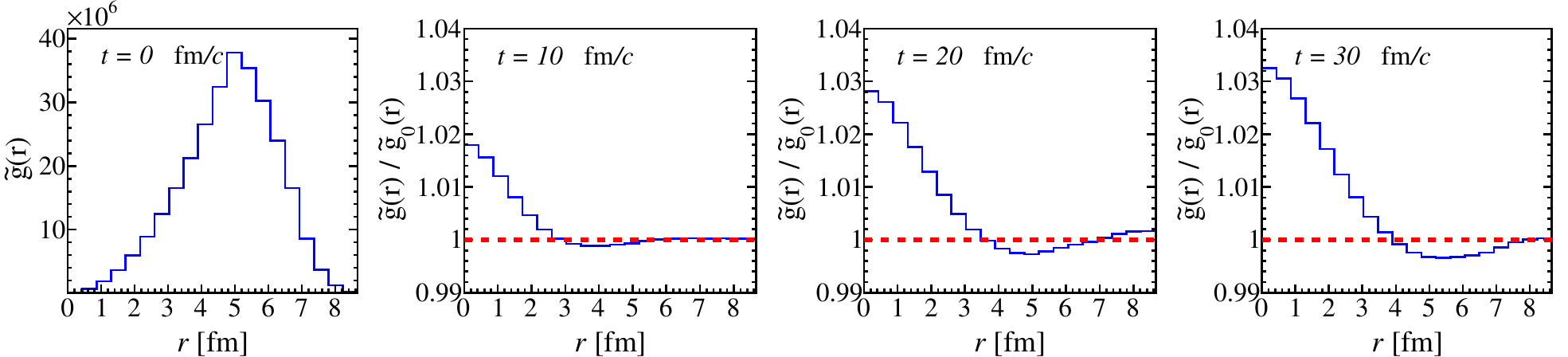}
	\caption{(Color online) Time evolution of the pair distribution function for a system initialized inside the QGP-like spinodal region (at baryon number density $n_B = 3.0 \ n_0$ and temperature $T = 1 \ \txt{MeV}$), averaged over $N_{\txt{ev}} = 500$ events. The $t = 0$ plot (first panel) shows the pair distribution at initialization, while plots at $t = 10, \ 20, \ 30\ \txt{fm}/c$ (second, third, and fourth panels) show normalized pair distributions. Spontaneous spinodal decomposition occurs at $t>0$ and leads to a formation of two coexisting phases: a less dense and a more dense nuclear liquid. The increased relative concentration of particles in the more dense phase results in an elevated normalized pair distribution at small distances.
	}
	\label{spinodal_decomposition_hadron_correlations_full}
\end{figure*}

In Fig.\ \ref{spinodal_decomposition_hadron_histograms_full}, we show the evolution of the baryon number distribution (see Sec.\ \ref{Continuous_baryon_number_distribution_function}). The cell width is chosen at $\Delta l = 2\ \txt{fm}$, and the histogram entries are scaled by the volume of the cell in order to be given in units of the baryon number density; we then further scale the results to express them in units of the saturation density, $n_0 = 0.160 \ \txt{fm}^{-3}$. In the figure, the red curve corresponds to the distribution at time $t = 0$, while the blue curves delineate the distribution at times $t>0$. At $t=0$, the distribution is peaked at the initialization density $n_B = 3\ n_0$, with its width reflecting the finite number statistics. In the course of the evolution the system separates into two coexisting phases, a ``less dense'' and a ``more dense'' nuclear liquid (see section \ref{parametrization} for more discussion). As a result, the baryon distribution displays two peaks largely coinciding with the theoretical values of the coexistence region boundaries, $n_L = 2.13\ n_0$ and $n_R = 3.57\ n_0$, indicated by the green arrows. We find that the prominence of the peaks depends slightly on the choice of the EOS. For example, an equation of state with the same value of critical density $n_c^{(Q)}$ and the same spinodal region $(\eta_L, \eta_R)$, but a higher critical temperature $T_c^{(Q)}$ will correspond to a more negative slope of the pressure in the spinodal region and, correspondingly, to stronger mean-field forces inside the spinodal region, leading to more prominent peaks.

Next, in Fig.\ \ref{spinodal_decomposition_hadron_correlations_full} we show the evolution of the pair distribution function. Similarly as in the case of nuclear spinodal decomposition, the ``hadron-quark'' spinodal decomposition leads to a pair distribution function indicating the formation of two phases of different densities. Unlike in nuclear spinodal decomposition, where drops of a ``nuclear liquid'' form in vacuum, in this case we have drops of a ``more dense liquid'' submerged in a ``less dense liquid'' (for a detailed discussion, see section \ref{parametrization}). Consequently, the absolute values of the normalized pair distribution function, $\widetilde{g}(r)/\widetilde{g}_0(r)$, are much smaller for the case of the ``hadron-quark'' spinodal decomposition, as the difference between the number of test particle pairs occupying the dense and dilute regions is less pronounced in this case. Nevertheless, the effect, although small, is clearly distinguishable and statistically significant.

We note here that a phase separation is such a distinct behavior of the system that the baryon distribution function and the pair distribution function as shown in Figs.\ \ref{spinodal_decomposition_hadron_histograms_full} and \ref{spinodal_decomposition_hadron_correlations_full}, respectively, can be largely recovered even in the case of minimal statistics, that is for one event. However, effects at and around the critical point, as discussed below, are much more subtle and require a relatively large number of events.

\begin{figure*}[t]
	\includegraphics[width=\textwidth]{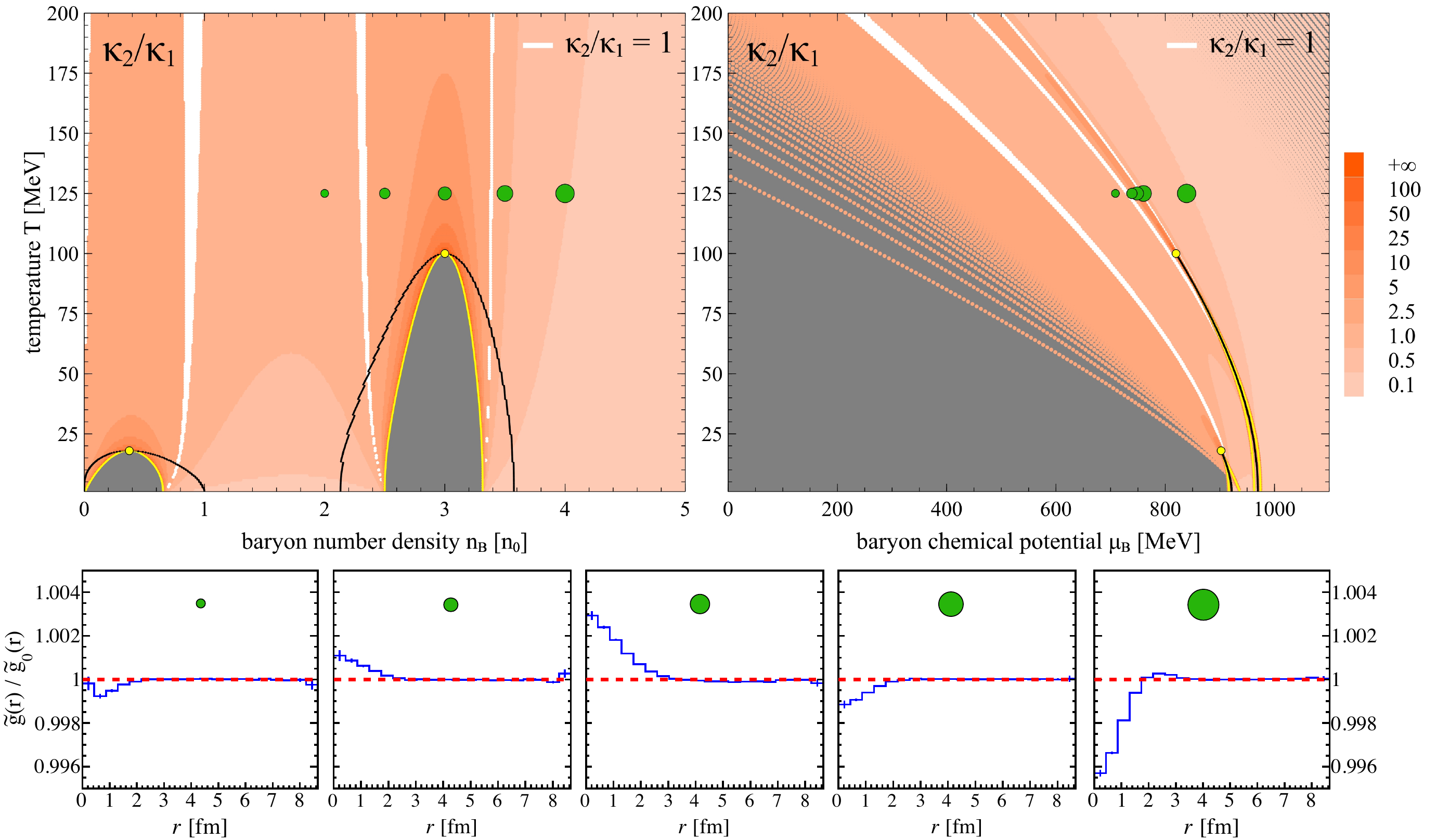}	
	\caption{(Color online) Comparison of the cumulant ratio $\kappa_2/\kappa_1$, calculated within the VDF model, and normalized pair distribution function at $t = 30\ \txt{fm}/c$, for a series of chosen initialization points. The description of the cumulant diagrams (upper panel) is the same as in Fig.\ \ref{Cumulants_diagrams}. The dots on the cumulant diagrams mark the points at which we initialize the system, specified in Eq.\ (\ref{chosen_densities}), and are intended to guide the eye toward the corresponding normalized pair distribution plots (lower panel). The deviation of the normalized pair distributions from the normalized pair distributions of a perfectly uncorrelated system (red line) directly follows the deviation of values of the cumulant ratio $\kappa_2/\kappa_1$ from the Poissonian limit of 1. See text for more details.}
	\label{correlations_vs_cumulants}
\end{figure*}

To conclude our study of dense nuclear matter in \texttt{SMASH}, we want to investigate the behavior of systems initialized at various points of the phase diagram above the critical point, inspired by possible phase diagram trajectories of heavy-ion collisions at different beam energies. Specifically, we initialize the system at one chosen temperature and a series of baryon number densities 
\begin{eqnarray}
\hspace{-5pt}T = 125~\txt{MeV}, \hspace{5pt} n_B \in \{ 2.0, 2.5, 3.0, 3.5, 4.0 \}~n_0~.
\label{chosen_densities}
\end{eqnarray}
In contrast with most of the previous examples, systems initialized in this region of the phase diagram are thermodynamically stable, and there are specific predictions for the behavior of thermodynamic observables such as ratios of cumulants of baryon number (see Fig.\ \ref{Cumulants_diagrams}). In the upper panel of Fig.\ \ref{correlations_vs_cumulants}, we show values of the second-order cumulant ratio, $\frac{\kappa_2}{\kappa_1}$, as calculated from the VDF model, both in the $T$-$n_B$ and the $T$-$\mu_B$ plane. The dots on the cumulant diagrams mark the points at which we initialize the system, specified in Eq.\ (\ref{chosen_densities}), and are intended to guide the eye toward the corresponding normalized pair distribution plots at the end of the evolution, $t = t_{\txt{end}}$, displayed in the lower panel of the same figure. The deviation of values of the normalized pair distributions at small distances from 1 (where 1 corresponds to a system of non-interacting particles) directly follows the deviation of values of the second-order cumulant ratio $\frac{\kappa_2}{\kappa_1}$ from the Poissonian limit of 1,
\begin{eqnarray}
&& \frac{\widetilde{g}~(0, \Delta r)}{\widetilde{g}_0(0, \Delta r)} > 1  ~ \Leftrightarrow ~  \frac{\kappa_2}{\kappa_1} > 1  ~, 
\label{pair_distribution_kappa_2_1} \\
&& \frac{\widetilde{g}~(0, \Delta r)}{\widetilde{g}_0(0, \Delta r)} < 1    ~\Leftrightarrow ~ \frac{\kappa_2}{\kappa_1} < 1 ~.
\label{pair_distribution_kappa_2_2}
\end{eqnarray}
We show a detailed derivation of this fact in Appendix \ref{pair_distribution_function_and_the_second-order_cumulant}. It is clear that a two-particle correlation corresponds to a value of the cumulant ratio $\frac{\kappa_2}{\kappa_1} > 1$, and a two-particle anticorrelation corresponds to a value of the cumulant ratio $\frac{\kappa_2}{\kappa_1} < 1$. This behavior is exactly reflected in Fig.\ \ref{correlations_vs_cumulants}. 

We want to stress that the pair distributions shown in Fig.\ \ref{correlations_vs_cumulants} develop relatively fast. In Fig.\ \ref{spinodal_decomposition_hadron_correlations_full}, where we explored the behavior of a system initialized at a temperature $T = 1 \ \txt{MeV}$, one can see by comparing the second and the fourth panels that already at $t = 10 \ \txt{fm}/c$ a significant part of the pair distribution has developed. This effect is further magnified at higher temperatures, where relatively larger momenta of the test particles result in a faster propagation of effects related to mean fields. For systems shown in Fig.\ \ref{correlations_vs_cumulants}, we have verified that the majority of the pair distribution function development occurs within $\Delta t = 3 \ \txt{fm}/c$.

These results show not only that hadronic transport is sensitive to critical behavior of systems evolving above the critical point, but also that this behavior is exactly what is expected based on the underlying model. Moreover, we note that the behavior of both the second-order cumulant and the pair distribution function across the region of the phase diagram affected by the critical point is remarkably distinct. It is evident that an equilibrated system traversing the phase diagram through the series of chosen points, Eq.\ (\ref{chosen_densities}), follows a clear pattern: first displaying anticorrelation, then correlation, and then again anticorrelation. Thus already the second-order cumulant ratio presents sufficient information to explore the phase diagram, and, provided that correlations in the coordinate space are transformed into correlations in the momentum space during the expansion of the fireball, this pattern may be utilized to help locate the QCD critical point, in addition to signals carried by the third- \cite{Asakawa:2009aj} and fourth-order \cite{Stephanov:2011pb} cumulant ratios. This may prove to be especially important given that the quantity observed in heavy-ion collision experiments is not the net baryon number, but the net proton number. In calculations of the net baryon number cumulants based on the net proton number cumulants, the higher order observables are increasingly more affected by Poisson noise \cite{Kitazawa:2011wh}. In view of this, the second-order cumulant ratio (or equivalently the two-particle correlation) could be considered among the key observables utilized in the search for the QCD critical point, and it remains to be seen if this somewhat smaller signal (as compared to higher order cumulant ratios) is nevertheless noteworthy due to the much higher precision with which it can be measured in experiments.

\begin{figure}[t]
	\includegraphics[width=\columnwidth]{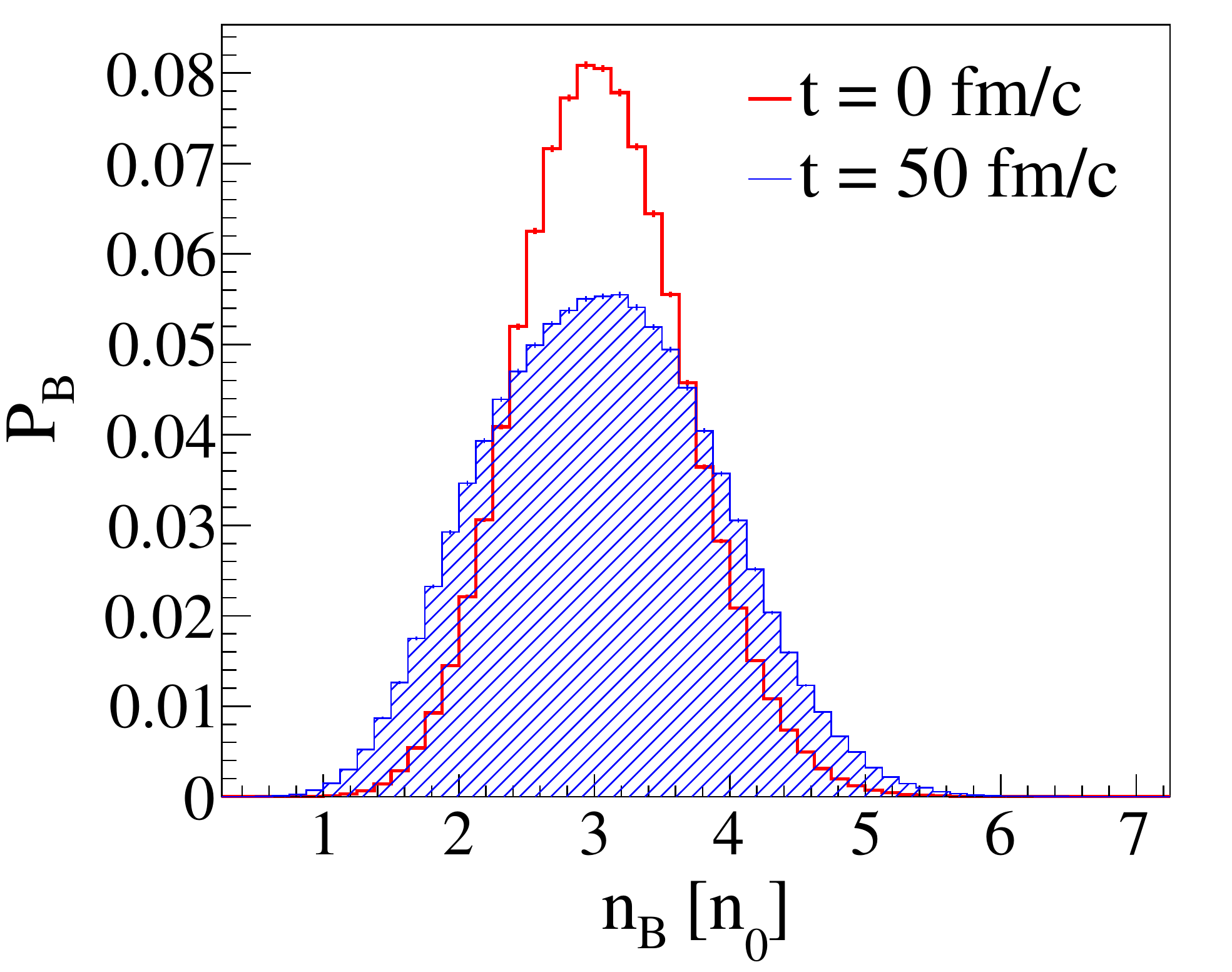}
	\caption{(Color online) Time evolution of the continuous baryon number distribution, scaled by the volume of the cell and shown in units of the saturation density of nuclear matter, $n_0 = 0.160\  \txt{fm}^{-3}$, for the same system as described in Fig.\ \ref{spinodal_decomposition_hadron_histograms_full}, averaged over $N_{\txt{ev}} = 500$ events. Here, cell width is chosen at $\Delta l = 1\ \txt{fm}$. The histogram delineated with the red curve corresponds to the baryon distribution at initialization ($t = 0$), while the histogram delineated and shaded with the blue curve corresponds to the distribution at the end of the evolution ($t_{\txt{end}} = 50 \ \txt{fm}/c)$. Nuclear matter, initialized in a mechanically unstable region of the phase diagram, undergoes a spontaneous separation into a less dense and a more dense nuclear liquid (see Sec.\ \ref{parametrization} for more discussion). Correspondingly, the distribution function becomes wider with time; however, due to the size of the binning cell, the average number of test particles in a cell is small and consequently the double-peaked structure, clearly seen on the right panel in Fig.\ \ref{spinodal_decomposition_hadron_histograms_full}, is washed out by Poissonian fluctuations. See text for more details.}
	\label{spinodal_decomposition_hadron_histograms_full_smaller_cell}
\end{figure}

\subsection{Effects of finite number statistics}
\label{effects_of_finite_number_statistics}

Qualitative and quantitative features of observables are influenced by the finite number of particles in analyzed samples. When analyzing observables such as the baryon distribution, one has to keep in mind that fluctuations due to finite number statistics may wash out the expected signals. This is not only a numerical problem but, as we shall discuss below, is also an issue relevant for experiments.

\begin{figure}[t]
	\includegraphics[width=\columnwidth]{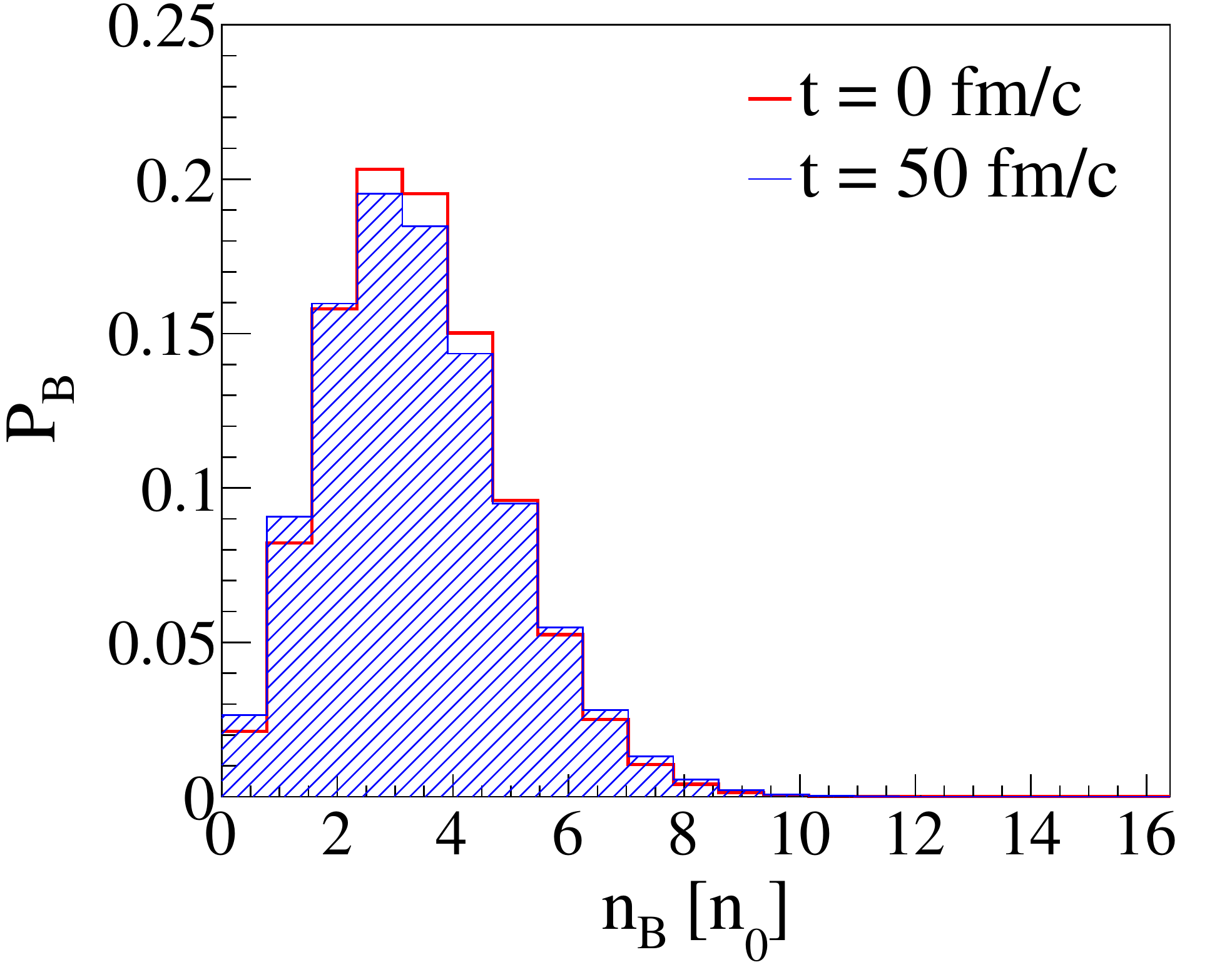}
	\caption{(Color online) Time evolution of the continuous baryon distribution, scaled by the volume of the cell and shown in units of the saturation density of nuclear matter, $n_0 = 0.160\  \txt{fm}^{-3}$, for the same system as described in Figs.\ \ref{spinodal_decomposition_hadron_histograms_full} and \ref{spinodal_decomposition_hadron_histograms_full_smaller_cell}, but calculated using the parallel ensembles method; the results are averaged over $N_{\txt{ev}}^{(\txt{parallel})} = N_T \times N_{\txt{ev}} = 25,000$ events. Cell width is chosen at $\Delta l = 2\ \txt{fm}$. The red curve corresponds to the distribution at initialization ($t = 0$), while the blue curve corresponds to the distribution at the end of the evolution ($t_{\txt{end}} = 50 \ \txt{fm}/c)$. Nuclear matter, initialized in a mechanically unstable region of the phase diagram, undergoes a spontaneous separation into a less dense and a more dense nuclear liquid (see Sec.\ \ref{parametrization} for more discussion). Correspondingly, the distribution function becomes wider with time; however, small numbers of particles in cells used to construct the histogram and corresponding finite number statistics effects wash out the structure clearly seen on the right panel in Fig.\ \ref{spinodal_decomposition_hadron_histograms_full}. See text for more details.
	}
	\label{spinodal_decomposition_hadron_histograms_parallel}
\end{figure}

First, we discuss this subject in the context of the choice of binning width. In particular, the double-peak structure in the baryon number distribution shown in the right panel of Fig.\ \ref{spinodal_decomposition_hadron_histograms_full} depends on the size of the cell used to construct the histogram, chosen to be $\Delta l = 2 \ \txt{fm}$. In this case, the Poissonian finite number statistics superimposed on the underlying baryon distribution is characterized by a certain width $\sigma_{(\txt{2~fm})}$. If we reduce the cell width $\Delta l$ by a factor of 2, the average number of particles in a cell is reduced by a factor of 8. Consequently, the width of the Poissonian fluctuations will be $\sigma_{(\txt{1~fm})} =2\sqrt{2}  \sigma_{(\txt{2~fm})}$, which is considerably larger than previously and which in fact washes out the double-peak structure. This can be seen in Fig.\ \ref{spinodal_decomposition_hadron_histograms_full_smaller_cell}, where we show the baryon number distribution for a sampling cell width of $\Delta l = 1 \ \txt{fm}$ for the same events as used to create Fig.\ \ref{spinodal_decomposition_hadron_histograms_full}; the red and blue lines correspond to the distribution at time $t=0$ and $t_{\txt{end}} = 50 \ \txt{fm}/c$, respectively. For the system at hand, the Poissonian widths in the two cases, in terms of baryon density, were $\sigma_{(\txt{2~fm})} = 0.22 \ n_0$ and $\sigma_{(\txt{1~fm})} = 0.62 \ n_0$. If we then estimate the full width at half maximum as approximately given by $2.355\sigma$ (the full width at half-maximum of a normal distribution), it is clear that in the case of the cell width $\Delta l = 1 \ \txt{fm}$, the full width is comparable with the separation of the peaks given by the width of the coexistence region, $n_R - n_L = 1.44 n_0$. As a result, the two-peak structure cannot be resolved for this sampling statistics. Let us note here that decreasing the volume of the cells, $(\Delta l)^3$, can be done without penalty if one proportionally increases the number of test particles per particle, $N_T$. Conversely, decreasing the number of test particles per particle $N_T$ exacerbates the effects of finite number statistics.

\begin{figure}[t]
	\includegraphics[width=\columnwidth]{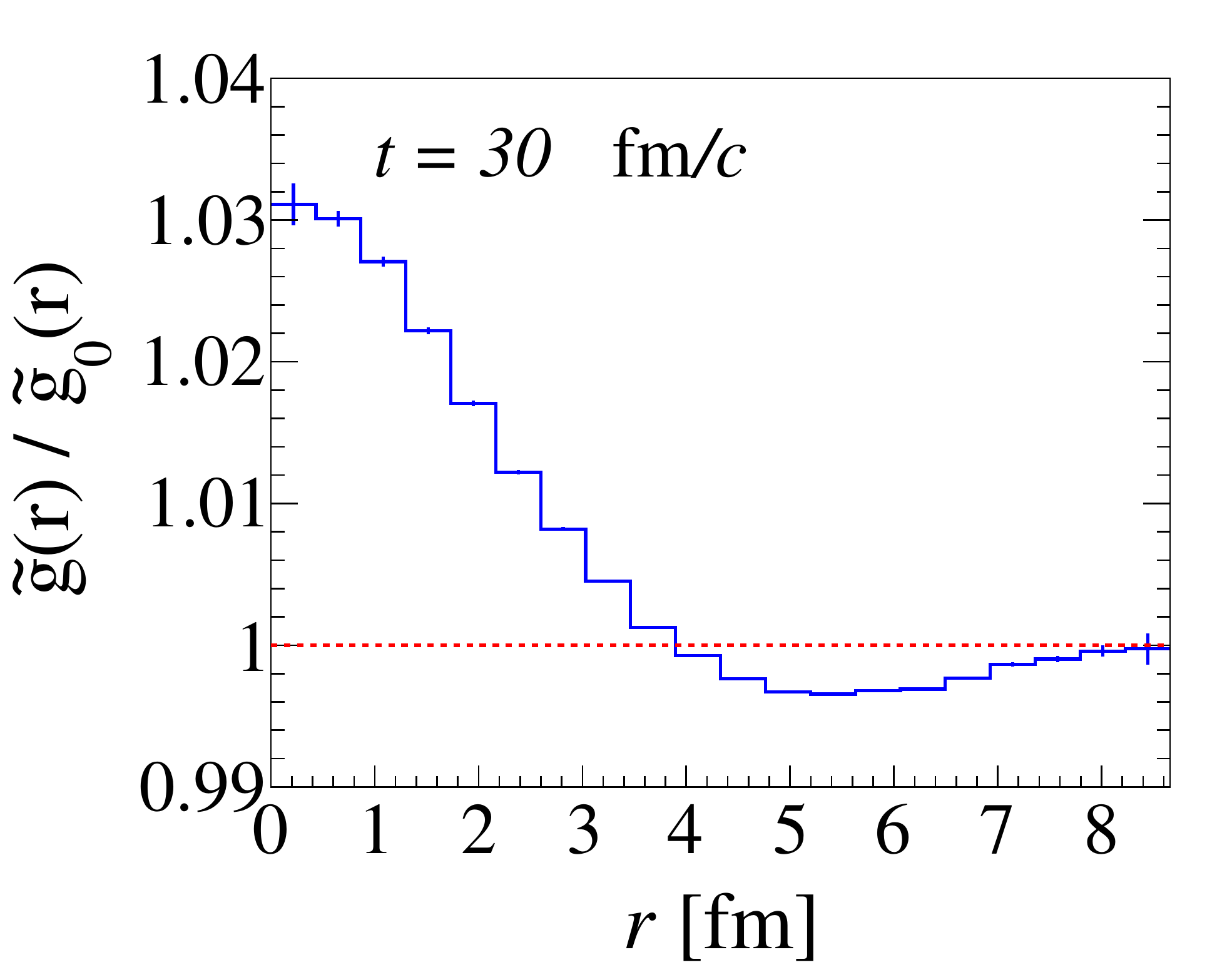}
	\caption{(Color online) Pair correlation function at $t = 30 \ \txt{fm}/c$ for a system initialized inside the quark-hadron spinodal region (at baryon number density $n_B = 3.0 \ n_0$ and temperature $T = 1 \ \txt{MeV}$), calculated within the parallel ensembles method; the results are averaged over $N_{\txt{ev}}^{(\txt{parallel})} = N_T \times N_{\txt{ev}} = 25,000$ events. Spontaneous spinodal decomposition leads to a formation of two coexisting phases: a ``less dense'' and a ``more dense'' nuclear liquid. The increased relative concentration of particles in the ``more dense'' phase results in an elevated normalized pair correlation at small distances. The correlation is exactly the same as shown on the rightmost panel in Fig.\ \ref{spinodal_decomposition_hadron_correlations_full}. See text for more details.
	}
	\label{spinodal_decomposition_hadron_correlations_parallel}
\end{figure}

While this discussion may appear to be of purely numerical nature, experimental data are similarly affected by finite number statistics. In experiments, one always deals with exactly $N_B$ particles per event, which in our simulations corresponds to $N_T = 1$. Naturally, it must lead to a distribution in which any possible peaks are even more washed out. This can be seen in Fig.\ \ref{spinodal_decomposition_hadron_histograms_parallel}, where we show results for the case of $N_T = 1$ and $\Delta l = 2\ \txt{fm}$; the red and blue lines correspond to the distribution at time $t=0$ and $t_{\txt{end}} = 50 \ \txt{fm}/c$, respectively. Here, in order to ensure that we are comparing systems with identical dynamics, we used the same simulation data as in Figs.\ \ref{spinodal_decomposition_hadron_histograms_full} and \ref{spinodal_decomposition_hadron_histograms_full_smaller_cell}, but this time we accessed the baryon number distribution corresponding to $N_T = 1$ using the parallel ensembles method (for details, see Sec.\ \ref{implementation_in_SMASH} and Appendix \ref{parallel_ensembles_in_SMASH}). Not surprisingly, the signal is almost entirely washed out and only a slight broadening of the distribution is discernible. We note that increasing the number of events does not resolve this issue, as the resolution is determined by Poissonian fluctuations in individual events. Consequently, one needs to devise other methods to extract the information about the underlying baryon distribution, one of which will be presented in a forthcoming work.

Finally, we note that the pair distribution function is less affected by finite number statistics. In Fig.\ \ref{spinodal_decomposition_hadron_correlations_parallel}, we show the pair distribution function calculated within the parallel ensembles method, which is nearly identical to the pair distribution function calculated in the full ensemble, Fig.\ \ref{spinodal_decomposition_hadron_correlations_full}. Indeed, the normalized pair distribution function is not determined by the total number of test particles in an event or in a given subvolume of the system, but by relations between any two test particles. The only difference between the pair distribution functions obtained within the two methods is in the error bars, which are larger in the parallel ensembles case due to smaller statistics: the number of pairs in the full ensemble is given by $N_{\txt{ev}} ( N_B N_T )^2$, while in the parallel ensembles it's equal $N_{T} N_{ \txt{ev}} ( N_B)^2$. Obtaining the same pair distribution function demonstrates that the physics accessible in the full ensemble and the parallel ensembles approach is the same.

\section{Summary and outlook}
\label{summary_and_outlook}

In this paper we have presented a flexible vector density functional (VDF) model, which allows one to construct a parameterized dense nuclear matter EOSs (Sec.\ \ref{formalism}). The model, based on the relativistic Landau Fermi-Liquid theory, obeys Lorentz covariance, preserves conservation laws, and is shown to be thermodynamically consistent. The constructed family of EOSs describes two first-order phase transitions: the experimentally observed nuclear liquid-gas phase transition, and a postulated high-temperature, high-density phase transition intended to model the QGP phase transition  (Sec.\ \ref{theoretical_results}). 

To study the dynamical evolution of dense nuclear matter, the model has been implemented in the hadronic transport code \texttt{SMASH} \cite{Weil:2016zrk} through solving the relativistic mean-field equations of motion derived from the VDF EOS (Sec.\ \ref{implementation_in_SMASH}). For investigating the qualitative features of the behavior of dense nuclear matter, we have concentrated on one specific realization of the dense matter EOS, keeping in mind that the ultimate motivation behind creating the VDF model and its supporting framework within \texttt{SMASH} is to enable large-scale comparisons between experimental data and simulations spanning a broad family of EOSs.

Results from simulations in \texttt{SMASH} (Sec\ \ref{simulation_results}) demonstrate that critical behavior in dense nuclear matter can be studied within a hadronic transport approach equipped with interactions corresponding to a chosen EOS. In particular, we have shown that systems initialized in unstable regions of the phase diagram undergo spontaneous spinodal decomposition, followed by an evolution towards an equilibrated mixture of two coexisting phases with compositions matching the predictions from the underlying EOS. Likewise, an investigation of equilibrated uniform nuclear matter in the vicinity of a critical point shows that the thermodynamic behavior expected from the underlying theory is reproduced. The correct description of both thermodynamics and non-equilibrium phenomena implies that hadronic transport can be used as a tool with unique capabilities to investigate the dynamic evolution of matter created in heavy-ion collisions.

We have also shown that for systems initialized at various points of the phase diagram, the pair distribution functions calculated from hadronic transport simulation data follow theoretical expectations based on the second-order cumulant ratio, $\frac{\kappa_2}{\kappa_1}$ (Sec.\ \ref{quark-hadron_phase_transition}). In particular, as the baryon number density (and, consequently, baryon chemical potential) is increased in the region of the phase diagram affected by the critical point, the pair distribution function follows a clear pattern: displaying first anticorrelation, then correlation, and then again anticorrelation. This behavior of two-particle correlations (and, on the theoretical side, of the second-order cumulant ratio $\frac{\kappa_2}{\kappa_1}$) is a clear signature of crossing the phase diagram above the critical point. This is especially important in view of the experimental search for the QCD critical point, as lower order statistical observables, such as $\frac{\kappa_2}{\kappa_1}$, are more likely to be measured with accuracy sufficient for discerning signals of critical behavior.

Multiple future research directions are possible, with a couple of them considered below. 

To start, possible generalizations of the VDF model include adding interactions of scalar type, which will allow for an even greater flexibility in postulating the position of the QCD critical point. While such interactions will be computationally much more demanding, their addition will ultimately allow for a more robust comparison with experimental data. This generalization of the VDF model is a subject of an ongoing work.

Further, finite number statistics affects both the qualitative and quantitative features of statistical observables. We have shown that within two complementary simulation paradigms, hadronic transport gives access to both the continuous baryon number distribution, employed in theoretical calculations, and the physical baryon number distribution relevant to experimental results (Secs.\ \ref{implementation_in_SMASH} and \ref{number_distribution_functions}). Though driven by the same physics, these distributions lead to starkly different values for integrated statistical observables (Secs.\ \ref{quark-hadron_phase_transition} and \ref{effects_of_finite_number_statistics}). A direct link between these two simulation paradigms and its consequence for comparisons with experimental data is the subject of an ongoing work.

\vspace{5mm}

\begin{center}
	\textbf{ACKNOWLEDGMENTS}
\end{center}

A.S.\ thanks Dmytro Oliinychenko for many fruitful discussions about hadronic transport in general, and \texttt{SMASH} in particular. A.S.\ also thanks Hannah Elfner and the \texttt{SMASH} team for access to the \texttt{SMASH} development branch. Last but not least, A.S.\ thanks Huan Zhong Huang for arranging her research opportunity at LBNL and for continued advice and guidance.

This work received support through the U.S.\ Department of Energy,  Office of Science, Office of Nuclear Physics, under Contract No.\ DE-AC02-05CH11231231 and received support within the framework of the Beam Energy Scan Theory (BEST) Topical Collaboration.

\appendix

 \section{The quasiparticle distribution function}
\label{quasiparticle_distribution_function}

Quasiparticles are understood as emergent phenomena occurring when a microscopically complex system of ``real'' particles can be described as if it was made of different, weakly interacting ``quasiparticles'' in free space. This concept is well known, among others, when applied to the behavior of an electron traveling through a semiconductor, which can be described as a motion of a free electron with a different, ``effective'' mass.

The bulk behavior of a system of quasiparticles is described by a quasiparticle distribution function, constructed based on a one-to-one correspondence between quasiparticles and ``real'' particles. In the non-relativistic limit, this correspondence can be understood as follows (for a more complete introduction see \cite{BaymPethickLandauFermiLiquidTheory}). One begins by considering an ideal Fermi gas (neglecting spin), in which the dispersion relation is
\begin{eqnarray}
\eps_{\bm{p}}^{\txt{free}} = \frac{\bm{p}^2}{2m}~.
\label{dispersion_relation_free_gas}
\end{eqnarray}
The state of the system as whole can be specified by giving the number of particles $N_{\bm{p}} = \{0, 1\}$ in each of the single-particle states defined by a specific value of the momentum $\bm{p}$. Thus, for example, in the ground state each of the states with momenta less than the Fermi momentum, $p_F$, is occupied ($N_{\bm{p}} = 1$), and all other states are empty ($N_{\bm{p}} = 0$). One can then imagine that interactions in the system are slowly turned on in such a way that the process is adiabatic. Quantum mechanics shows that while such an adiabatic change will lead to a distortion of the energy levels, it will preserve their number. This means that the distribution function $N_{\bm{p}}$, while also smoothly distorted, preserves its functional form. Now, however, the dispersion relation $\eps_{\bm{p}}^{\txt{int}} $ takes interactions into the account, and it is \textit{different} than that of a free particle, Eq.\ (\ref{dispersion_relation_free_gas}).

We stress that the construction of the quasiparticle distribution is based on the assumption that there exists a one-to-one correspondence between quasiparticles and ``real'' particles. This means, for example, that this formalism is not appropriate for describing phenomena in which the number of particles in the system changes throughout the evolution, such as formation or dissolution of bound states. 

For describing the macroscopic properties of a Fermi liquid, it is sufficient to use a mean or smoothed quasiparticle distribution function, often denoted by $f_{\bm{p}}$, which is an average of $N_{\bm{p}}$ over a group of neighboring single-particle states. While $N_{\bm{p}}$ is a discontinuous function of $\bm{p}$, $f_{\bm{p}}$ and is a smooth function of $\bm{p}$.

\section{Model derivations}
\label{model_derivations}

\subsection{Quasiparticle energy}
\label{quasiparticle_energy_derivation}

To obtain the quasiparticle energy, we calculate a functional differential of the energy density, $\delta \mathcal{E}_{(1)}$, where $\mathcal{E}_{(1)}$ is given by Eq.\ (\ref{energy_density_postulated}). Taking into the account that the kinetic energy $\epsilon_{\txt{kin}}$, Eq.\ (\ref{kinetic_energy}), is also a functional of the quasiparticle distribution function  through the dependence of $\epsilon_{\txt{kin}}$ on baryon current, we get 
\begin{eqnarray}
\delta \mathcal{E}_{(1)} &=&  - C_1 \left(  b_1 - 2\right) \big(j_{\mu} j^{\mu}\big)^{\frac{b_1}{2} - 2}  ~ j_{\mu} \delta j^{\mu}  ~ \bm{j} \cdot \bm{j} \non \\
&& - ~ C_1 \big( j_{\mu} j^{\mu} \big)^{\frac{b_1}{2}  - 1}  ~ \bm{j} \cdot \delta \bm{j}  \non \\
&& + ~  g \int \frac{d^3p}{(2\pi)^3} ~  \epsilon_{\txt{kin}} ~  \delta f_{\bm{p}} \non \\
&& +~ C_1 \left( b_1 - 2 \right) \big( j_{\mu} j^{\mu}\big)^{\frac{b_1}{2} - 2} ~  j_{\mu} \delta j^{\mu} \big( j^0\big)^2 \non \\
&& + ~ 2C_1 \big( j_{\mu} j^{\mu} \big)^{\frac{b_1}{2} - 1} ~ j_0 \delta j^0  \non \\
&& + ~ C_1 \big( b_1 - 1\big) \big( j_{\mu} j^{\mu} \big)^{\frac{b_1}{2}  - 1} ~ j_{\mu} \delta j^{\mu}~,
\end{eqnarray}
where in the first two terms we have used the definition of the vector baryon current $\bm{j}$, Eq.\ (\ref{current_spatial_postulated}). The first, fourth, and sixth terms can be combined using $j_0j^0 - \bm{j} \cdot \bm{j} = j_{\mu} j^{\mu}$, so that 
\begin{eqnarray}
\delta \mathcal{E}_{(1)} &=&  - C_1 \big(j_{\mu} j^{\mu}\big)^{\frac{b_1}{2} - 1}  ~ j_{\mu} \delta j^{\mu}   \non \\
&& - ~ C_1 \big( j_{\mu} j^{\mu} \big)^{\frac{b_1}{2}  - 1}  ~ \bm{j} \cdot \delta \bm{j}  \non \\
&& + ~  g \int \frac{d^3p}{(2\pi)^3} ~  \epsilon_{\txt{kin}} ~  \delta f_{\bm{p}} \non \\
&& + ~ 2C_1 \big( j_{\mu} j^{\mu} \big)^{\frac{b_1}{2} - 1} ~ j_0 \delta j^0  ~.
\end{eqnarray}
Then we also note that $j_{\mu} \delta j^{\mu} = j_0 \delta j^{0}  - \bm{j} \delta \bm{j}$, which further reduces the above equation to 
\begin{eqnarray}
\delta \mathcal{E}_{(1)} &=&   g \int \frac{d^3p}{(2\pi)^3} ~  \epsilon_{\txt{kin}} ~  \delta f_{\bm{p}} \non \\
&& + ~ C_1 \big( j_{\mu} j^{\mu} \big)^{\frac{b_1}{2} - 1} ~ j_0 \delta j^0  ~.
\end{eqnarray}
Using the definition of baryon density $j^0$, Eq.\ (\ref{current_temporal_postulated}), we arrive at
\begin{eqnarray}
\hspace{-5mm}\delta \mathcal{E}_{(1)} &=&   g \int \frac{d^3p}{(2\pi)^3} ~ \left[ \epsilon_{\txt{kin}} + C_1 \big( j_{\mu} j^{\mu} \big)^{\frac{b_1}{2} - 1} ~ j_0 \right]~  \delta f_{\bm{p}} ~,
\end{eqnarray}
from which we immediately obtain the quasiparticle energy,
\begin{eqnarray}
\eps_{\bm{p}} \equiv \frac{\delta \mathcal{E}}{\delta f_{\bm{p}}} = \epsilon_{\txt{kin}} + C_1 \big(j_{\mu} j^{\mu} \big)^{\frac{b_1}{2} - 1} j_0 ~.
\end{eqnarray}

\subsection{Relativistic covariance of the equations of motion}
\label{EOMs_covariant}

With the definition of the kinetic momentum $\Pi^{\mu}$, Eq.\ (\ref{kinetic_momentum}), the Hamilton's equations, Eqs.\ (\ref{equation_of_motion_x}) and (\ref{equation_of_motion_p}), can be rewritten as
\begin{eqnarray}
\frac{dx^i}{dt} = \frac{\Pi^i}{\Pi_0} 
\label{Hamilton_x_i}
\end{eqnarray}
and
\begin{eqnarray}
\frac{dp^i}{dt}  = \frac{\sum_k \Pi_k}{\Pi_0}\parr{A^k}{x_i} + \parr{A_0}{x_i} ~.
\label{Hamilton_p_i}
\end{eqnarray}
Using the fact that $H_{(1)} = \eps_{\bm{p}} = p_0$, we can see that for the temporal component of $x^{\mu}$ we have trivially
\begin{eqnarray}
\frac{dx^0}{dt} = \parr{H_{(1)}}{p_0}  = 1 = \frac{\Pi_0}{\Pi_0}~,
\label{Hamilton_x_0}
\end{eqnarray}
which allows us to write Eqs.\ (\ref{Hamilton_x_i}) and (\ref{Hamilton_x_0}) together as
\begin{eqnarray}
\frac{dx^{\mu}}{dt}  = \frac{\Pi^{\mu}}{\Pi_0}   ~.
\label{x_relativistic}
\end{eqnarray}
For the temporal part of $p^{\mu}$ we can likewise write
\begin{eqnarray}
\frac{dp^0}{dt} = \frac{dp^0}{dx_0} = \frac{\sum_{k} \Pi_k}{\Pi_0} \parr{A^k}{x_0} + \parr{A^0}{x_0} ~,
\label{Hamilton_p_0}
\end{eqnarray}
where on the right-hand side we have simply carried out the differentiation with respect to $x_0$, and it follows that Eqs.\ (\ref{Hamilton_p_i}) and (\ref{Hamilton_p_0}) can be jointly written as
\begin{eqnarray}
\frac{dp^{\mu}}{dt} &=& \frac{\sum_{k} \Pi_k}{\Pi_0} \parr{A^k}{x_{\mu}}  + \parr{A^0}{x_{\mu}} = \non \\
&=& \frac{\sum_{k} \Pi_k}{\Pi_0} \parr{A^k}{x_{\mu}}  + \frac{\Pi_0}{\Pi_0}\parr{A^0}{x_{\mu}} = \non \\
&=& \sum_{\nu} \frac{\Pi_{\nu}}{\Pi_0} \parr{A^{\nu}}{x_{\mu}}  ~.
\label{eq345}
\end{eqnarray}

Let us note that from the definition of the kinetic momentum $\Pi^{\mu}$ we have
\begin{eqnarray}
\frac{d \Pi^{\mu}}{dt} = \frac{dp^{\mu}}{dt} - \frac{dA^{\mu}}{dt}~.
\label{eq_EOM_3}
\end{eqnarray}
Using Eq.\ (\ref{eq345}), the above equation becomes
\begin{eqnarray}
\frac{d \Pi^{\mu}}{dt} = \sum_{\nu}  \frac{\Pi_{\nu}}{\Pi_0} \parr{A^{\nu}}{x_{\mu}}  - \frac{dA^{\mu}}{dt}~.
\end{eqnarray}
We can always write
\begin{eqnarray}
\frac{dA^{\mu}}{dt} = \parr{A^{\mu}}{x^{\nu}} \frac{dx^{\nu}}{dt}  = \parr{A^{\mu}}{x^{\nu}} \frac{\Pi^{\nu}}{\Pi_0} ~ ,
\label{eq_EOM_4}
\end{eqnarray}
so that in the end
\begin{eqnarray}
\hspace{-5mm}\frac{d \Pi^{\mu}}{dt} &=& \sum_{\nu}  \frac{\Pi_{\nu}}{\Pi_0} \parr{A^{\nu}}{x_{\mu}}  - \frac{\Pi^{\nu}}{\Pi_0} \parr{A^{\mu}}{x^{\nu}}  \non  \\ 
&=& \sum_{\nu}  \frac{\Pi_{\nu}}{\Pi_0} \parr{A^{\nu}}{x_{\mu}}   - \frac{\Pi_{\nu}}{\Pi_0} \parr{A^{\mu}}{x_{\nu}} \non   \\ 
&=& \sum_{\nu} \frac{\Pi_{\nu}}{\Pi_0} \Big( \partial^{\mu} A^{\nu} - \partial^{\nu} A^{\mu}  \Big) =\sum_{\nu} \frac{\Pi_{\nu}}{\Pi_0} F^{\mu \nu} ~,
\label{p_relativistic}
\end{eqnarray}
where $F^{\mu\nu}$ is defined similarly as the field strength in EM. 

Both Eq.\ (\ref{x_relativistic}) and Eq.\ (\ref{p_relativistic}) are written in a relativistically covariant form.

\begin{table*}[t]
	\caption{Parameter sets corresponding to the EOSs reproducing sets of the QGP-like phase transition characteristics $\big( T_c^{(Q)}, n_c^{(Q)}, \eta_L, \eta_R\big)$, listed in Table \ref{example_characteristics}.}
	\label{parameters}
	\begin{center}
		\bgroup
		\def\arraystretch{1.4}
		\begin{tabular}{c r r r r l l l l}
			\hline
			\hline
			%\cline{2-6}
			\hspace{3mm}set\hspace{3mm} & \hspace{6mm}$b_1$\hspace{6mm} &  \hspace{6mm} $b_2$  \hspace{6mm} &  \hspace{6mm}
			$b_3$  \hspace{6mm} &  \hspace{6mm} $b_4$  \hspace{6mm} &  \hspace{7mm} $\tilde{C}_1$ [MeV]  \hspace{4mm} &  \hspace{4mm} $\tilde{C}_2$ [MeV]  \hspace{4mm} &   \hspace{4mm}
			$\tilde{C}_3$ [MeV]  \hspace{4mm} &  \hspace{4mm}  $\tilde{C}_4$ [MeV]  \hspace{4mm}\\ 
			\hline
			I & 1.7614679  & 3.8453863  & 4.4772660  & 6.7707861  &\hspace{3mm} -8.315987$\times 10^1$&  6.144706$\times 10^1$   &  -3.108395$\times 10^1$   & 3.127069$\times 10^{-1}$  \\  
			II & 1.8033077  & 3.0693813  & 7.9232548  & 10.7986978  &\hspace{3mm} -9.204350$\times 10^1$& 3.968766$\times 10^{1}$  & -1.306487$\times 10^{-1}$  & 2.434034$\times 10^{-3}$ \\  
			III & 1.8042024  & 3.0631798  & 6.6860893  & 20.7276154  &\hspace{3mm} -9.224000$\times 10^1$& 3.986263$\times 10^{1}$  & -1.066766$\times 10^{-1}$  & 2.160279$\times 10^{-11}$ \\  
			IV & 1.7681391  & 3.5293515  & 5.4352787  & 6.3809823   &\hspace{3mm} -8.450948$\times 10^1$& 3.843139$\times 10^{1}$  & -7.958557  & 1.552593\\  
			V & 1.7782362  & 3.4936863  & 4.2528897  & 10.3240297 &\hspace{3mm} -8.627959$\times 10^1$& 4.786488$\times 10^{1}$  & -1.406946$\times 10^{1}$ & 1.182795$\times 10^{-4}$ \\  
			VI & 1.7989835  & 3.1098389 & 6.3017683  & 8.0937872 &\hspace{3mm} -9.101665$\times 10^1$& 3.899891$\times 10^{1}$ &-4.856681$\times 10^{-1}$ & 1.935808$\times 10^{-2}$ \\
			\hline
			\hline
		\end{tabular}
		\egroup
	\end{center}
\end{table*}

\subsection{Form of the quasiparticle distribution function}
\label{form_of_the_quasiparticle_distribution_function}

To obtain the functional form of the quasiparticle distribution function $f_{\bm{p}}$ of a thermal Fermi system, we use fundamental thermodynamic relations. We know that any variation in the energy density is connected to a variation in entropy density, $s$, and particle density, $n$, through
\begin{eqnarray}
\delta \mathcal{E} = T ~\delta s + \mu ~\delta n~,
\label{energy_variation}
\end{eqnarray}
where $T$ is the temperature and $\mu$ is the chemical potential. We already know that the dependence of $\delta \mathcal{E}$ on the distribution function is given by the definition of the quasiparticle energy $\eps_{\bm{p}}$, $\delta \mathcal{E}\mathcal \ \equiv \  \eps_{\bm{p}} ~ \delta f_{\bm{p}}$, but we need to establish the dependence of $\delta s$ and $\delta n$ on $f_{\bm{p}}$.

It is possible to calculate the entropy of a given state of the system by combinatorial considerations only, and in view of the one-to-one correspondence between the states of the Fermi liquid and the free Fermi gas (see Appendix \ref{quasiparticle_distribution_function}), it is natural to assume that the entropy density must have the same form as in the case of the free Fermi gas,
\begin{eqnarray}
s = - \frac{1}{V} \sum_{\bm{p}} \Big[ f_{\bm{p}} \ln f_{\bm{p}} + (1 - f_{\bm{p}}) \ln (1 - f_{\bm{p}})  \Big]~. 
\end{eqnarray}
(We note that we use the natural units in which the Boltzmann constant $k_B = 1$.) Consequently,
\begin{eqnarray}
\delta s = - \frac{1}{V} \sum_{\bm{p}} \left[\delta f_{\bm{p}}  \ln \frac{f_{\bm{p} }}{1 - f_{\bm{p}}}  \right]~.
\end{eqnarray}

The number of quasiparticles in the interacting system directly corresponds to the number of particles in the corresponding state of the free Fermi gas. Furthermore, the interaction between the particles conserves the particle number, and so the total number of particles in a state of the interacting system must be the same as in the non-interacting system. In consequence, we can express the quasiparticle density using the quasiparticle distribution function,
\begin{eqnarray}
n = \frac{1}{V} \sum_{\bm{p}} f_{\bm{p}} ~,
\label{number_density_from_distribution_function}
\end{eqnarray}
from which we have
\begin{eqnarray}
\delta n = \frac{1}{V} \sum_{\bm{p}} \delta f_{\bm{p}} ~. 
\end{eqnarray}

With all this, we can rewrite Eq.\ (\ref{energy_variation}) as
\begin{eqnarray}
\frac{1}{V} \sum_{\bm{p}} \eps_{\bm{p}} ~\delta f_{\bm{p}} &=& -\frac{T}{V} \sum_{\bm{p}} \ln \frac{f_{\bm{p} }}{1 - f_{\bm{p}}} ~\delta f_{\bm{p}}  \non \\
&& \hspace{5mm} + ~\frac{\mu}{V} \sum_{\bm{p}} \delta f_{\bm{p}} ~,
\end{eqnarray}
which can be further rearranged as
\begin{eqnarray}
\frac{1}{V} \sum_{\bm{p}} \left[ \eps_{\bm{p}} + T \ln \frac{f_{\bm{p} }}{1 - f_{\bm{p}}}  - \mu \right]~ \delta f_{\bm{p}} = 0 ~.
\end{eqnarray}
The above equality will hold for any variation $\delta f_{\bm{p}}$ if and only if the term in the square bracket vanishes for any $\bm{p}$, and we can immediately use this fact to solve for the quasiparticle distribution function,
\begin{eqnarray}
f_{\bm{p}} = \frac{1}{\exp\left( \frac{\eps_{\bm{p}} - \mu }{T} \right) + 1} ~. 
\end{eqnarray}
Note that, because the quasiparticle energy $\eps_{\bm{p}}$ itself depends on the quasiparticle distribution $f_{\bm{p}}$, the above equation is in fact a rather complicated implicit equation for $f_{\bm{p}}$, in contrast to the free Fermi gas case.

\section{Parameter sets}
\label{parameter_sets}

Here we provide parameters corresponding to the EOSs reproducing sets of the QGP-like phase transition characteristics $\big( T_c^{(Q)}, n_c^{(Q)}, \eta_L, \eta_R\big)$, listed in Table \ref{example_characteristics}. It is important to note that the values of the coefficients of the interaction terms, $\{C_1, C_2, C_3, C_4 \}$, depend on a chosen system of units. Here, we adopt a convention used in many Skyrme-like parametrizations, in which the single-particle potential is written in the form
\begin{eqnarray}
U = \sum_{i=1}^N  \tilde{C}_i \left( \frac{n_B}{n_0} \right)^{b_i - 1} ~,
\end{eqnarray}
where $n_0$ is the saturation density, so that $\tilde{C}_i$ must have a dimension of energy.  Naturally, $\tilde{C}_i$ and $C_i$ are related by
\begin{eqnarray}
C_i = \frac{\tilde{C}_i }{n_0^{b_i - 1}} ~. 
\end{eqnarray}

In Table \ref{parameters}, we list coefficients $\{\tilde{C}_1,  \tilde{C}_2, \tilde{C}_3, \tilde{C}_4\}$ in units of MeV. Note that in particular, the sum of all coefficients yields the (rest frame) value of the single-particle potential at $n_B = n_0$, $\sum_{i=1}^N \tilde{C}_i = -52.484 \ \txt{MeV}$.

\section{Symmetric spinodal regions}
\label{symmetric_spinodal_regions}

The spinodal region is the range of baryon number densities between two local extrema of pressure, a maximum at $\eta_L$ and a minimum at $\eta_R$, with $\eta_L < \eta_R$. A curve exhibiting two extrema will most naturally have an inflection point approximately in between them. We can see this by considering the following polynomial:
\begin{eqnarray}
f(x) = ax^3 + bx^2 + cx + d ~,
\end{eqnarray}
which is a ``minimal'' polynomial needed to produce two local extrema. The condition for an extremum at some point $x_0$ is
\begin{eqnarray}
\frac{df}{dx}\bigg|_{x = x_0} = 3ax^2 + 2bx + c \bigg|_{x = x_0}  = 0 ~.  
\end{eqnarray}
We can solve this equation to yield the positions of the extrema $x_L$ and $x_R$,
\begin{eqnarray}
&& x_L = \frac{-b - \sqrt{b^2 - 3ac}}{3a} ~, \\
&& x_R = \frac{-b + \sqrt{b^2 - 3ac}}{3a}  ~. 
\end{eqnarray}
The position of the inflection point is established through the condition
\begin{eqnarray}
\frac{d^2f}{dx^2}\bigg|_{x = x_{\txt{infl}}} = 6ax + 2b \bigg|_{x = x_{\txt{infl}}} = 0 ~,
\end{eqnarray}
from which we get 
\begin{eqnarray}
x_{\txt{infl}} = - \frac{-b}{3a}~.
\end{eqnarray}
It is immediately apparent that 
\begin{eqnarray}
x_{\txt{infl}} = \frac{x_L + x_R}{2}~,
\end{eqnarray}
placing the inflection point exactly in the middle between the two extrema. This result is only exact for a third-order polynomial, and will be changed if the polynomial includes additional terms with which one is able to manipulate the behavior of the curve between the extrema.

We will now argue that in a model with vector-type interactions only, the inflection point of the pressure curve at zero temperature, 
\begin{eqnarray}
\frac{d^2P(T=0)}{dn_B^2} \bigg|_{n_B = n_{\txt{infl}}} = 0~,
\end{eqnarray}
will coincide with the location of the critical point on the $n_B$ axis. Let us first write the pressure as a sum of an ideal gas term and an interaction term,
\begin{eqnarray}
P = P_{\txt{ideal}} + P_{\txt{int}} ~.
\end{eqnarray}
In particular, at $T=0$ the ideal part of the pressure is given by the ideal Fermi gas, $P_{\txt{ideal}}(T=0) = P^{\txt{FG}}_0$. Because the Fermi gas at zero temperature depends on the baryon density as $P^{\txt{FG}}_0 \propto n_B^{4/3}$, for large densities we can safely assume that
\begin{eqnarray}
\frac{d^2 P^{\txt{FG}}_0 }{dn_B^2} = \frac{4}{9} n_B^{-2/3} \approx 0 ~.
\label{eq457}
\end{eqnarray}
It then follows that at the inflection point we must have
\begin{eqnarray}
\frac{d^2P_{\txt{int}}}{dn_B^2} \bigg|_{n_B = n_{\txt{infl}}} \approx 0 ~.
\end{eqnarray}
At the same time, the condition for the position of the critical point at some location $(T_c, n_c)$ leads to
\begin{eqnarray}
\frac{d^2 P_{\txt{int}}}{dn_B^2} \bigg|_{n_B = n_c} = - \frac{d^2 P_{\txt{ideal}}}{dn_B^2} \bigg|_{\substack{n_B = n_c\\ T = T_c }} ~.
\label{critical_point_two_pressure_parts}
\end{eqnarray}
For large enough temperatures, the ideal Fermi gas is well approximated by the ideal Boltzmann gas, and we can write the ideal part of the pressure as 
\begin{eqnarray}
P_{\txt{ideal}} \approx T n_B ~.
\end{eqnarray}
As a result, Eq.\ (\ref{critical_point_two_pressure_parts}) becomes
\begin{eqnarray}
\frac{d^2 P_{\txt{int}}}{dn_B^2} \bigg|_{n_B = n_c} = 0~,
\end{eqnarray}
which immediately confirms that in this case, the location of the critical density $n_c$ coincides with the location of the inflection point $n_{\txt{infl}}$ of the pressure at zero temperature. Moreover, going beyond the approximation used in Eq.\ \eqref{eq457}, we see that at zero temperature the pressure at $n_B= n_c$ will have a very small and positive curvature, which means that the critical density is somewhat larger than the inflection point density, $n_c \gtrsim n_{\txt{infl}} $.

The VDF model largely reproduces the behavior described above. First, due to the fact that the pressure fits in the VDF model are ``minimal'' fits reproducing (among other constraints) two local extrema, a maximum at $\eta_L$  and a minimum at $\eta_R$, the inflection point of the pressure lies roughly in the middle between $\eta_L$ and $\eta_R$. Second, due to the thermal part of the pressure being just like that of an ideal gas, the location of the critical point $n_c$ and the location of the inflection point of the pressure at zero temperature $n_{\txt{infl}}$ are related by $n_c = n_{\txt{infl}} + \delta n$, where $\delta n$ is a small positive correction. This explains why in the VDF model the critical baryon number density $n_c$ lies roughly in the middle of the spinodal region $(\eta_L, \eta_R)$.

 \section{The method of test particles}
 \label{the_method_of_test_particles}

The function $f(t, \bm{x},\bm{p})$ is a continuous distribution function for a given total number $A$ of nucleons. Solving the Boltzmann equation is equivalent to obtaining the time evolution of the distribution function. Numerically, given the initial condition in form of the distribution function at some time $t_0$, $f(t_0, \bm{x}_0,\bm{p}_0)$, we solve for the distribution $f(t, \bm{x},\bm{p})$ at a slightly later time $t = t_0 + \delta t$, and repeat the process until a final time $t = t_{\txt{end}}$ is reached. In more detail, the numerical solution of the VUU equation is achieved through the method of test particles \cite{Wong:1982zzb}, which is based on the assumption that the continuous $f(t,\bm{x},\bm{p})$ distribution can be approximated by the distribution of a large number $N$ of discrete test particles with phase space coordinates $\big(\bm{x}_i (t), \bm{p}_i(t)\big)$, see Eq.\ \eqref{test_particle_approximation}. If we demand that these test particles are propagated according to 
\begin{eqnarray}
\frac{d \bm{x}}{dt} = \parr{H_{(1)}}{\bm{p}}~, \hspace{5mm}\frac{d \bm{p}}{dt} = -\parr{H_{(1)}}{\bm{x}} ~,
\end{eqnarray}
then the Vlasov equation, which is the left-hand side of Eq.\ (\ref{BUU_equation}), immediately follows from the Liouville theorem. 

For an evolution without mean-fields, it is most natural to take the number of test particles exactly corresponding to the actual number of nucleons present in the system, $N = A$, $N_T = 1$. However, employing mean-fields dependent on local density and its gradients requires adopting an approach in which statistical noise due to a finite number of test particles is suppressed. This is especially important in the case of models with competing repulsive and attractive potentials of large magnitudes (as is often the case in relativistic models), where relatively small numerical fluctuations can produce significant errors in the mean-field potential calculations. Thus, for example, for studies of nuclear matter with average density around the saturation density $n_0$, a number of test particles per nucleon $N_T = 100$ is often used.

The local baryon current is then defined on a lattice, where at a given lattice point, the current is a sum of contributions from all test particles which are in the volume element $V_i$ corresponding to that lattice point,
\begin{eqnarray}
j^{\mu} (\bm{r}_i) = \frac{1}{N_T}\frac{1}{V_i} \sum_{k \in V_i} \frac{\Pi^\mu (k) }{\Pi^0 (k)} ~. 
\end{eqnarray} 
This prescription naturally reproduces the baryon number in a given volume element,
\begin{eqnarray}
B(i) = j^0 (i) V_i = \frac{ N(i) }{N_T}~,
\end{eqnarray}
where $N(i)$ is the number of test particles in $V_i$. In practice, in order for the local densities and currents to be smooth enough, a prescription is used in which currents at a given lattice point $i$ are weighted sums of contributions from all test particles in some chosen volume $V_s$ around the lattice point $i$, which is larger than the volume element $V_i$, $V_s > V_i$,
\begin{eqnarray}
j^{\mu} (\bm{r}_i)  = \sum_{k \in V_s} \frac{\Pi^\mu (k) }{\Pi^0 (k)}  S(\bm{r}_i - \bm{r}_k)~,
\end{eqnarray}
where  the weight $S(\bm{r}_i - \bm{r}_k)$ is known as the smearing function, normalized such that
\begin{eqnarray}
V_i \sum_{i} S (\bm{r}_i - \bm{r}_k) = \frac{1}{N_T}~.
\end{eqnarray}
Various smearing functions are being employed in existing transport codes. In our approach, we employ a triangular smearing function, originating from the lattice Hamiltonian method of solving nuclear dynamics \cite{Lenk:1989zz}.

\section{Pair distribution function and the second-order cumulant}
\label{pair_distribution_function_and_the_second-order_cumulant}

The procedure to compute the radial distribution function $g_i(r)$, given by Eq.\ \eqref{radial_distribution_function}, can be generalized to the case of a continuous system described by a particle density distribution $n(\bm{r}')$, 
\begin{eqnarray}
\hspace{-8mm} g_i (r, \Delta r) &=&  \int d\bm{r}'  \Bigg(n(\bm{r}') - 1 ~\delta(\bm{r}_i - \bm{r}')     \Bigg) \non \\
&&\hspace{-15mm} \times ~\theta\Big(  r + \Delta r - |\bm{r}_i - \bm{r}' |   \Big) \theta\Big( |\bm{r}_i - \bm{r}' | - ( r - \Delta r )   \Big) ~,
\end{eqnarray}
where care must be taken to subtract the self-contribution from the reference particle. Similarly, the pair distribution function $\widetilde{g}(r)$, Eq.\ \eqref{pair_distribution_function}, can be rewritten as
\begin{eqnarray}
\widetilde{g} (r, \Delta r) &=&\frac{\mathcal{N}}{2}   \int d \bm{r}'  \int d\bm{r}''  ~   n(\bm{r}') \Big(n(\bm{r}'')  - \delta( \bm{r}' - \bm{r}'') \Big)  \non  \\
&& \hspace{-10mm} \times ~\theta\Big(  r + \Delta r - |\bm{r}' - \bm{r}'' |   \Big) \theta\Big( |\bm{r}' - \bm{r}'' | - ( r - \Delta r )   \Big)\non  \\
&=& \frac{\mathcal{N}}{2}   \int d \bm{r}'  \int d\bm{r}''  ~   n(\bm{r}') n(\bm{r}'')    \non \\
&&\hspace{-10mm} \times ~\theta\Big(  r + \Delta r - |\bm{r}' - \bm{r}'' |   \Big) \theta\Big( |\bm{r}' - \bm{r}'' | - ( r - \Delta r )   \Big) \non  \\
&& \hspace{5mm} - ~ \frac{\mathcal{N}}{2}   \int d \bm{r}' ~  n(\bm{r}')    ~ \theta\Big( \Delta r  - r    \Big) ~.
\label{pair_distr_kappa_2_1}
\end{eqnarray}
We note that the second term is only non-zero when $r < \Delta r$, which is correct given that the self-contribution only needs to be subtracted if we consider the pair distribution function within a distance $\Delta r$ around the reference particles. 

It is possible to establish a connection between the pair distribution function and the second-order cumulant $\kappa_2$. For this, we consider the pair distribution function $\widetilde{g}(r)$ at distances close to the reference particle, that is we put $r =0$, by means of which Eq.\ \eqref{pair_distr_kappa_2_1} becomes
\begin{eqnarray}
\hspace{-0mm} \widetilde{g} (0, \Delta r) &=& \frac{\mathcal{N}}{2} \bigg[  \int d \bm{r}'  \int d\bm{r}''  ~   n(\bm{r}') n(\bm{r}'')   \non \\
&& \hspace{-10mm} \times ~\theta\Big(  \Delta r - |\bm{r}' - \bm{r}'' |   \Big)  -    \int d \bm{r}' ~  n(\bm{r}')    \bigg] ~.
\label{pair_distr_kappa_2_2}
\end{eqnarray}
Let us assume that $\Delta r$ is small and that within the distance $\Delta r$ from $\bm{r}'$ the density is smooth enough for $n(\bm{r}'')  \approx n(\bm{r'})$ to hold, in which case
\begin{eqnarray}
\hspace{-7mm} \widetilde{g} (0, \Delta r) &=& \frac{\mathcal{N}}{2} \bigg[  \int d \bm{r}' ~ \big[n(\bm{r}') \big]^2 \non \\
&& \hspace{-10mm}\times ~  \int d\bm{r}''    ~\theta\Big(  \Delta r - |\bm{r}' - \bm{r}'' |   \Big)  -    \int d \bm{r}' ~  n(\bm{r}')    \bigg] \non \\
&=&  \frac{\mathcal{N}}{2}  \left[ V_{\Delta} \int d \bm{r}' ~ \big[n(\bm{r}') \big]^2  -    \int d \bm{r}' ~  n(\bm{r}')    \right] ~,
\label{pair_distr_kappa_2_3}
\end{eqnarray}
where $V_{\Delta} = (4/3)\pi r_{\Delta}^3$. Furthermore, let us divide the volume of the system $V$ into cubes of volume $V_{\Delta}$, $N_{\txt{cubes}} = V / V_{\Delta}$, and assume that we can safely discretize the remaining integrals in Eq.\ \eqref{pair_distr_kappa_2_3} according to $\int d \bm{r}' ~ f(\bm{r'})  \to  \sum_{i=1}^{N_{\txt{cubes}}} V_{\Delta}  f(\bm{r}_i)$, where $\bm{r}_i$ points to the center of each cube. With this and taking the number of particles in the $i$-th cube to be $N_i(\bm{r}_i) \equiv V_{\Delta} n(\bm{r}_i)$, Eq.\ (\ref{pair_distr_kappa_2_3}) becomes
\begin{eqnarray}
\hspace{-5mm}\widetilde{g} (0, \Delta r) &\approx& \mathcal{N}~ \frac{1}{2} \left[  \sum_{i=1}^{N_{\txt{cubes}}}   \big[ N(\bm{r}_i)\big]^2  -   \sum_{i=1}^{N_{\txt{cubes}}}  N(\bm{r}_i)    \right] ~.
\end{eqnarray}
Since the normalization can be freely chosen given that $\widetilde{\rho} (0, \Delta r) $ should be compared to a reference distribution for an ideal gas $\widetilde{\rho}_0(0, \Delta r)$, in particular we can take $\mathcal{N} = 2/N_{\txt{cubes}}$, so that finally 
\begin{eqnarray}
\hspace{-7mm}\widetilde{g} (0, \Delta r) &=& \frac{1}{N_{\txt{cubes}}} \left[  \sum_{i=1}^{N_{\txt{cubes}}}   \big[ N(\bm{r}_i)\big]^2  -   \sum_{i=1}^{N_{\txt{cubes}}}  N(\bm{r}_i)    \right] ~,
\label{pair_distr_kappa_2_4}
\end{eqnarray}
where $N_{\txt{cubes}}$ is determined by $\Delta r$.

It is clear from Eq.\ \eqref{pair_distr_kappa_2_4} that the radial distribution function of all distinct particle pairs at distances close to the reference particles is
\begin{eqnarray}
\widetilde{g}  (0, \Delta r) = M_2 - M_1 = F_2 = \langle N (N-1) \rangle ~,
\label{pair_distr_kappa_2_5}
\end{eqnarray}
where $M_i$ and $F_i$ are moments and factorial moments of the distribution, respectively. Moreover, assuming that the pair distribution function for uncorrelated pairs $\widetilde{g}_0 (0, \Delta r)$ is described by the Poisson distribution, for which $\langle N \rangle = \lambda$ and $\langle N^2 \rangle = \lambda^2 + \lambda$ (where $\lambda$ is the mean), we have
\begin{eqnarray}
\widetilde{g}_0(0, \Delta r) = \langle N \rangle^2 ~. 
\label{pair_distr_kappa_2_6}
\end{eqnarray}

Let us consider the deviation of the behavior of the pair distribution function $\widetilde{g}(0, \Delta r)$ from the ideal case of $\widetilde{g}_0(0, \Delta r)$, which can be conveniently done by considering the measure
\begin{eqnarray}
R = \frac{\widetilde{g}~(0, \Delta r)}{\widetilde{g}_0(0, \Delta r)} - 1 ~.
\end{eqnarray}
Using Eqs.\ \eqref{pair_distr_kappa_2_5} and \eqref{pair_distr_kappa_2_6} we can immediately rewrite this as
\begin{eqnarray}
R = \frac{\langle N^2 \rangle - \langle N \rangle  - \langle N \rangle^2}{\langle N \rangle^2} = \frac{\kappa_2 - \kappa_1  }{\kappa_1^2}  ~.
\end{eqnarray}
In particular, provided that $\kappa_1 > 0$, we immediately obtain that $R$ is bigger (smaller) than 0 if and only if the second-order cumulant ratio $\kappa_2/\kappa_1$ is bigger (smaller) than 1, which can be alternatively expressed as in Eqs.\ (\ref{pair_distribution_kappa_2_1}) and (\ref{pair_distribution_kappa_2_2}).

We would like to stress that the above relations hold for an arbitrary distribution of particles, without any assumptions on the underlying physics, provided that the corresponding uncorrelated system can be described by the Poisson distribution. In any such system the sign of $[\widetilde{g}(r, \Delta r)/\widetilde{g}_0(r, \Delta r)] - 1$ at $r \to 0$ is the same as the sign of $(\kappa_2/\kappa) - 1$. In particular, it follows that $\widetilde{g}(r, \Delta r)/\widetilde{g}_0(r, \Delta r) < 1$ for systems where a repulsive interaction dominates at short distances (leading to a distribution more uniform than that of an ideal gas), while $\widetilde{g}(r, \Delta r)/\widetilde{g}_0(r, \Delta r) > 1$ for systems where an attractive interaction dominates at short distances (which leads to a distribution that is less uniform than that of an ideal gas).

\section{Parallel ensembles in \texttt{SMASH}}
\label{parallel_ensembles_in_SMASH}

The version of \texttt{SMASH} that we used did not have the option to run in a parallel ensembles mode (this option has been recently added to \texttt{SMASH} and is currently being tested). However, for simulations with all collision and decay channels turned off (such as we study in this paper), we can still employ the concept of parallel ensembles \textit{a posteriori}, that is at the analysis stage. Specifically, in each event we divide the $N_T N_B$ test particles obtained from a full ensemble \texttt{SMASH} simulation (where $N_B$ is the baryon number evolved in the simulation and $N_T$ is the number of test particles per particle) into $N_T$ separate groups. We then treat these groups as separate events. Each of these \textit{a posteriori} constructed events is governed by $P_{N_B}(N_i)$ (see Sec.\ \ref{Physical_baryon_number_distribution_function}). 

We note that for a \texttt{SMASH} simulation run in the full ensemble mode with $N_{\txt{ev}}$ events and $N_T$ test particles per particle, the corresponding calculation in the parallel ensembles mode will be characterized by $N_T N_{\txt{ev}}$ events with $N_T = 1$ test particles per particle.

\begin{figure*}[t]
	\includegraphics[width=\textwidth]{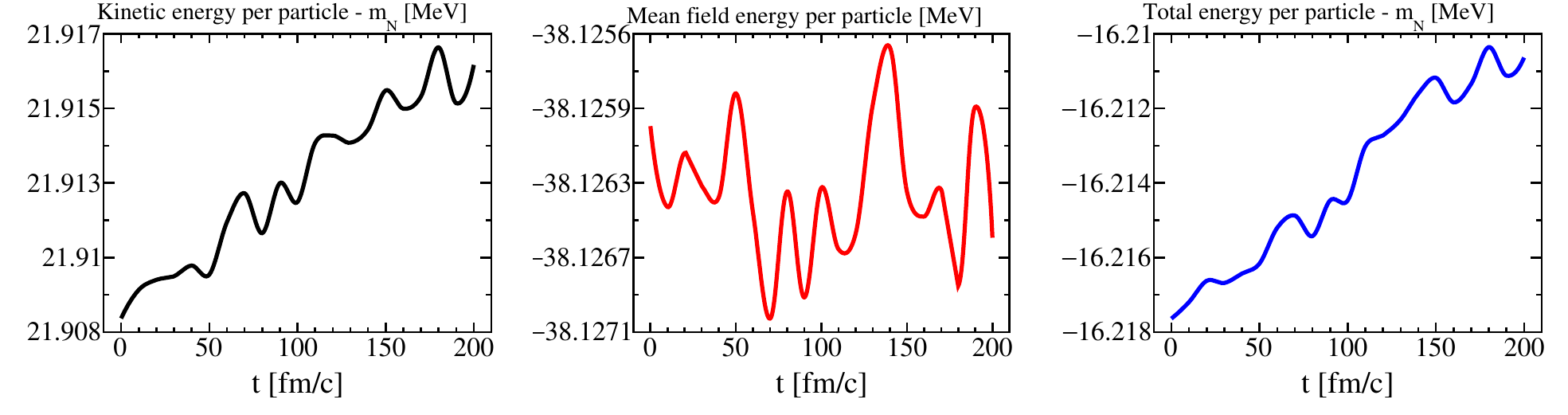}
	\caption{Time evolution of kinetic energy per particle (left panel), mean-field energy per particle (middle panel), and total (binding) energy per particle (right panel) for a system initialized at nuclear saturation density $n_0 = 0.160 \ \txt{fm}^{-3}$ and temperature $T = 1 \ \txt{MeV}$. The binding energy per particle at initialization, $E_B (t=0) \approx - 16.218 \ \txt{MeV}$, is within 0.1\% from the value expected from model calculations, and the readout of these quantities from the mean-field lattice. The mean-field energy oscillates slightly throughout the evolution, reflecting local fluctuations in density, but its average value remains the same. The increase in kinetic energy per particle in time, which also causes the increase of total energy per particle in time, is an unwanted feature of the simulation. Slight violation of the conservation of energy is a common feature of many hadronic transport codes, and is connected to the choice of the integration method for the equations of motion, as well as to details of density and density gradient calculations (see text for more details).}
	\label{energy_graphs}
\end{figure*}

\section{Energy evolution}
\label{energy_evolution}

Theory predicts that the total (binding) energy per particle at the saturation point should be $E_B = \frac{\mathcal{E}}{n_B} \bigg|_{\substack{T =1~[\txt{MeV}] \\n_B=n_0  }} - m_N = -16.23~\txt{MeV}$, on average. Further, conservation of energy demands that the total energy in the system, and consequently the total energy per particle, be conserved. In Fig.\ \ref{energy_graphs}, we show the energy evolution of a system initialized at the saturation density of nuclear matter, $n_B = n_0$, and temperature $T = 1 \ \txt{MeV}$. The left panel shows the kinetic energy per particle, the middle panel shows the mean-field energy per particle, and the right panel shows the total (binding) energy per particle. The binding energy per particle at initialization is found to be within 0.1\% from the expected value, $E_B (t=0) \approx -16.218~\txt{MeV}$. The mean-field energy is found to oscillate slightly throughout the evolution, reflecting local fluctuations in density, but its average value remains the same, which is what we expect. An unwanted feature of the simulation is the increase in kinetic energy per particle in time, which is also what causes the increase of total energy per particle in time. This unphysical gain in energy is a feature of many hadronic transport codes, and is connected to the choice of the integration method for the equations of motion, as well as to details of density and density gradient calculations, and the readout of these quantities from the mean-field lattice. The spurious contributions to the kinetic energy depend particularly strongly on statistical noise fluctuations in the magnitude of local density gradients, and one of the main reasons for using a significant number of test particles per particle, $N_T$, is suppressing unphysical density fluctuations due to the finite number of particles. While there exist methods of ensuring exact energy conservation in non-relativistic systems \cite{Wang:2019ghr}, we are unaware of generalizations of such methods applicable to relativistic transport codes. In view of this, some level of energy conservation violation will always be present in our simulations. 

\begin{table}[b]
	\caption{A summary of average unphysical gains in energy per particle, $\Delta \left(\frac{E}{N} \right)$, for infinite matter simulations pertaining to different points on the phase diagram. For each simulation, the side length of the box was set at $L = 10 \ \txt{fm}$ and the lattice spacing was chosen at $a = 1 \ \txt{fm}$. The results were averaged over ten events. The dependence of $\Delta \left(\frac{E}{N} \right)$ on the initialization point is evident. Additionally, the number of test particles per particle $N_T$ and the time step $\Delta t$ are also shown to play a role. See text for more details.}
	\label{energy_non-conservation}
	\begin{center}
		\bgroup
		\def\arraystretch{1.4}
		\begin{tabular}{c c c c c c}
			\hline
			\hline
			%\cline{2-6}
			$n_B\ [n_0]$ & $T\ [\txt{MeV}]$  & $t_{\txt{end}}\ [\txt{fm}/c]$ & $N_T$ & $\Delta t\ [\txt{fm}/c]$   & 
			$\Delta\left(\frac{E}{N} \right)\ [\txt{MeV}]$\\ 
			\hline
			0.25 & 1 & 200 & 20 & 0.1 &  2.291 \\
			0.25 & 1 & 200 & 200 & 1.0  &  1.516 \\
			0.25 & 1 & 200 & 200 & 0.1 &  1.411 \\
			0.25 & 1 & 200 & 200 & 0.01  & 1.393  \\
			0.25 & 1 & 200 & 500 & 1.0  &  1.315 \\
			0.25 & 25 & 200 & 200 & 1.0  &  1.135$\times 10^{-4}$ \\
			1.0 & 1 & 200 & 200 & 1.0  &  5.684$\times 10^{-6}$ \\
			3.0 & 1 & 50 & 10  & 0.1  & 1.615 \\  
			%3.0 & 1 & 50 & 50 & 1.0   & 0.626 \\  
			3.0 & 1 & 50 & 50 & 0.1    & 0.542   \\  
			%3.0 & 1 & 50 & 50 & 0.01   & 0.530 \\
			3.0 & 1 & 50 & 100& 0.1  & 0.420 \\  
			3.0 & 125 & 50 & 50 & 0.1  & 1.373$\times 10^{-4}$  \\  
			\hline
			\hline
		\end{tabular}
		\egroup
	\end{center}
\end{table}

The degree of energy conservation violation shown in Fig.\ \ref{energy_graphs} is negligible; this is the case because the system in question is initialized in equilibrium, where mean-field forces are small. However, in general the issue can become much more troublesome. A summary of average unphysical gains in energy per particle, $\Delta \left(\frac{E}{N} \right)$, for simulations pertaining to different points on the phase diagram, is included in Table \ref{energy_non-conservation}. Generally, contributions to $\Delta \left(\frac{E}{N} \right)$ are larger for systems initialized in regions of the phase diagram where forces acting on test particles are large, e.g., inside the spinodal region of a phase transition (and especially in the spinodal region of the nuclear phase transition, where density gradients tend to be very large). Conversely, energy conservation is very satisfactory when forces acting on test particles are small, e.g., in regions of the phase diagram where nuclear matter is thermodynamically stable, and in particular at the saturation point of nuclear matter. Additionally, $\Delta \left(\frac{E}{N} \right)$ depends on the number of test particles $N_T$ and time step $\Delta t$. In general, a larger number of test particles per particle and a smaller time step lead to a better energy conservation; however, they also lead to a significant increase in the simulation time. Thus greater accuracy needs to be balanced with practical considerations.

\bibliography{SMASH_BoxModus_paper_1_PRC_v3,non_inspire_AS}

\end{document}